\documentclass[aps,twocolumn,groupaddress,usenatbib,showkeys,showpacs]{revtex4-1}
\usepackage{graphicx}
\usepackage{dcolumn}
\usepackage{amssymb}
\usepackage{amsfonts}
\usepackage{amsbsy}
\usepackage{color}
\usepackage{rotating}
\usepackage[english]{babel}
\usepackage{epsfig}

\begin{document}

\title{A comprehensive cosmographic analysis by Markov Chain Method}

\author{S. Capozziello,}
\affiliation{Dipartimento di Scienze Fisiche, Universita' degli Studi di
Napoli ``Federico II" and  INFN, Sezione di Napoli, Complesso
Universitario di Monte S. Angelo, Via Cinthia, Edificio N, 80126
Napoli, Italy}
\author{R. Lazkoz, and V. Salzano,}
\affiliation{Fisika Teorikoaren eta Zientziaren Historia Saila, Zientzia eta Teknologia Fakultatea, \\ Euskal
Herriko Unibertsitatea, 644 Posta Kutxatila, 48080 Bilbao, Spain}

\begin{abstract}
We study the possibility to extract model independent information
about the dynamics of the universe by using Cosmography. We intend
to explore it systematically, to learn about its
limitations and its real possibilities. Here we are sticking  to
the series expansion approach on which Cosmography is based. We
apply it to different data sets: Supernovae Type Ia (SNeIa),
Hubble parameter extracted from differential galaxy ages,
Gamma Ray Bursts (GRBs) and the Baryon Acoustic Oscillations (BAO)
data. We go beyond past results in the
literature extending the series expansion up to the fourth order
in the scale factor, which implies the analysis of the deceleration,
$q_{0}$, the jerk, $j_{0}$ and the snap,  $s_{0}$. We use the
Markov Chain Monte Carlo Method (MCMC) to analyze the data
statistically. We also try to relate direct results from
Cosmography to dark energy (DE) dynamical models parameterized by
the Chevalier-Polarski-Linder (CPL) model, extracting clues about
the matter content and the dark energy parameters. The main results
are: $a.$ even if relying on a mathematical approximate assumption
such as the scale factor series expansion in terms of time,
cosmography can be extremely useful in assessing
dynamical properties of the Universe; $b.$ the deceleration parameter
clearly confirms the present acceleration phase; $c.$ the MCMC method
can help giving narrower constraints in parameter estimation, in
particular for higher order cosmographic parameters (the jerk and
the snap), with respect to the literature; $d.$ both the
estimation of the jerk and the DE parameters, reflect the
possibility of a deviation from the $\Lambda$CDM cosmological
model.
\end{abstract}


\maketitle

\section{Introduction}
\label{sec:intro}

Cosmography is a branch of cosmology which has not been
explored so much yet in spite of its promising prospects towards a
deeper  understanding of  the acceleration of the universe.
Thus, there remain caveats to be filled and misunderstandings
to be rectified before one can take full advantage of the
possibilities offered by Cosmography.

This approach proceeds by making minimal dynamical assumptions,
namely, it does not assume any particular form of the Friedmann
equations and only relies on the assumption that the spacetime
geometry is well described  on large
scales by the Robertson-Walker metric. This makes cosmography fully model and setting independent, in the
sense that it is only at the late stages of interpretation of the
results offered by this approach that one has to care for the
theoretical framework in which the conclusions fit best (dark
energy, dark gravity, modified gravity, etc.).

This is clearly important, for instance, when
one attempts at comparing the $\Lambda$CDM model (or any dark
energy model) with alternative theories of gravity (such as $f(R)$
gravity). In the traditional approach a parameterized model is
assumed at the beginning, and then, by contrasting it against the
data, it is possible to check its viability and to constrain its
characterizing parameters. But this approach is clearly a model
dependent one so that some doubts always remain on the validity of
the constraints on derived quantities as the present day values of
the deceleration parameter and the age of the universe, just to
mention a couple of them. Conversely, the constraints that one can
infer from cosmography are more universal since they are
completely model-free. If well understood and analyzed in full details,
weighting its pros and cons, results from Cosmography can be rightly
considered as milestones in the study of properties of universe dynamics,
which any theoretical model has to consider and satisfy, and can help
pointing in the right direction in connection with the properties of the Universe. \\

Of course, everything is not so clearly cut. The fundamental rule
of cosmography is based on the series expansion of the scale
factor, $a(t)$. From it one obtains the cosmographic parameters
usually referred to as the \textit{Hubble} ($H$),
\textit{deceleration} ($q$), \textit{jerk} ($j$), \textit{
snap} ($s$) and \textit{lerk} ($l$) parameters, which we define
below. However, the series of $a(t)$ places difficulties almost
from the very beginning. As discussed in \cite{Cattoen07}, there
are two main problems when working with series expansions: the
convergence and the truncation of the series. Those authors show
it is possible to solve the convergence problem defining a new
redshift variable, the so called \textit{y-redshift}:
\begin{equation}
z \rightarrow y = \frac{z}{1+z}~.
\end{equation}
For a series expansion in the classical z-redshift the
convergence radius is equal to $1$. This represents a drawback
when one wants to extend the application of cosmography to
redshift $z>1$, which is not such a big past extension, if we
consider that current supernovae data go up to $z = 1.4$, or that
the CMB analysis involves the decoupling redshift $z \approx
1089$. It would be pleasant to be able to work correctly at least
up to the highest redshift demanded by the observational data set
one wants to use.

The y-redshift could potentially solve this problem because the z-interval
$[0,\infty]$ corresponds to the y-interval $[0,1]$, so that we are
mainly inside the convergence interval of the series, even for CMB
data ($z = 1089 \rightarrow y = 0.999$). So, in principle, we
could extend the series up to the decoupling redshift value, and
one could place CMB related constraints within the cosmographic
approach. Unfortunately,  some problems arise and we will discuss them
below. Leaving aside this limitation, the possibilities offered by
the y-redshift can be considered as a step in the right direction, because,
obviously, defining a new redshift does not change the physical
content of the data.

But the problem of the truncation of the series is not solved; of
course, we expect that for low redshifts a low order series
expansion will work well, while for higher redshifts it is likely
that higher order expansions of any physical quantity of relevance
will be required. But, is this true? And what is the good order
for any redshift? Is there an even approximate relation which
correlates redshift with series order?

This paper represents an improvement in this direction. As
considering a higher  order of expansion enforces taking into account
more free parameters, it seems a priori it will be difficult to obtain tight
constraints on them if the discriminating power of the data sets
used is weak or not fully exploited statistically. Despite this, we
successfully put bounds on the parameters of the third order series in a
statistically  consistent way, and so make a considerable improvement with respect to
\cite{Cattoen07b}, where for the same redshift extent the authors only managed to put
constraints on the coefficients of a second order series. This is one of the major contributions
of our paper.

In addition, there is  another matter on  which this paper enhances the worth of
the cosmographic approach:
 the values and the errors of the
cosmographic parameters. Estimations of that sort can be found in
the literature, not only in \cite{Cattoen07b}, but also in
\cite{Visser04} and \cite{John05}. While the former used
supernovae data, in related works like \cite{Poplawski06,Poplawski07,Salzano09,Mariam10,Mariusz10}
one can find theoretical estimations (in the
context of $f(R)$ theories in both metric and Palatini approaches)
which could be related to estimations from observational data.  Some
 recent contributions to this topic are the works
\cite{Capozziello08,Izzo09},  where  gamma ray burst data
have been used. Here we analyze at the same time data from SNeIa and
GRBs, and combine them with a sample of Hubble parameters values extracted
by means of the differential galaxy ages, and BAO data.
The use of the Hubble parameter data, in particular, introduces additional
advantages we will refer to later.

The main problem with earlier contributions to the topic is
that the error bars obtained were considerably large. Due to this fact, the
statistical conclusions to be inferred are rather weak and, in
consequence, they hardly provide valid priors for future analysis.
For example, from \cite{John05} we have $q_0 = -0.90 \pm 0.65$, $j_0 = 2.7 \pm 6.7$, $s_0 = 36.5 \pm
52.9$, $l_0 = 142.7 \pm 320$; some of the relative errors
are as high as about a  $200\%$ (as usual, the $0$ subindex denotes the $z=0$ values of
the parameters). Now, we can expect that errors on
the coefficients will get bigger when higher order expansions are
considered. As those parameters are correlated among them,
errors in the low order series coefficients propagate to the
additional coefficients included in the higher order  series. If
these errors turn out to be big, the positive aspects of
cosmography will vanish. For instance, if the error on $q_{0}$
allows for a positive value of the deceleration parameter in the
$3\sigma$ confidence interval, it will not be so obvious that the
Universe is accelerating (see again \cite{Cattoen07b}). Moreover,
the $\Lambda$CDM model enforces $j_{0} = 1$ (this can be easily
derived from the definition of $j_0$, which is to be found in \S~(\ref{sec:comparing}));
but if the error bar is as big as reported above (from
\cite{John05}), it will not be possible to confront this model
with competing ones.

Given those problems we have just discussed,  an
important question arises: is it not possible to give narrower
statistical constraints? To this end we apply a Markov Chain Monte Carlo
(MCMC) method: it allows us obtaining marginalized
likelihoods on the series coefficients from which we infer
rather tight constraints on those parameters. The reason for
this considerable improvement with respect to earlier works is
that in our code we have implemented several controls which give us power over
any physical requirement we expect from our theoretical
apparatus. Actually,  setting  restrictions on the Hubble
parameter is possible because we use data related to this quantity
that reveal to allow a considerable improvement in the quality of constraints.

We have organized the paper as follows: in \S~(\ref{sec:cosmopar}) and Appendix)
we define the cosmographic parameters and give all the relations needed;
in \S~(\ref{sec:comparing}) we describe some caveats that have to be taken in mind when working
with Cosmography; in \S~(\ref{sec:Obstest}) we
describe the observational data used for the analysis; in \S~(\ref{sec:cosmo_analysis}) and \S~(\ref{sec:CPL})
we present our main results and discuss their
meaning and consequences; finally, in \S~(\ref{sec:Conclusion})
we summarize.

\section{Cosmographic Parameters}
\label{sec:cosmopar}

The key rule in cosmography is the Taylor series expansion of the
scale factor with respect to the cosmic time. To this aim, it is
convenient to introduce the following functions:
{\setlength\arraycolsep{0.2pt}
\begin{eqnarray}
H(t) &=& \frac{1}{a}\frac{da}{dt} \; , \nonumber
\\
q(t) &=& - \frac{1}{a}\frac{d^{2}a}{dt^{2}}\frac{1}{H^{2}} \; , \nonumber
\\
\label{eq:cosmopar}
j(t) &=&  \frac{1}{a}\frac{d^{3}a}{dt^{3}}\frac{1}{H^{3}} \; ,
\\
s(t) &=&  \frac{1}{a}\frac{d^{4}a}{dt^{4}}\frac{1}{H^{4}} \; , \nonumber
\\
l(t) &=&  \frac{1}{a}\frac{d^{5}a}{dt^{5}}\frac{1}{H^{5}} \; . \nonumber
\end{eqnarray}}
They are usually referred to as the \textit{Hubble},
\textit{deceleration}, \textit{jerk}, \textit{snap} and
\textit{lerk} parameters, respectively (see \cite{Dunajski08} for
an historical account of these names). Using these definitions it
is easy to write the fifth order Taylor expansion of the scale
factor: {\setlength\arraycolsep{0.2pt}
\begin{eqnarray}\label{eq: aseries}
\frac{a(t)}{a(t_{0})} & = & 1 + H_{0} (t-t_{0}) -
\frac{q_{0}}{2} H_{0}^{2} (t-t_{0})^{2} + \frac{j_{0}}{3!}
H_{0}^{3} (t-t_{0})^{3}  \nonumber \\ & + & \frac{s_{0}}{4!}
H_{0}^{4} (t-t_{0})^{4}+ \frac{l_{0}}{5!} H_{0}^{5}
(t-t_{0})^{5} +\emph{O}[(t-t_{0})^{6}]
\end{eqnarray}}
with $t_0$ being the current age of the universe. Note that
Eq.~(\ref{eq: aseries}) is also the fifth order expansion of $(1 +
z)^{-1}$, as the definition for the redshift $z$ is $z :=
a(t_0)/a(t) - 1$. This is what we need for developing
the cosmographic apparatus; for sake of clearness we remind all the
detailed calculations to the appendix section, while giving here only the
main results.

\subsection{Luminosity distance series}
\label{sec:dL series}

Since we are going to use SNeIa and GRBs data, it
will be useful to give the Taylor series of the expansion
of the distance modulus, which is the quantity about which those
observational data typically inform. The final expression for the distance
modulus based on the Hubble free luminosity distance (Eqs.~(\ref{eq: dlseries})~-~(\ref{eq: dlseries2})),
$\mu(z) = 5 \log_{10} d_{L}(z)+\mu_0$, is the following:
{\setlength\arraycolsep{0.2pt}
\begin{eqnarray}\label{eq:museries}
\mu(z) &=& \frac{5}{\log 10} \left( \log z + \mathcal{M}^{1} z
+ \mathcal{M}^{2} z^2 + \mathcal{M}^{3} z^{3}  \mathcal{M}^{4}
z^{4} \right)\nonumber\\
&&+\mu_0\, ,
\end{eqnarray}}
with {\setlength\arraycolsep{0.2pt}
\begin{eqnarray}
\mathcal{M}^{1} &=& - \frac{1}{2} \left[ -1 + q_{0} \right] \, ,
\nonumber \\
\mathcal{M}^{2} &=& - \frac{1}{24} \left[7 - 10 q_{0} - 9q_{0}^{2}
+ 4j_{0} \right] \, ,\nonumber \\
\mathcal{M}^{3} &=& \frac{1}{24}\left[5 - 9 q_{0} - 16 q_{0}^{2} -
10 q_{0}^{3} + 7 j_{0} + 8 q_{0} j_{0} + s_{0} \right] \, ,
\nonumber
\\
\mathcal{M}^{4} &=& \frac{1}{2880}\left[-469 + 1004 q_{0} + 2654
q_{0}^{2} + 3300 q_{0}^{3} + 1575 q_{0}^{4}  \right. \nonumber \\
&+& \left. 200 j_0^{2} - 1148 j_{0} -2620 q_{0} j_{0} - 1800 q_{0}^{2}
j_{0} - 300 q_{0} s_{0}  \right. \nonumber \\
&-& \left. 324 s_{0} - 24 l_{0} \right] \, . \nonumber
\end{eqnarray}}
Of course we have also derived the same relations for the y-redshift,
but we relegate them to the appendix.

\subsection{Hubble parameter series}
\label{sec:H series}

The definition of the luminosity distance given by
Eq.~(\ref{eq:dldef}), can be presented in this other way:
\begin{equation}\label{eq:dl_H}
D_{L}(z) = c \ (1+z) \ \int_{0}^{z} \mathrm{d}z' \frac{1}{H(z')}.
\end{equation}
It is interesting to study the possibility of obtaining the same
final expression for the  distance defined starting from a Taylor
series expansion of the Hubble parameter instead of the scale factor,
namely:
{\setlength\arraycolsep{0.2pt}
\begin{eqnarray}\label{eq:Hseriesdef}
H(z) &=& H_{0} + \frac{dH}{dz}\Bigg{|}_{z=0} z + \frac{1}{2!}
\frac{d^{2}H}{dz^{2}}\Bigg{|}_{z=0} z^{2} + \frac{1}{3!}
\frac{d^{3}H}{dz^{3}}\Bigg{|}_{z=0} z^{3} \nonumber \\ &+&
\frac{1}{4!} \frac{d^{4}H}{dz^{4}}\Bigg{|}_{z=0} z^{4} +
\emph{O}(z^{5}) \, .
\end{eqnarray}}
This series expansion will be also an important rule in the
definition of our Markov chains algorithm given the observational data we use. To compute all the terms
of this series we have to keep in mind the derivation rule (for a
clearer notation we will suppress the redshift dependence of the Hubble
parameter setting $H \equiv H(z)$):
\begin{equation}\label{eq:firstHt}
\frac{d}{dt} = -(1 + z) H \frac{d}{dz}\,.
\end{equation}
It is an easy but cumbersome task to obtain from the latter the
higher-order time-derivatives of the Hubble parameter (they are given
in \S~(\ref{sec:derivativesH}). Then, from the definitions of the cosmographic
parameters, Eq.~(\ref{eq:cosmopar}), it is easy to demonstrate the
following relations:
{\setlength\arraycolsep{0.2pt}
\begin{eqnarray}\label{eq: hdot}
\dot{H} = -H^2 (1 + q) \, ,
\end{eqnarray}
\begin{eqnarray}\label{eq: h2dot}
\ddot{H} = H^3 (j + 3q + 2)\, ,
\end{eqnarray}
\begin{eqnarray}\label{eq: h3dot}
\frac{d^{3}H}{dt^{3}} = H^4 \left [ s - 4j - 3q (q + 4) - 6 \right ]\, ,
\end{eqnarray}
\begin{eqnarray} \label{eq: h4dot}
\frac{d^{4}H}{dt^{4}}& = & H^5 \left [ l - 5s + 10 (q + 2) j\right.\nonumber\\
&&\left. + 30 (q + 2) q +
24 \right ]
\end{eqnarray}}
If we convert time-derivatives into derivatives with respect to redshift
using Eqs.~(\ref{eq:firstHt})~-~(\ref{eq:secondHt})~-~(\ref{eq:thirdHt})~-~(\ref{eq:fourthHt})
we have:{\setlength\arraycolsep{0.2pt}
\begin{eqnarray}\label{eq:dH2z1}
\frac{dH}{dz} &=& \frac{H}{1 + z} \left(1 + q\right)\, ,
\end{eqnarray}
\begin{eqnarray}\label{eq:dH2z2}
\frac{d^{2}H}{dz^{2}} &=& \frac{H}{(1 + z)^{2}} \left(- q^{2} +
j\right) \, ,
\end{eqnarray}
\begin{eqnarray}\label{eq:dH2z3}
\frac{d^{3}H}{dz^{3}} &=& \frac{H}{(1 + z)^{3}} \left(3 q^{2} + 3
q^{3} -4 q j - 3 j -s \right) \, ,
\end{eqnarray}
\begin{eqnarray}\label{eq:dH2z4}
\frac{d^{4}H}{dz^{4}} &=& \frac{H}{(1 + z)^{4}} \left(- 12 q^{2} -
24 q^{3} - 15 q^{4} + 32 q j + 25 q^{2} j  \right. \nonumber \\
&+& \left. 7 q s + 12 j - 4 j^{2} + 8 s + l\right) \, .
\end{eqnarray}}
Exactly the same luminosity distance formula as given in Eq.~(\ref{eq: dlseries})
can be recovered after a series of steps: after having evaluated Eqs.~
(\ref{eq:dH2z1})~-~(\ref{eq:dH2z4}) at $z = 0$ (namely $t_{0}$), we
must insert them in Eq.~(\ref{eq:Hseriesdef}), and then insert
this one  in Eq.~(\ref{eq:dl_H}). Then we must Taylor expand the
integration function, integrate it, and we will have reached the
final sought result and the coincidence will be complete.

This is all, so far, in what the basic setup is concerned, in the
next section we move on to applications of these definitions.

\section{Series orders comparison}
\label{sec:comparing}

As we have said in the introductory section, the first question we
would try to answer is whether there is a relation between the
highest expansion order in the Taylor series and the redshift
range where this series can be applied. What one could expect is
the rough rule that we would need a series expansion truncated at
higher orders when increasing up the redshift range. This would be
a mathematical requirement, of course. But what about physics, our
main interest, and the fit to the Hubble SNeIa diagram; namely,
the reproduction of the luminosity distance or its corresponding
distance modulus? We have found that the answer seems to be not so
obvious.

Let us start by writing some useful relations for successive results.
To make the discussion more specific we will use the Chevallier\,-\,Polarski\,-\,Linder (CPL)
(\cite{Chevallier01},\cite{Linder03}) parametrization:
\begin{equation}
w = w_0 + w_{a} (1 - a) = w_0 + w_{a} z (1 + z)^{-1} \,
,\label{eq:cpleos}
\end{equation}
so that, in a spatially flat universe filled with cold dark matter and dark energy,
the dimensionless Hubble parameter $E(z) = H/H_0$ reads\,:
\begin{equation}
E^2(z) = \Omega_m (1 + z)^3 + \Omega_X (1 + z)^{3(1 + w_0 + w_{a})}
{\rm e}^{-\frac{3 w_{a} z}{1 + z}}\, . \label{eq: ecpl}
\end{equation}
with $\Omega_X = 1 - \Omega_m$ because of the flatness assumption.
In order to determine the cosmographic parameters for such a
model, we avoid integrating $H(z)$ to get $a(t)$ by noting that
$d/dt = -(1 + z) H(z) d/dz$. We can use such a relation to
evaluate $(\dot{H}, \ddot{H}, d^3H/dt^3, d^4H/dt^4)$ and convert
Eqs.~(\ref{eq: hdot})~-~(\ref{eq: h4dot}) to redshift derivatives, i.e.:
{\setlength\arraycolsep{0.2pt}
\begin{eqnarray}\label{eq:cosmo_CPL}
q(z) &=& \frac{(1+z)}{2\,H^{2}(z)}\frac{dH^{2}(z)}{dz} - 1\, , \nonumber \\
j(z) &=& \frac{(1+z)^{2}}{2\,H^{2}(z)}\frac{d^{2}H^{2}(z)}{dz^{2}}
- \frac{(1+z)}{H^{2}(z)}\frac{dH^{2}(z)}{dz} + 1 \, .
\end{eqnarray}}
For the snap and the lerk the expressions are much longer and involve respectively
the third and the fourth order derivatives of $H^{2}(z)$. We underline that these
expressions for cosmographic parameters are exact.
Given an analytical expression for $H(z)$, we can evaluate them at $z= 0$,
and solve them with respect to the parameters of interest. \\
Some algebra finally gives:
{\setlength\arraycolsep{0.2pt}
\begin{equation}
q_0 = \frac{1}{2} + \frac{3}{2} (1 - \Omega_m) w_0 \ ,
\label{eq:qzcpl}
\end{equation}
\begin{equation}
j_0 = 1 + \frac{3}{2} (1 - \Omega_m) \left [ 3w_0 (1 + w_0) + w_{a}
\right ] \ , \label{eq:jzcpl}
\end{equation}
\begin{eqnarray}
s_0 & = & -\frac{7}{2} - \frac{33}{4} (1 - \Omega_m) w_{a} \nonumber
\\ ~ & - & \frac{9}{4} (1 - \Omega_m) \left [ 9 + (7 - \Omega_m)
w_{a} \right ] w_0 - \frac{9}{4} (1 - \Omega_m)  \nonumber \\
~ & \times & (16 - 3\Omega_m) w_0^2 \frac{27}{4} (1 - \Omega_m) (3
- \Omega_m) w_0^3 \ , \label{eq:szcpl}
\end{eqnarray}
\begin{eqnarray}
l_0 & = & \frac{35}{2} + \frac{1 - \Omega_m}{4} \left [ 213 + (7 -
\Omega_m) w_{a} \right ] w_{a} \nonumber \\ ~ & + & \frac{(1 -
\Omega_m)}{4}
\left [ 489 + 9(82 - 21 \Omega_m) w_{a} \right ] w_0 \nonumber \\
~ & + & \frac{9}{2} (1 - \Omega_m) \left [ 67 - 21 \Omega_m +
\frac{3}{2} (23 - 11 \Omega_m) w_{a} \right ] w_0^2 \nonumber \\ ~ &
+ & \frac{27}{4} (1 - \Omega_m) (47 - 24 \Omega_m) w_0^3 \nonumber
\\ ~ & + & \frac{81}{2} (1 - \Omega_m) (3 - 2\Omega_m) w_0^4 \ .
\label{eq:lzcpl}
\end{eqnarray}}
From Eq.~(\ref{eq: ecpl}) we can derive the exact analytical
expression for the Hubble free luminosity distance and the
distance modulus. Then, we will compare it with its expression as
a cosmographic series, i.e. a series depending on the cosmographic
parameters. We have considered the cosmographic series with two
($q_{0}$,\,$j_{0}$), three ($q_{0}$,\,$j_{0}$,\,$s_{0}$) and four
parameters ($q_{0}$,\,$j_{0}$,\,$s_{0}$,\,$l_{0}$). Of course,
from any CPL scenario we will be able to derive the corresponding
set of cosmographic parameters using
Eqs.~(\ref{eq:qzcpl})~-~(\ref{eq:lzcpl}).

\begin{figure}
\centering
\includegraphics[width=84mm]{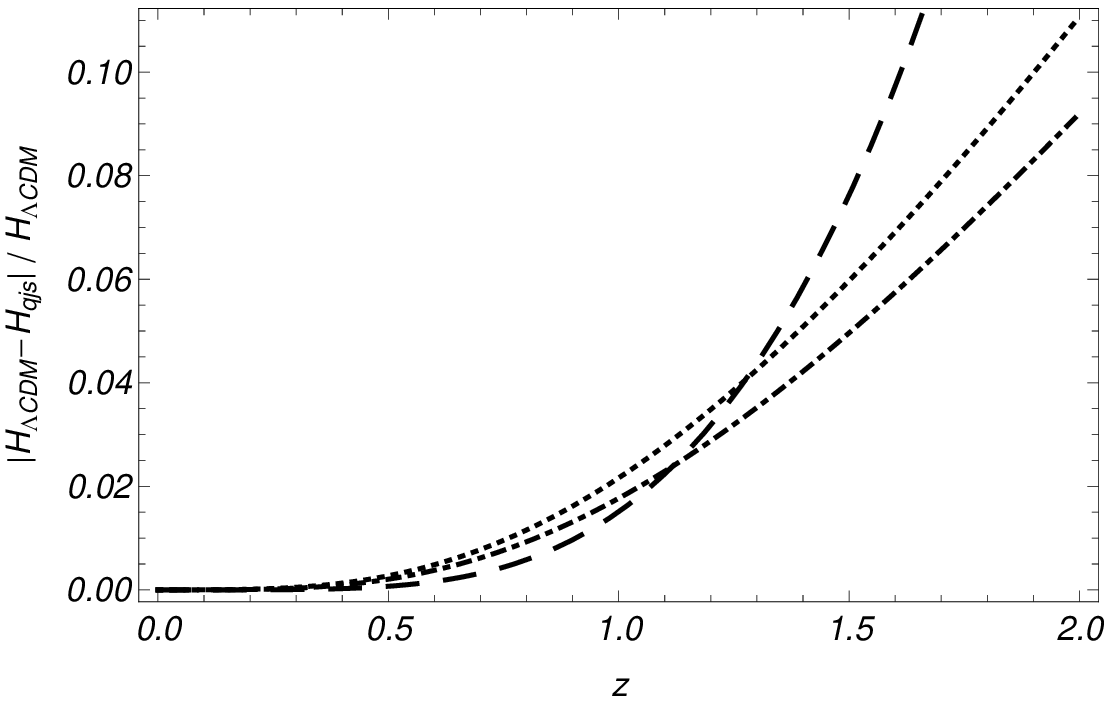}
\includegraphics[width=84mm]{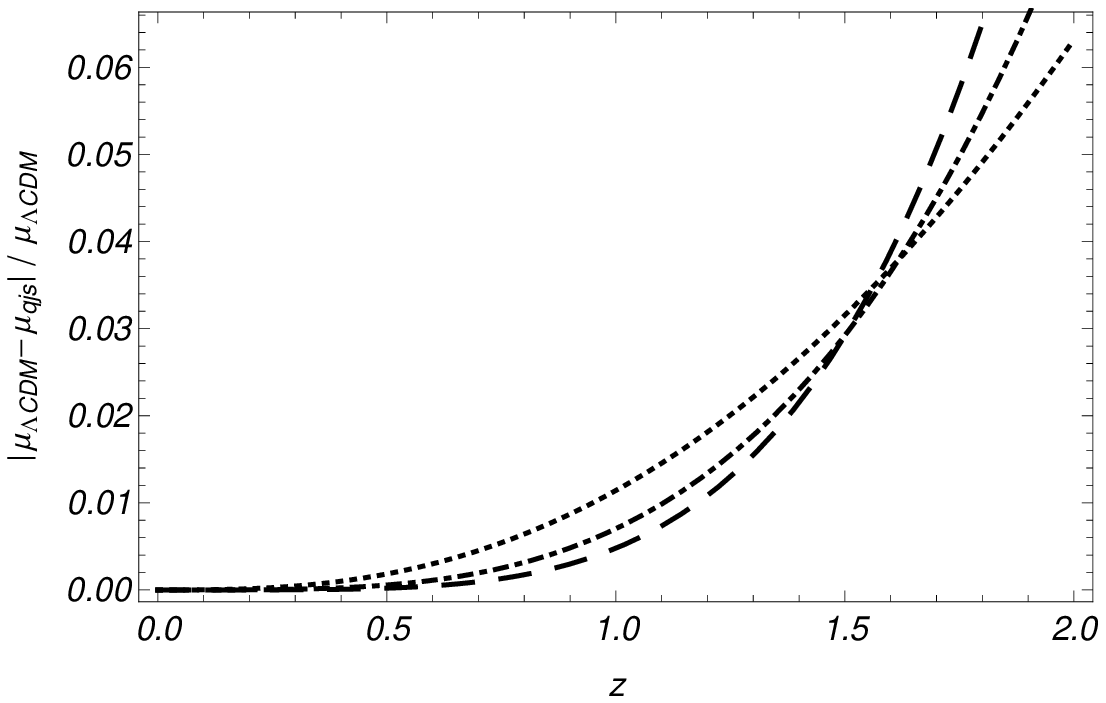}
\caption{(\textit{Top.}) Residuals between the CPL exact Hubble
function and different-order series expansions of the same
quantity in the $\Lambda$CDM case. The redshift range is $0<z<2$.
(\textit{Bottom.}) The same as above picture but for distance modulus.
One might except that the value of the denominator,
$\mu_{\Lambda CDM}$, depends on the nuisance parameter $\mu_0$, i.e. on the value
of the Hubble constant, $H_{0}$. But we have checked that changing $H_{0}$
in the range $[0.65,0.75]$, which is large enough to contain all the acceptable
physical values of $H_{0}$, does not produce any sensible change in the figures.}  \label{fig:Lambda_H_Mu_Delta}
\end{figure}

\begin{figure}
\includegraphics[width=84mm]{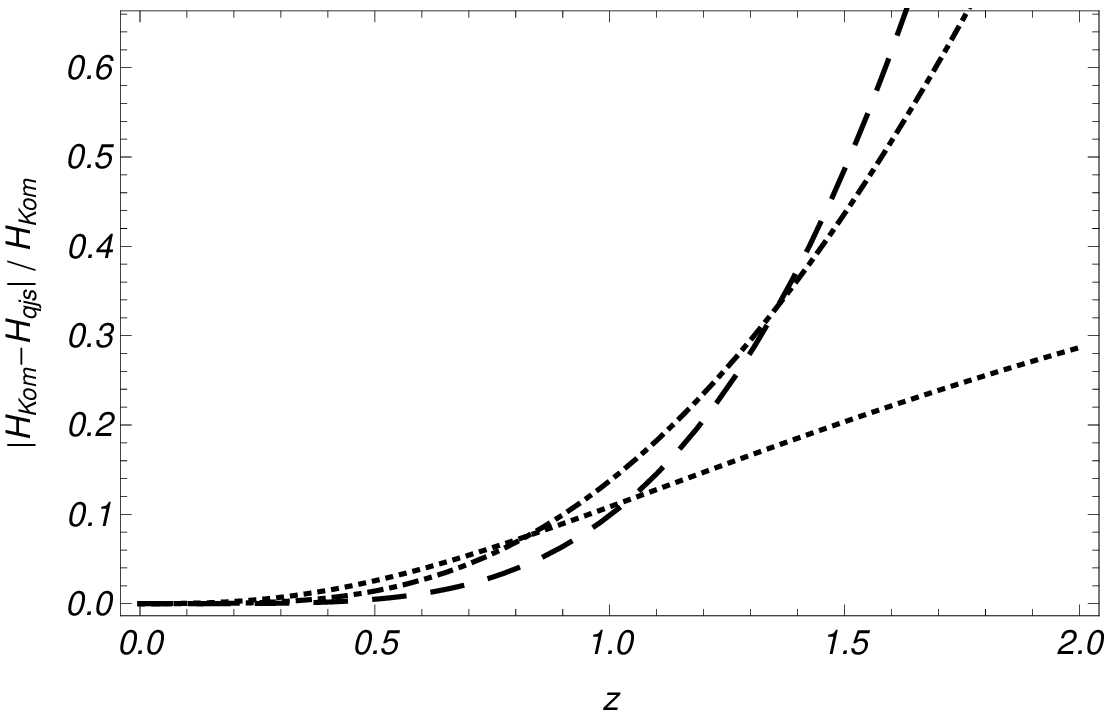}
\includegraphics[width=84mm]{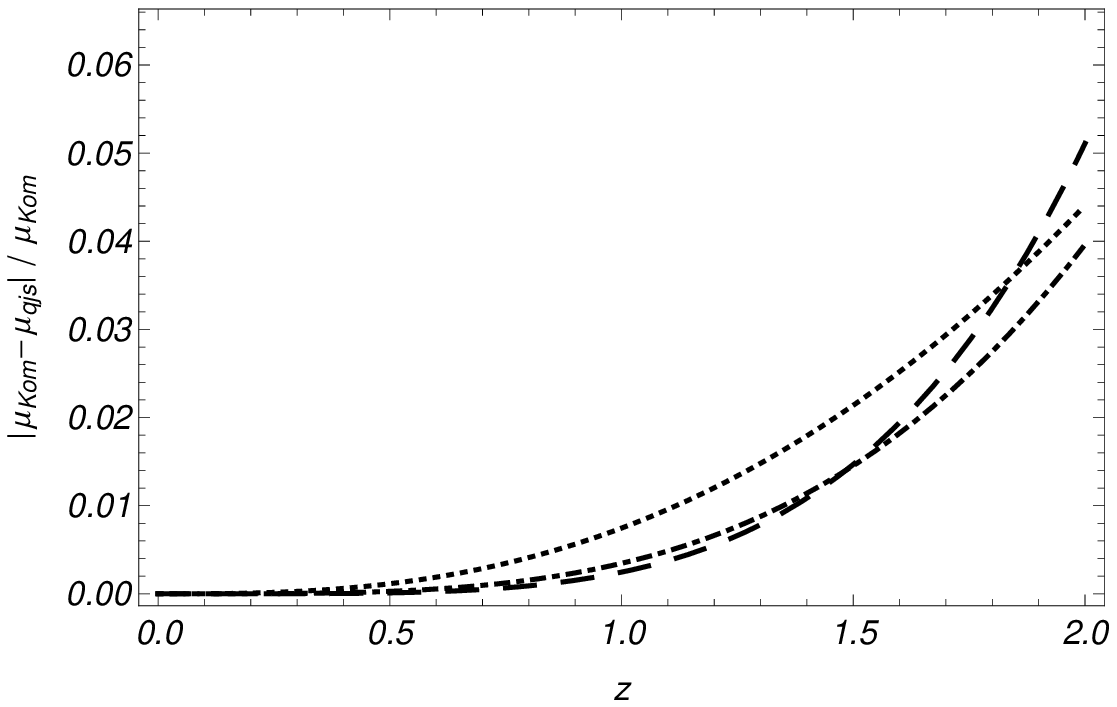}
\caption{(\textit{Top.}) Residuals between the CPL exact Hubble
function and different-order series expansions of the same
quantity in the model from \cite{Komatsu10}. The redshift range is $0<z<2$.
(\textit{Bottom.}) The same as above picture but for distance
modulus.} \label{fig:Komatsu_H_Mu_Delta}
\end{figure}

We have tested two different toy models: a $\Lambda$CDM model
($\Omega_{m} = 0.3 \, , \,w_{0} = -1 \, , \,w_{a} = 0$); and a
dynamical dark energy model with $\Omega_{m} = 0.245 \, , \,w_{0} = -0.93 \, ,
\,w_{a} = -0.41$ coming from \cite{Komatsu10}. In
Figs.~(\ref{fig:Lambda_H_Mu_Delta})~-~(\ref{fig:Komatsu_H_Mu_Delta})
we show the comparison in the redshift range $0<z<2$, which is the
range where cosmography has been mainly applied in past works,
regarding to the maximum available redshift of the SNeIa surveys
used in the literature.

As we can see in Fig.~(\ref{fig:Lambda_H_Mu_Delta}), it is not so
obvious that the series expanded at higher orders can describe
well the underlying Hubble parameter or distance modulus when applying
cosmography out of the convergence radius. There we
show the relative residuals with respect to the exact CPL
functions (i.e. the Hubble parameter and the distance modulus) coming from
the different order series in the $\Lambda$CDM case. As we can
see, inside the redshift convergence radius ($z<1$) the
four-parameter series effectively gives a better approximation to the
Hubble parameter and the distance modulus than other orders expansions.

But it is also clear that the rough rule (higher redshift, higher order series)
is somewhat not followed when out of the convergence interval, namely,
above $z = 1$. In this case residuals from
the four-parameter expansion grow with redshift faster than the
ones coming from the three and two-parameters ones.

The same analysis made with the dynamical model lead to similar
final conclusions and adds further interesting caveats which have
to be kept in mind when working with cosmography and using its
results. Looking at Fig.~(\ref{fig:Komatsu_H_Mu_Delta}) we can see
some consequences from changing the cosmological model:
\begin{itemize}
  \item the inversion in goodness between the order series
  expansions even inside the convergence radius. In the
  $H(z)$ case, we can see that in the $\Lambda$CDM model, for $z>1$,
  the two and the tree-parameters expansions were quite
  similar, with a slight preference for the latter one. In the
  dynamical dark energy model things go different, and the two-parameters
  expansion is strongly preferred;
  \item different magnitudes for residuals, notably in the Hubble
  function case, slightly in the distance modulus.
\end{itemize}

The first point leads us to think about the real effectiveness of
cosmography approach as a \textit{``model independent"} one.
Cosmography \textit{is} model independent, of course, starting
from its basic assumptions, and the information pieces one can extract
from it are model independent too; but data are tracers of a
cosmological model.

From the previous considerations it seems that there is an implicit
relation between the series expansion and the underlying
\textit{unknown} cosmological model which we want to read out from
data. If cosmography, by its order series expansion, is sensible
enough to detect and discriminate a model against another one, then the
choice of working with two, three or four parameters can be critical.

The second consideration is important for defining the main lesson
of this section. All the analysis we have done until now could be
useless without relating it to the experimental possibilities we
have nowadays. Namely: are our present data and their related
errors able to make us distinguish between a two, three, or
four-parameter expansion for the measured physical quantities we
are working with?

To answer this question we need a look at the data we are going
to use. If we consider the data from \cite{Stern09}, we see that
the relative errors on $H(z)$ range in the interval $[0.10,0.62]$,
which is clearly larger than the residual we have plotted in
Figs.~(\ref{fig:Lambda_H_Mu_Delta})~-~(\ref{fig:Komatsu_H_Mu_Delta}).
So any possibility to discriminate what the right expansion order
is in cosmography is completely out of question. Things are
slightly better with the SNeIa data set, which shows relative errors
in a range $[0.002,0.025]$. But even now results have to be taken
with a pinch of salt. For example, in the redshift range $z<1$, SNeIa relative
errors are $\leq 0.010$, which means that they are of the same order of the
residuals we have found. In this case, we can say we are in a
``border-line" situation.

\begin{figure}
\includegraphics[width=84mm]{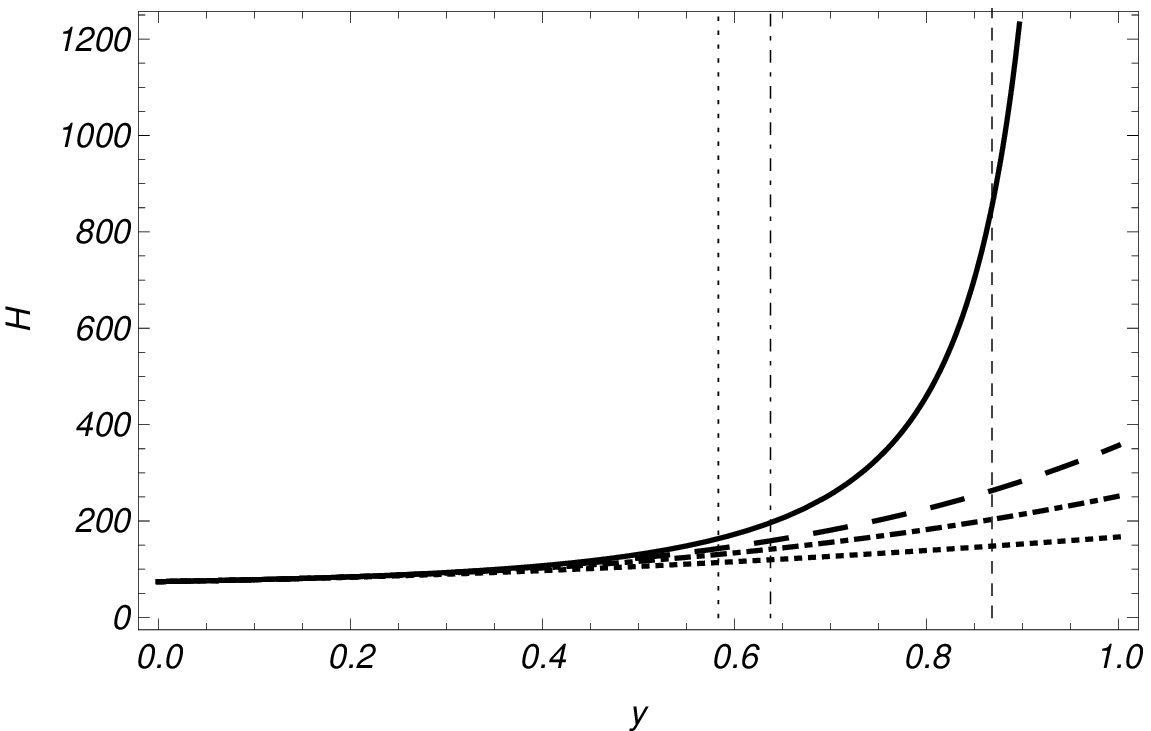}
\includegraphics[width=84mm]{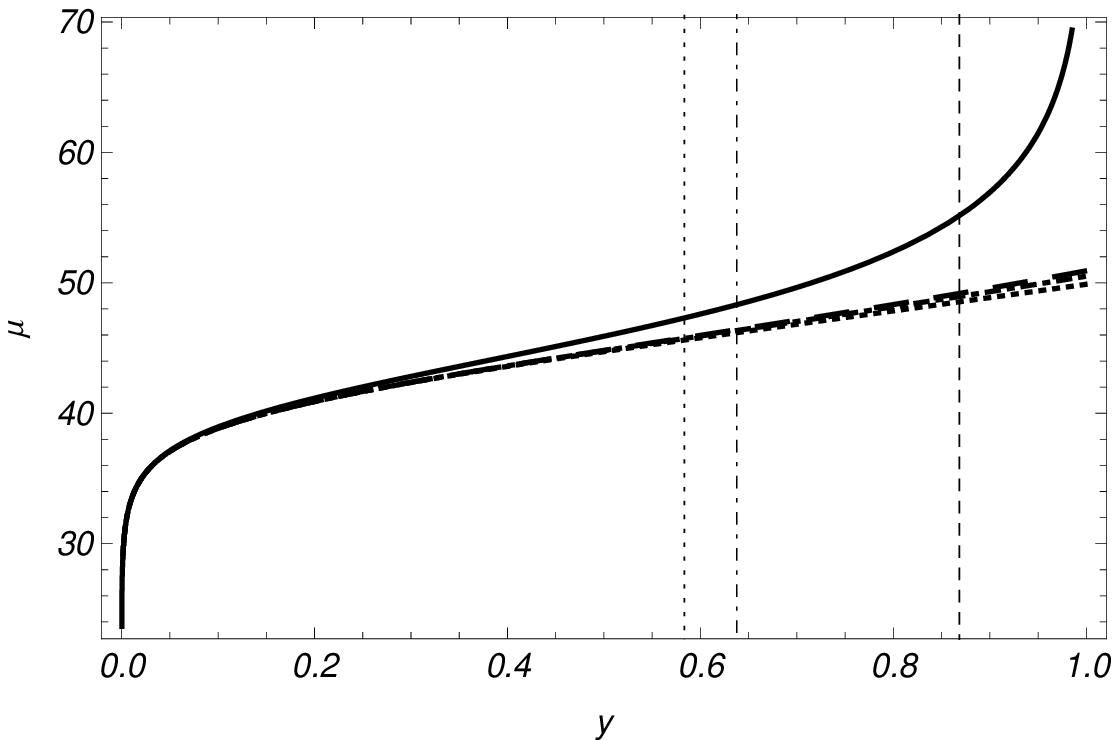}
\caption{(\textit{Top.}) Hubble parameter in the $\Lambda$CDM
model. The y-redshift range is $0<y<1$. The solid black line is
the exact analytical expression; the dashed line is the fourth
order series; the dot-dashed line is the third order series; and
the dotted line is the second order series. The vertical dotted
line is the maximum redshift from SNeIa sample; the vertical
dotdashed line is the maximum redshift from Hubble data; the
vertical dashed line is the maximum redshift from GRBs sample.
(\textit{Bottom.}) The same as above picture but for distance
modulus.} \label{fig:Lambda_H_Mu_Y}
\end{figure}

Last but not least, we have done the same analysis working with
the y-redshift. In this case we have found out that the higher
order series give a better approximation to the exact relation than
the lower order ones, which is quite expected, because we are always
well inside the convergence radius, i.e. $0<y<1$. On the contrary, the most
important thing that emerges, is that the y-redshift can be useful
only up to a certain redshift. Above a certain limit it cannot be
used to constrain the Hubble parameter or distance modulus, for some
intrinsic mathematical properties coming from its
definition. As it is showed in the Fig.~(\ref{fig:Lambda_H_Mu_Y}),
above $y \approx 0.4\div0.5$, there is a clear deviation between the
series expansions and the exact expression of our physical
quantities, larger than observational errors. And the y-redshift
series are unable to follow the right trend.

For all these reasons, we have chosen to work in the reminder with
second and third order series expansions, giving at most
constrains to the deceleration, the jerk and the snap parameters.
We have seen that it is impossible to extend the analysis to four
parameters, the lerk being completely undetermined.

\section{Observational tests}
\label{sec:Obstest}

\subsection{Hubble parameter}
\label{sec:Hubbledata}

Recently in \cite{Stern09} an update of the Hubble parameter $H(z)$ data extracted
from differential ages of passively evolving galaxies previously
published in \cite{Simon05} was presented. In \cite{Mortsell11} this method
was analyzed and some caveats about its use are exposed; anyway, data in \cite{Stern09}
are considered greatly improved with respect of previous ones derived in \cite{Simon05}.
Constraining the background evolution of the universe using these data is
interesting for several reasons. Firstly, they can be used together with
other cosmological tests in order to get useful consistency checks
or tighter constraints on models. Secondly, it is well known that detailed
and reliable information in the behavior of $H(z)$ is
lost when using the luminosity distance or the correlated modulus
distance. As a result of the integration of this function, the fine
details cannot be analyzed and no information can be derived (if
there is any information, probing or refuting the previous ones).
For this reason, we have chosen to work with this data set: any
refinement in the Hubble parameter could give important details.

The Hubble parameter dependence on the differential age of the
Universe in terms of redshift is given by
\begin{equation}
H(z)=-\frac{1}{1+z}\frac{dz}{dt}.
\end{equation}
Thus, $H(z)$ can be determined from measurements of $dt/dz$. As
reported in \cite{Stern09}, \cite{Simon05}, \cite{Jimenez02} and \cite{Jimenez03},
values of $dt/dz$ can be computed using absolute ages of passively evolving galaxies.

The galaxy spectral data used by \cite{Stern09} come from
observations of bright cluster galaxies done with the Keck/LRIS
instrument (see \cite{Stern09B} for a detailed description
of the observations, reductions and the catalog of all the
measured redshifts). The purposely planned Keck-survey
observations have been extended with other datasets: SDSS
improvements in calibration available in the Public Data Release 6
(DR6) have been applied to data in \cite{Jimenez03}; the SPICES
infrared-selected galaxies sample in \cite{Stern00}; and the VVDIS
survey by the VLT/ESO telescope in \cite{LeFevre05}.

The authors of these references bin together galaxies with a
redshift separation which is small enough so that the galaxies in
the bin have roughly the same age; then, they calculate age
differences between bins which have a small age difference which
is at the same time larger than the error in the age itself
(\cite{Stern09}). The outcome of this process is a set of 12 values
of the Hubble parameter versus redshift. A particularly nice
feature  of this test is that the sensitivity of differential ages
to systematic error is lower than in the case of absolute ages
(\cite{Jimenez04}).

Observed values of $H(z)$ can be used to estimate cosmographic
parameters by minimizing the quantity
\begin{equation}\label{eq: Hub_chi}
\chi^{2}_{\mathrm{H}}(H_{0}, \{\theta_{i}\}) = \sum^{12}_{j = 1}
\frac{(H(z_{j}; \{\theta_{i})\} -
H_{obs}(z_{j}))^{2}}{\sigma^{2}_{\mathrm{H}}(z_{j})}
\end{equation}
where $H_{0} \doteq 100 \, h$ will be fixed as $h= 0.742$
(\cite{Riess09}), while $\theta_{i}$ is the vector of model
parameters, namely in our case $\theta_{i}=(q_{0}, j_{0}, s_{0})$.

To minimize the $\chi^2$ function we will use Markov Chain Monte Carlo
(MCMC) methods (fully described in \cite{Berg}, \cite{MacKay},
\cite{Neal} and references therein) testing their convergence with the
method developed and fully described in \cite{Dunkley05}.

\subsection{Supernovae}
\label{sec:SNdata}

We use the most updated SNeIa sample we have now, the Union2
sample described in \cite{Amanullah10}. The Union2 SNeIa
compilation is the result of a new low-redshift nearby-Hubble-flow
SNeIa and new analysis procedures to work with several
heterogeneous SNeIa compilations. It includes the Union data set
from \cite{Kowalski08} with six SNeIa first presented in
\cite{Amanullah10}, with SNeIa from \cite{Amanullah08}, the low-z
and the intermediate-z data from \cite{Hicken09a} and
\cite{Holtzman08} respectively. After various selection cuts were
applied in order to create a homogeneous and high signal-to-noise
data set, we have final $557$ SNeIa events distributed over the
redshift interval $0.015 \leq z \leq 1.4$.

The statistical analysis of Union2 SNeIa sample rest on the definition
of the distance modulus,
\begin{equation}
\mu(z_{j}) = 5 \log_{10} ( d_{L}(z_{j}, \{\theta_{i}\}) )+\mu_0
\end{equation}
where $d_{L}(z_{j}, \{\theta_{i}\})$ is the Hubble free luminosity
distance, Eq.~(\ref{eq: dlseries}), expressed as a series and
depending on the cosmographic parameters, $\theta_{i}=(q_{0},
j_{0}, s_{0})$. The best fits were obtained by minimizing the
quantity
\begin{equation}\label{eq: sn_chi}
\chi^{2}_{\mathrm{SN}}(\mu_{0}, \{\theta_{i}\}) = \sum^{557}_{j =
1} \frac{(\mu(z_{j}; \mu_{0}, \{\theta_{i})\} -
\mu_{obs}(z_{j}))^{2}}{\sigma^{2}_{\mathrm{\mu},j}}
\end{equation}
where the $\sigma^{2}_{\mathrm{\mu},j}$ are the measurement
variances. The nuisance parameter $\mu_{0}$ encodes the Hubble
constant and the absolute magnitude $M$, and has to be
marginalized over. Giving the heterogeneous origin of Union data
set, and the procedures described in \cite{Kowalski08} for
reducing data, we have worked with an alternative version of
Eq.~(\ref{eq: sn_chi}), which consists in minimizing the quantity
\begin{equation}\label{eq: sn_chi_mod}
\tilde{\chi}^{2}_{\mathrm{SN}}(\{\theta_{i}\}) = c_{1} -
\frac{c^{2}_{2}}{c_{3}}
\end{equation}
with respect to the other parameters. Here
\begin{equation}
c_{1} = \sum^{557}_{j = 1} \frac{(\mu(z_{j}; \mu_{0}=0,
\{\theta_{i})\} -
\mu_{obs}(z_{j}))^{2}}{\sigma^{2}_{\mathrm{\mu},j}}\, ,
\end{equation}
\begin{equation}
c_{2} = \sum^{557}_{j = 1} \frac{(\mu(z_{j}; \mu_{0}=0,
\{\theta_{i})\} -
\mu_{obs}(z_{j}))}{\sigma^{2}_{\mathrm{\mu},j}}\, ,
\end{equation}
\begin{equation}
c_{3} = \sum^{557}_{j = 1}
\frac{1}{\sigma^{2}_{\mathrm{\mu},j}}\,.
\end{equation}
It is trivial to see that $\tilde{\chi}^{2}_{SN}$ is just a
version of $\chi^{2}_{SN}$, minimized with respect to $\mu_{0}$.
To that end it suffices to notice that
\begin{equation}
\chi^{2}_{\mathrm{SN}}(\mu_{0}, \{\theta_{i}\}) = c_{1} - 2 c_{2}
\mu_{0} + c_{3} \mu^{2}_{0} \,
\end{equation}
which clearly becomes minimum for $\mu_{0} = c_{2}/c_{3}$, and so
we can see $\tilde{\chi}^{2}_{\mathrm{SN}} \equiv
\chi^{2}_{\mathrm{SN}}(\mu_{0} = 0, \{\theta_{i}\})$. Furthermore,
one can check that the difference between $\chi^{2}_{SN}$ and
$\tilde{\chi}^{2}_{SN}$ is negligible.

\subsection{Gamma Ray Bursts}
\label{sec:GRBdata}

Working and interpreting results from a GRBs analysis is not an
easy task, being the errors on their observable quantities much
larger than those ones for SNeIa, and being their source mechanism
still not well understood. For this reason choosing a good GRBs
sample is crucial; we have chosen to work with the sample
described in \cite{Cardone09}. There the authors perform a new
calibration procedure on the widely used GRBs sample from
\cite{Schaefer07} which perfectly matches our requirements for
using them with cosmography. The possibility of a joint analysis
of SNeIa and GRBs is strictly related to the building of a Hubble
diagram for GRBs too, being this extremely difficult because of
GRBs are not standard candles as SNeIa. To create an Hubble
diagram for GRBss, one has to look for a correlation between a
distance dependent quantity and a directly observable property.
Starting from some of the many correlations that have been
suggested in the last years, \cite{Schaefer07} created a sample of
69 GRBs in the redshift range $0.17<z<6.6$ whose Hubble diagram is
well settled.

In \cite{Cardone09} the authors have updated such a sample in many
aspects, the main one being the test of a new method for the
calibration of GRBs based one the assumption of none \textit{a
priori} cosmological model. Such a model independent calibration
is built on the idea that SNeIa and GRBs at the same redshift
should exhibit the same distance modulus. In this way,
interpolating the SNeIa Hubble diagram gives the value of $\mu(z)$
for a sub-sample of GRBs which lies in the same redshift range.
This sub-sample can be finally used for calibrating the well know
GRBs correlations and, assuming that this calibration is redshift
independent, it can be extended to high redsfhit GRBs. With this
procedure, in \cite{Cardone09} the authors were able to convert the
\cite{Schaefer07} sample to a new one, with the same number of
objects but with SNeIa-calibrated GRBs distance modulus. These
data can then be used for minimizing the corresponding $\chi^2$:
\begin{equation}
\chi^{2}_{\mathrm{GRB}}(\{\theta_{i}\}) = \sum^{69}_{j =
1} \frac{(\mu(z_{j}; \{\theta_{i})\} -
\mu_{obs}(z_{j}))^{2}}{\sigma^{2}_{\mathrm{\mu},j}} \; .
\end{equation}

\subsection{Baryonic Acoustic Oscillations}
\label{sec:BAOdata}

In \cite{Percival10} the authors analyze the clustering of galaxies within the
spectroscopic Sloan Digital Sky Survey (SDSS) Data Release 7 (DR7) galaxy
sample, including both the Luminous Red Galaxy (LRG) and Main samples, and
also the 2-degree Field Galaxy Redshift Survey (2dFGRS) data. In total,
the sample comprises $893319$ galaxies over $9100$ deg$^2$. Baryon Acoustic
Oscillations are observed in power spectra measured for different slices in
redshift; this allows constraining the distance-redshift relation at
multiple epochs. A distance measure at redshift $z=0.275$ was achieved; but
what is more important for our application of cosmography is the almost
independent constraint on the ratio of distances,
\begin{equation}
\mathcal{B}=\frac{D_{V}(0.35)}{D_{V}(0.2)} = 1.736 \pm 0.065 \; ,
\end{equation}
which is consistent at $1.1\sigma$ level with the best fit $\Lambda$CDM model
obtained when combining the $z=0.275$ derived distance constraint with the WMAP
5-year data. This measurement is particularly well suited for our purpose, because
without it we could not apply BAO constraints on cosmographic approach.
The usually derived distance measure from BAO are in the form $r_{s}(z_{d})/D_{V}$,
where $r_{s}(z_{d})$ is the comoving sound horizon at the baryon drag epoch, while
$D_{V}$ is defined as:
\begin{equation}
D_{V}(z_{BAO}) = \left[ \left( \int_{0}^{z_{BAO}} \frac{c \, \mathrm{d}z}{H(z)}\right)^2
\frac{c \, z_{BAO}}{H(z_{BAO})} \right]^{1/3}
\end{equation}
with $H(z)$ the Hubble parameter. It is clearly obvious that the
sound horizon quantity is incompatible with cosmography, being it
evaluated at at the drag epoch redshift which is $z \approx 1000$,
well out of the possible redshift applicability range of
cosmography. On the contrary, the quantity $D_{V}$ can be easily
used for cosmography by substituting for $H(z)$ the appropriate
series expression and evaluating it at $z=0.2$ and $0.35$, well
inside the convergence redshift radius of cosmographic series. The
quantity $\mathcal{B}$ can then be used for estimating
cosmographic parameters by minimizing the quantity
\begin{equation}\label{eq: bao_chi}
\chi^{2}_{\mathrm{BAO}}(\{\theta_{i}\}) =
\frac{(\mathcal{B}(\{\theta_{i})\} -
\mathcal{B}_{obs})^{2}}{\sigma^{2}_{\mathcal{B}}} \; .
\end{equation}

\section{Cosmographic analysis}
\label{sec:cosmo_analysis}

As said in the previous sections, we have used an MCMC algorithm
to perform the analysis of the multi-dimensional space of the
cosmographic parameters.

Our main purpose is to explore the possibility of obtaining better
constraints on the cosmographic parameters than offered by the
literature. This is our motivation to use MCMC methods; as opposed to
the approaches used in past examples in the same field, our choice offers
the interesting possibility to introduce and manage easily possible physical
constraints, thus making it possible to discover
underlying relations and to realize a systematic report of the
results.

\subsection{Preliminary discussion}
\label{sec:preliminary}

First of all, we try to answer a question: are there any physical
constrains that can be imposed as priors to our fitting procedure?
If yes: how do these constraints work on the estimation of parameters?

The accuracy of the fitting parameters is of course a property
derived from the likelihood function and it is quite independent of the
method used for its maximization. The point that we want to investigate here is
if there is any physical reasonable prior that can be introduce in such analysis
for reducing the historically large uncertainties on cosmographic parameters
(as described in the introduction). While a great part of these uncertainties surely depend
on the series approach which cosmography is based on, we think that probably one could improve
the analysis imposing some physical basic and general requirements. For example:
does it make any sense to perform a minimization of $\chi^2$
without considering if the best fit parameters give well based physical information, i.e.
without considering how physics is mapped in the parameter space (in our case: cosmographic
parameter space)?

This is a matter of great importance (and also their strong point in our idea)
just while using statistical methods like the MCMC: they are based on an algorithm
that moves ``randomly'' in the parameter space: do all the points (even around the minimum
$\chi^2$ location) tested by this algorithm satisfy some general physical requirements,
as a positive $d_{L}$ or a positive $H^2$? These questions would be of course useless
if we were working with exact analytical expressions but, as log as cosmography is built
on series expansions, we think it is necessary to analyze this aspect.

As we will show below, even if we will be able to sensibly reduce uncertainties
on cosmographic parameters, other problems arise (as, for example, the dependence on the
maximum order series expansion, and on the data redshift range) that are intrinsic to the
cosmographic apparatus.

The main quantities we are going to fit are the Hubble parameter and the distance modulus;
we are quite sure that these quantities will be
physically well-based in the best fit location and around it. However, we have to take into account also
quantities that are not tested directly.  For example, eventual constrains have to be
given to the series expansions of $H^{2}(z)$ and $d_{L}(z)$.
The most general and obvious limitation is the positiveness of these two functions:
for $H^{2}$ it is a natural mathematical property; and the same
consideration is true for $d_{L}$,  this being a distance. At the
same time there is not a direct connection between the
positiveness of $H$ and $d_{L}$: it is possible to have an $H(z)$
function with varying signature, for example a sinusoidal one, and
having a positive definite luminosity distance. We have also
verified that, obviously, when fitting models in the redshift range of
our data we automatically get a definite positive $H(z)$, at least in
the prescribed redshift range (thus excluding a bounce).

Then another question arises naturally: does it make any sense to
extend the priors (in our case the positiveness of the physical
functions) to any range of redshift well beyond the convergence
radius of the series? The answer to this question is not so
simple. Since we are working with a truncated series expression we
have to impose a \textit{``minimum requirement"} for the
positivity of our functions: it ranges only in the limited
convergence redshift interval, $0<z<1$ or in the redshift range
defined by the used data sets. For this reason we also perform a cosmographic analysis
both using the full (in a redshift sense) data samples and a cut sub-sample
limited to $z>1$.
We also test what happens when extending positivity to all the possible redshift values,
namely $z>0$ even if this would be rigorously right only if we were
working with \textit{exact} analytical functions, and to the
maximum redshift value coming from any observational sample. \\
Instead, in the y-redshift, things are different: this quantity
spans all the physical distances we are interested in, and it
remains always inside the series convergence radius, as $0<y<1$.
So we can impose positivity on all the y-redshift interval without
having forced final results and loss of information due to this
choice. Anyway we also test results when the y-redshift is
constrained to the range corresponding to the relative z-redshift
range for comparing results.

We have also found it is possible to add another constraint: the
requirement that $\Omega_{m}$ be positive and smaller than unity.
Using the procedure we will describe in next section, where we
will compare our cosmographic analysis with dark energy models
described by the CPL parametrization, it turns out that $\Omega_{m}$ is a function of
the cosmographic parameters. In the two dimensional case we have
\begin{equation}
\Omega_{m}(q_{0}, j_{0}) = \frac{2 (j_{0} - q_{0} - 2 q_{0}^{2})}{1 + 2 j_{0} -
6 q_{0}},
\end{equation}
while in the three dimensional case the expression becomes quite longer,
so we omit it for the sake of simplicity. Clearly, the above mentioned
constraint on $\Omega_{m}$ can be enforced while running the MCMC algorithm.\\
At the end, our priors are:
\begin{itemize}
 \item $d_{L}(z) > 0$ \, ;
 \item $H^{2}(z) > 0$ \, ;
 \item $0 < \Omega_{m} < 1$\,,
\end{itemize}
applied on both $z$ and $y$ redshift ranges.

Finally, for fully understanding and comparing our results with literature,
it is order now to underline the different use we have made of the jerk as
compared to previous references. In \cite{Cattoen07} and
\cite{Capozziello08}, the quantity $j_{0} + \Omega_{0}$ was estimated, where
\begin{equation} \Omega_{0} = 1 + \frac{k c^{2}}{H_{0}^{2}
a_{0}^{2}}\, .
\end{equation}
As stated in \S~(\ref{sec:appendix}), in our case the spatial curvature is assumed to be $k=0$ so that we can derive
$j_{0}$ values also from other different works. Of course when comparing
our results to these ones, we have
to consider that in those cases error bars correspond to the composite quantity
$j_{0} + \Omega_{0}$. But we are confident that, considering
independent estimations of the curvature parameter and its related
error, the largest contribution to errors is mainly
attributable to the jerk parameter. On the other side, in
\cite{Poplawski06}, \cite{Rapetti07} and \cite{John04} the jerk
parameter has the same definition as we use. Finally, in
\cite{Capozziello08} GRBs were the only data used (instead of SNeIa).

\subsection{z-redshift}

Finally, if we consider the analysis with the common definition of
z-redshift, looking at Tables~(\ref{lab:2P_Z})~-~(\ref{lab:3P_Z})
we can summarize the main results as follows:
\begin{itemize}
  \item with two cosmographic parameters, working with cut subsamples,
  namely considering only data with $z<1$, always gives better results
  than using the complete sample and results are mostly uniforme. This
  is a direct consequence of the SNeIa sample having the largest weight in
  the chi-square minimization because of the highest number of data points
  with respect to other data sets;
  \item with three cosmographic parameters the differences between using
  total and cut samples decrease. Best results always come from the cut
  samples but this time, differently from the previous point, they are
  related to the high-redshift extended priors, while assuming priors in
  the series convergence radius range does not satisfy the MCMC convergence
  requirements. This is somewhat expected from discussion in
  \S~(\ref{sec:comparing}), and possibly due to the fact that adding
  a third parameter in the restricted redshift range does not give any further
  improvement, while needing it for comparing cosmographic series with larger redshift
  ranges;
  \item adding GRBs to SNeIa (it was interesting to verify what happens
  when joining these data sets because they constitute an homogeneous sample as they are both
  based on the definition of the distance modulus) when considering two cosmographic parameters
  does not improve the analysis resulting in a very high chi-square value. When working
  with three cosmographic parameters the only statistically good results come from the
  cut samples. But in these cases the weight of GRBs is low (only $19$ GRBs have
  redshift less than one) and no statistically significative changes are detected in
  the estimation of cosmographic parameters;
  \item the contribution of BAO data is almost statistically insignificant;
  \item adding only Hubble data to SNeIa in the two parameters case, while considering
  the total sample case, slightly increases chi-square value and produces different
  values for cosmographic parameters. Introducing the prior on positiveness of physical
  quantities on the Hubble data redshift range makes cosmographic estimations more similar
  to the SNeIa-only case. On the other side, when working with three parameters we detect
  an higher influence of Hubble data in the chi-square balance, with even largest (and
  previously undetected) deviations when applying the prior on positiveness
  $z<1.76$. A further discussion on the reconstructed cosmological scenario is
  given in \S~(\ref{sec:CPL}).
\end{itemize}

With these considerations in mind, we can take a look to the
cosmographic parameters coming from using only SNeIa and
considering just two parameters: the deceleration and the jerk.
This is the most popular cosmographic parametrization in the
literature so that our results and the advantages of the MCMC
method can be easily judged and tested. We also underline that all
the results in the literature are taken from sample of SNeIa where
no redshift cut is applied. \\
As we have pointed before, there is a dependence of results from
priors and from using cut or total sample which gets clear when
looking to Fig.~(\ref{fig:Q_J_1}). If we move from the total
sample and the prior applied on the total redshift range $z>0$
(top panel), to the total sample with prior applied on the SNeIa
redshift range (middle panel), and finally to the cut sample with
prior in the z-redshift convergence series radius, we detect a
trend in the cosmographic parameters: the deceleration starts far
from usual values in literature (and $\Lambda$CDM one too) and
moves towards them, while jerk starts from a value very near to
the $\Lambda$CDM one and which agrees perfectly with most of the
results in literature, and moves to smaller values.

Considering the best chi-square values of the deceleration
parameter, we remember that we are just using the Union2 data set
from \cite{Amanullah10}, which mixes many independent SNeIa
observations so we may expect some differences with results like
the ones in \cite{Cattoen07}. It is also interesting to underline
that we agree with the analysis from \cite{Rapetti07}, which uses
a completely physically different data set, the X-ray observations
of galaxy clusters, together with Gold and Legacy supernovae.
Their method is also different: while in \cite{Cattoen07} they use
the same series expansion approach as we do, in \cite{Rapetti07}  an
alternative numerical way leading to analytically exact functions
was defined. The main difference in our work with respect to
\cite{Cattoen07} is the use of physical constrains. Giving our
agreement with the exact analytical approach from
\cite{Rapetti07}, we may argue that these constrains are an
irrefutable requirement when working with cosmography as a series
expansion tool. Finally, we perfectly agree at $1\sigma$
confidence level with the results in \cite{Poplawski06}, while
only slightly agree with results in \cite{Capozziello08} coming
from the application of cosmography to high redshifts using gamma
ray burst data.

About the jerk parameter it can be verified there is a good
agreement  with the analysis from X-ray emission from clusters of
galaxies described in \cite{Rapetti07} and with the value coming
from the Gold data set in \cite{Cattoen07}, just as in the
deceleration case.

At the same time we underline that it is also nude-eye visible
that the error bars coming out from our analysis are actually
narrower than in any other case (except for \cite{Poplawski06},
who derives cosmographic parameters from theoretical assumptions
without using any observational data). If we consider that we are
working with a physically homogeneous sample (it is made of SNeIa
only) but built from different (observationally and technically
reduced) data sets, this positive result may be strongly related
to the chosen statistical method, thus MCMC are showing their good
nature with respect to other popular fitting procedures.

When moving to the three dimensional analysis, including the
possibility of constraining the snap parameter, we find very
different results. If we consider only the SNeIa sample, now the
deceleration parameter is less negative and the jerk even changes
sign becoming negative. We find a very weak match with past
values from literature mainly due to their large errors. On the
contrary, using MCMC makes us able to narrow down the
confidence levels strongly.\\
The discussion cannot be complete without considering the value
of the snap. Unfortunately we have not many examples in the
literature we can compare with: from \cite{John04} and
\cite{Capozziello08} we have some strongly positive values with
percentage errors much larger than the $200\%$, while our analysis
gives a negative value with narrower constraints. We remind the
reader that the snap value in $\Lambda$CDM should be equal to
$-0.35$; in this sense, the sign we have found shows a clear trend
in this parameter and in the cosmological properties subtended by
it.

When adding Hubble data, we have two different situations related
to the different applied priors: when the positiveness is for
$z>0$ we find the main differences for the jerk, which is much
more negative, and for the snap, less negative; on the other side,
when the positiveness if for $z<1.76$, differences are more
pronounced: the deceleration is very near to $-1$, while the jerk
and the snap are positive and larger than before, $3.134$ and
$4.399$ respectively. It is not an easy task to understand why
this happens; moreover no correspondence with past values in
literature is found.

\subsection{y-redshift}

If we work with the y-redshift defined in \cite
{Cattoen07} to
solve the convergence problem of the series, looking at
Tables~(\ref{lab:2P_Y})~-~(\ref{lab:3P_Y}) we can find some
interesting changes:
\begin{itemize}
  \item as expected from the definition of the y-redshift, differences
  both among total and cut samples and using different y-redshift prior ranges
  are strongly softened because we are always well inside the convergence radius
  of the series;
  \item results are always consistent and uniform, showing only small statistical
  variations, when considering two and three cosmographic parameters
  separately, but cosmographic estimations are different between the
  two and the three dimensional approach;
  \item adding high order y-redshift data (GRBs) to SNeIa does not change
  results in a significant statistical way. We have to remind that the y-redshift
  series show problem in reconstructing the distance modulus for
  $y \approx 0.4 \div 0.5$, which is well under the SNeIa maximum redshift even.
  So, we feel that this feature mostly depends on the intrinsic
  impossibility of y-redshift series to match the analytically exact physical
  behavior of distance modulus;
  \item BAO has no weight on the estimation of cosmographic parameters;
  \item as happened in the z-redshift case, here too Hubble data are the
  only exception in the uniformity of results exhibiting a stronger weight
  in the cosmographic estimations with respect of GRBs and BAO. While the
  contribution of the cut Hubble sample is not significant, the total one, extended up to
  $z=1.76$, i.e. $y=0.638$, produces the largest deviation.
  As the Hubble sample does not extend the y-redshift range in a much sensible
  way for explaining such a deviation, we can only argue that some peculiarities
  in Hubble high redshift data are present also inducing the difficulties
  in fitting them as discussed in \S~(\ref{sec:CPL}). A corroborating test
  for this explanation is the not reached convergence for MCMC in the three dimensional
  analysis when the total Hubble data sample is considered.
\end{itemize}

When comparing our results with the literature, the only previous
examples where the y-redshift was used can be found in
\cite{Cattoen07}. As in our case, when moving from z-redshift to
y-redshift they found an increase in errors (which are bigger than
ours, anyhow) of the cosmographic parameters, more consistent in
the case of the jerk, as we have verified. This could be given by
the substantial down-stretching of the redshift data when using
this different parametrization; even if the physical content of
the data is not altered, it is more difficult to extract
cosmographic results when the data are flattened on the new
redshift axes. The z-redshift range of the Union2 data set,
$0<z<1.4$, goes over to the y-redshift interval $0<y<0.584$;
Hubble data convert from $0<z<1.76$ to $0<y<0.638$, and GRBs from
$0<z<6.6$ to $0<y<0.871$. It is clear that if we consider the
great contribution from SNeIa to the chi-square function, we are
fitting our physical distances on a limited range (low
$y$-redshift) with respect to the total convergence radius ($y =
1$) so that higher order parameters (as the jerk and the snap) can
be not as well estimated as lower order ones. Thus while for the
deceleration parameter errors still remain quite centrally based,
for the jerk case we have a long tail for high positive values
(nearer the $\Lambda$CDM value but only matching it at $2\sigma$
confidence level) while for the snap we have a tail for large
negative values matching the corresponding $\Lambda$CDM value.
This may be considered a trend which one eventually needs to test
and confirm if having higher order data.

The agreement is good for most of the past works we have
considered, with the greatest number of successful matchings for
the deceleration parameter than for the jerk. Anyway, the best
results at all, come from the case where we use the full data
samples and applied the prior in the range depicted by them. In
this case also the deceleration and the jerk match well with the
corresponding $\Lambda$CDM values.

\section{Equivalent CPL model}
\label{sec:CPL}

\begin{figure}
\includegraphics[width=85mm]{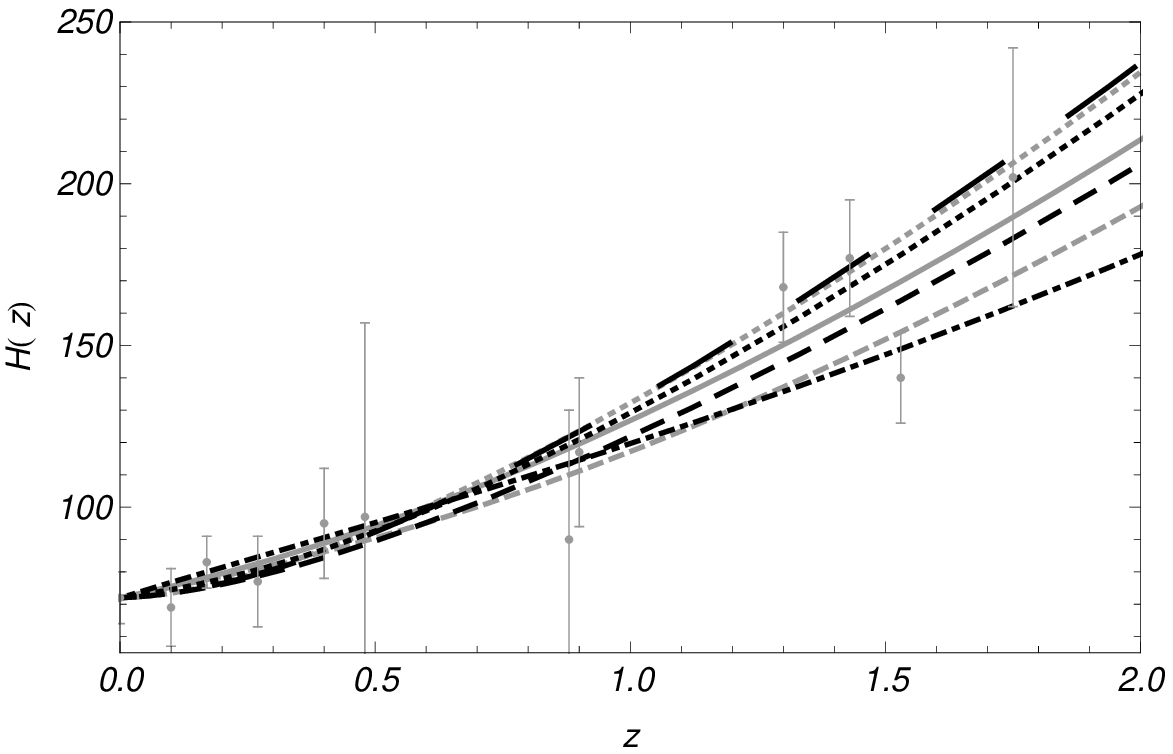}
\includegraphics[width=85mm]{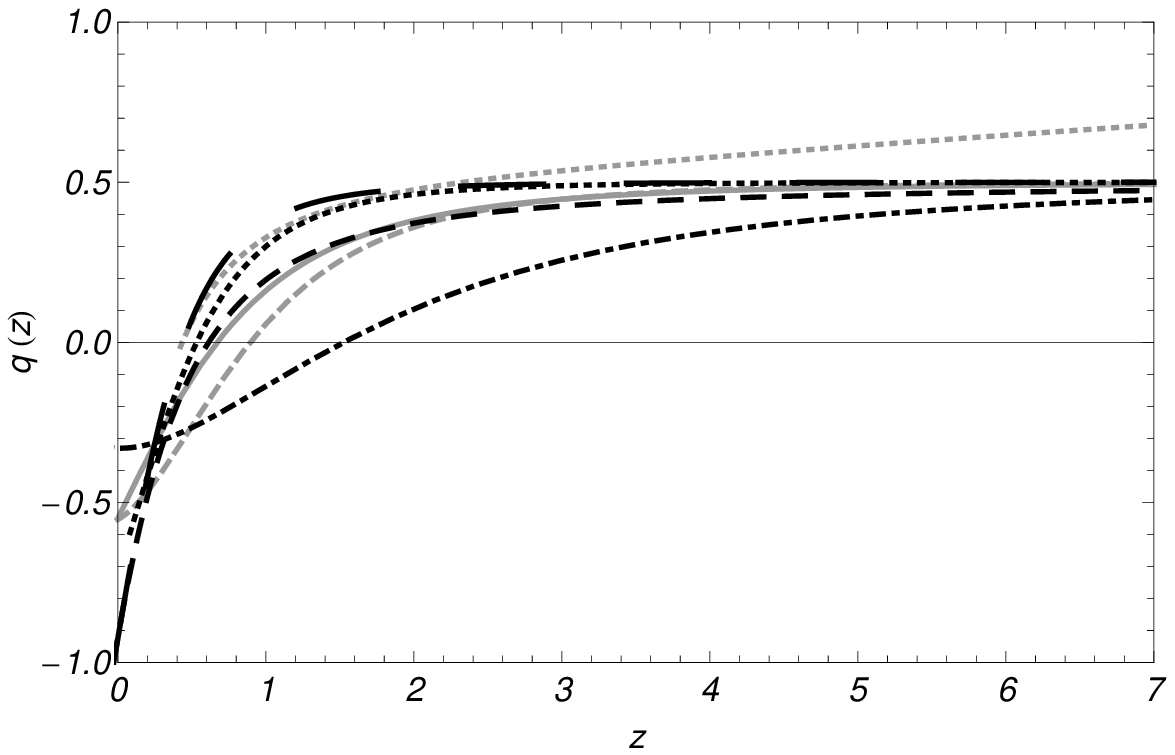}
\includegraphics[width=85mm]{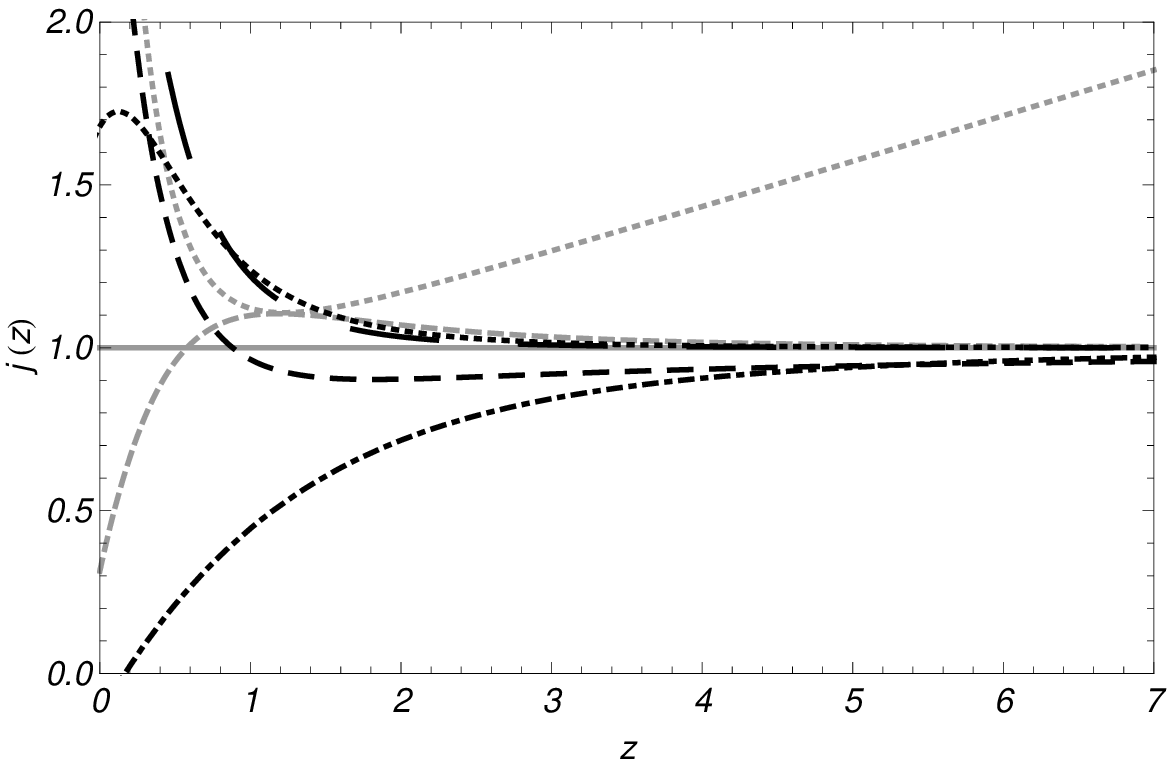}
\includegraphics[width=85mm]{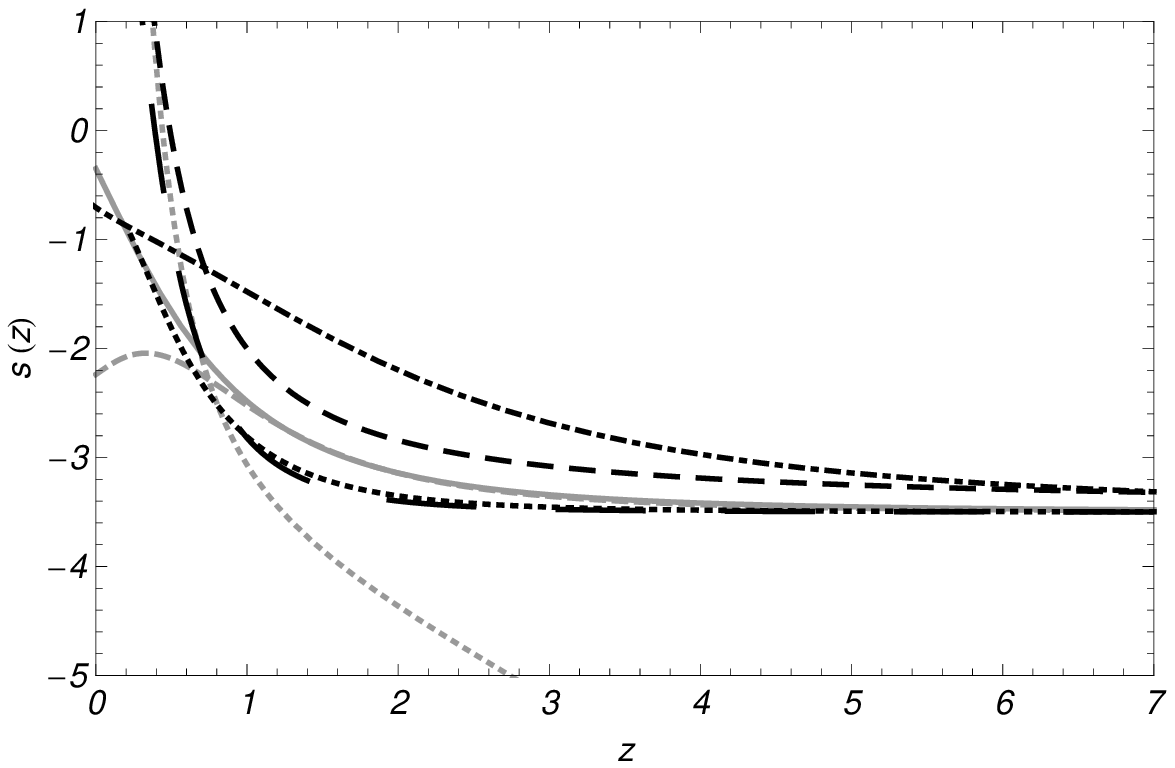}
\caption{Hubble parameter $H(z)$ and cosmographic parameters
$q(z)$, $j(z)$, $s(z)$ in different cosmological scenarios:
$\Lambda$CDM model (solid gray line); Komatsu et al. (2010) model (dashed gray
line); Perivolaropoulos et al. (2008) model (dotted gray line); Model I
(long-dashed black line); Model II (short-dashed black line);
Model III (dotdashed black line); Model IV (dotted black line).}
\label{fig:Summ_redshift}
\end{figure}

To round up  results from simulations on cosmographic parameters,
we can go deeper into the analysis and try to our results with a
specific  cosmological model. It is easy to check the link
between them. In this paper we have used the CPL parametrization
for dark energy given by Eq.~({\ref{eq: ecpl}}) which depends only
on three parameters, ($\Omega_{m} \, , \, w_{0}$ and $w_{a}$).
Using Eqs.~(\ref{eq:qzcpl})~-~(\ref{eq:jzcpl})~-~(\ref{eq:szcpl})
we can relate cosmographic parameters and CPL ones. We have
different possibilities depending on the number of cosmographic
parameters we are working with and on how many CPL parameters we
are going to consider as free. In the end we have:
\begin{itemize}
  \item with two cosmographic parameters, $(q_{0}, j_{0})$, we can
        derive news about a Constant Dark Energy (CDE) model (i.e. $w_{a} =
        0$), with:
        {\setlength\arraycolsep{0.2pt}
        \begin{eqnarray}
        \Omega_{m}(q_{0}, j_{0}) &=& \frac{2 (j_{0} - q_{0} - 2 q_{0}^{2})}{1 + 2 j_{0}
        - 6 q_{0}}\, , \nonumber \\
        w_{0}(q_{0}, j_{0}) &=& \frac{1 + 2 j_{0} - 6 q_{0}}{-3 + 6 q_{0}}\, ;
        \end{eqnarray}}
  \item with two cosmographic parameters, $(q_{0}, j_{0})$, we can
        derive news also about a Dynamical Dark Energy (DDE) model, (i.e. $w_{a} \neq
        0$), leaving $\Omega_{m}$ free, with:
        {\setlength\arraycolsep{0.2pt}
        \begin{eqnarray}
        w_{0}(q_{0}, j_{0}, \Omega_{m}) &=& \frac{1 - 2 q_{0}}{3 (-1 + \Omega_{m})}\, , \nonumber \\
        w_{a}(q_{0}, j_{0}, \Omega_{m}) &=& \frac{1}{3 (-1 + \Omega_{m})^{2}} \left(-2j_{0}(-1+\Omega_{m})-2q_{0} \right. \nonumber \\
        &\times& \left. (1+2q_{0}) + \Omega_{m}(-1+6q_{0})\right)\, ;
        \end{eqnarray}}
  \item with three cosmographic parameters, $(q_{0}, j_{0}, s_{0})$,
        we can derive news about a Dynamical Dark Energy (DDE) model,
        with $\Omega_{m}$ depending on cosmographic parameters, i.e.
        $\Omega_{m} \doteq \Omega_{m}(q_{0}, j_{0}, s_{0})$. The same holds true for
        $w_{0} \doteq w_{0}(q_{0}, j_{0}, s_{0})$ and $w_{a} \doteq w_{a}(q_{0}, j_{0},
        s_{0})$. These expressions are much longer than previous ones and we do not write
        them here for the sake of brevity.
\end{itemize}
The most important thing is that with these relations all the
statistical properties of these parameters (median, error bars, etc.) can
be easily extracted from the cosmographic samples we have obtained
from MCMC simulations. Final results of this dark energy analysis
are reported in the tables on previous pages.

The first main result concerns Hubble data and their relation with
the CDE model available from using cosmographic series with two
parameters: when using z-redshift, this dynamical model is always
unable to fit Hubble data. Even if we consider the model with
parameters directly obtained from using this kind of data, we are
only able to fit few points at very low redshift and not all the
sample. Moreover, when moving to the CDE models with y-redshift
and to the DDE models derived from three cosmographic parameters
using Hubble data, we can easily verify that the derived
cosmological parameters (and the cosmographic one too, as we
pointed out in \S.~(\ref{sec:cosmo_analysis})) are always very
different from the ones derived from other kinds of observational
data. This feature can be interpreted in two different ways:
\begin{itemize}
  \item it depends on the procedure which Hubble data are obtained from;
  \item there is an intrinsic information about the dynamical model
  in Hubble data.
\end{itemize}
Surely, for the two cosmographic parameters case, a better way of
proceeding may be to obtain $\Omega_{m}$ from alternative ways and
then join results with those one from cosmography as we discuss
later in this section.

If we consider the CDE model derived from two cosmographic
parameters and the DDE models derived from three cosmographic
parameters with y-redshift, we can see that:
\begin{itemize}
  \item $\Omega_{m}$ is lower than usual, being $\approx 0.07$, and
  matching at the top $1\sigma$ confidence level the usual estimation of $\Omega_{m}\sim 0.16$
  obtained considering plasma contribution;
  \item if we also add Hubble data, then the value for the matter
  content grows up to $\approx 0.40$ in the CDE case and to $\approx
  0.36$ in the DDE case. In both cases the bottom $1\sigma$ confidence
  limit matches the matter content value (plasma included) and the total
  $\Lambda$CDM one.
\end{itemize}

It is much intriguing to find such small values. Generally
from SNeIa one is able to extract a slightly overestimated value
of $\Omega_{m} \sim 0.30$ in \cite{BuenoSanchez09}; in
\cite{Komatsu10}, unifying SNeIa, BAO and WMAP7, a value for
$\Omega_{m} \sim 0.245$ was obtained; while in \cite{Rapetti07} it
is $\Omega_{m} = 0.306_{-0.040}^{+0.042}$. We underline that any
conclusion about the real possibility of such universe dynamics is
out of the purpose of this paper. We have to consider that we are
always working with an approximated approach (series expansion)
and that this conclusion should be corroborated by a more extended
analysis of all the other dynamical aspects of universe (formation
of any gravitational structure, explanation of acceleration of
universe). Then we need to refine the use and the approach of
cosmography (what we are intended to do with a series of papers).

In the same cases as before, for what concerns the dark energy
parameters, we have:
\begin{itemize}
  \item for the CDE models, $w_{0} \approx -0.82$, excluding the phantom line crossing
  at $1\sigma$ level; while adding Hubble data it becomes much more
  negative, $w_{0} \approx -1.59$;
  \item for the DDE models, $w_{0} \approx -0.715$ and $w_{a} \approx
  0.01$ ($\sim -0.16$ at $1\sigma$ error); adding Hubble data we
  get $w_{0} \approx -1.272$ and $w_{a} \approx -0.326$.
\end{itemize}

The DDE models from two cosmographic parameters have $\Omega_{m}$
fixed at the value obtained in \cite{Komatsu10}, while the dark
energy parameters are:
\begin{itemize}
 \item with the z-redshift, $w_{0} \approx -1$ and $w_{a} \approx
 -0.71$; adding Hubble data they change to $w_{0} \approx -1.4$ and $w_{a} \approx
 -1.2$, but they also have high chi-square values. About the $w_{a}$ parameter,
 the question is very subtle and tricky: we have a negative value
 for it very far from the usual theoretical expected values and almost of the same
 order of $w_{0}$.
 \item with the y-redshift, $w_{0} \approx -1$ and $w_{a} \approx
 -0.56$ which match well the results from \cite{Komatsu10}, $w_{0} =
 -0.93$ and $w_{a}  = -0.41$. Adding Hubble data they change to
 $w_{0} \approx -1.25$ and $w_{a} \approx 1.2$. In this case, on the contrary,
 the value of $w_{a}$ is of the same order of $w_{0}$ but positive so that $w_{0}+w_{a} \sim
 0$ and agrees with results from \cite{Perivolaropoulos08} where
 starting from Union SNeIa sample find $w_{0} = -1.4$ and $w_{a} = 2$.
\end{itemize}

To test all this results we can also verify that considering the
evolution of Hubble parameter or of the deceleration parameter, we
do not find particular problems, but rather some intriguing
properties. Using Eqs.~(\ref{eq:cosmo_CPL}) we can derive the
functions $q(z,\Omega_{m},w_{0},w_{a})$,
$j(z,\Omega_{m},w_{0},w_{a})$ and $s(z,\Omega_{m},w_{0},w_{a})$.
In this way, giving the cosmological model obtained by our
cosmographic analysis and parameterized by the CPL prescription,
we can infer the redshift dependence of the cosmographic
parameters. In Fig.~(\ref{fig:Summ_redshift}) we have plotted the
analytical behavior of $H(z)$, $q(z)$, $j(z)$ and $s(z)$ for five
different cosmological models:
\begin{itemize}
 \item $\Lambda$CDM: $\Omega_{m} = 0.30$, $w_{0} = -1$, $w_{a} = 0$;
 \item best cosmological model from \cite{Komatsu10}: $\Omega_{m} = 0.245$,
       $w_{0} = -0.93$, $w_{a} = -0.41$;
 \item best cosmological model from \cite{Perivolaropoulos08}: $\Omega_{m} = 0.30$,
       $w_{0} = -1.4$, $w_{a} = 2$;
 \item Model I, CDE model from two dimensional cosmographic analysis, in y-redshift
       applying priors only in the redshift range depicted by using all the data samples: $\Omega_{m} = 0.399$, $w_{0} = -1.581$;
 \item Model II, DDE model from two dimensional cosmographic analysis, in y-redshift
       applying priors only in the redshift range depicted by using all the data samples: $\Omega_{m} = 0.245$, $w_{0} = -1.256$, $w_{a} =
       1.220$;
 \item Model III, DDE model from three dimensional cosmographic analysis, in z-redshift
       applying priors in the full z-redshift range and using all the data samples: $\Omega_{m} = 0.151$, $w_{0} = -0.652$, $w_{a} = -0.207$;
 \item Model IV, DDE model from three dimensional cosmographic analysis, in y-redshift
       applying priors only in the redshift range depicted by using all the data samples: $\Omega_{m} = 0.364$, $w_{0} = -1.272$, $w_{a} = -0.326$;
\end{itemize}

The first thing we can say is that our models can be different
from the $\Lambda$CDM scenario for low redshifts, but they are
equivalent to it for high redshifts. This is not the case of the
model in \cite{Perivolaropoulos08}, which is completely different
from $\Lambda$CDM model at any redshift. Of course, in our case,
we have the pleasant property of that some properties at high
redshift (such as CMB theory and data) remain untouched; but, of
course, the necessity of re-discussing what happens at low
redshifts cannot be waived.

Considering Model I, for example, it has a transition from positive
to negative $q(z)$ at $z=0.44$ while this happens at $z=0.68$ in a
$\Lambda$CDM scenario. At the same time, transition from a matter
dominated epoch to a dark energy dominated one happens at $z \sim 0.1$
in our model, while it is at $z=0.32$ in $\Lambda$CDM.

Considering Model II, for example, it has a transition from positive
to negative $q(z)$ at $z=0.62$ and from dark matter to dark energy
domination at $z \sim 0.42$. Even if it has dynamical parameters different
from the $\Lambda$CDM ones, it also exhibits gross properties that are
very similar to the consensus model, but at the same time the jerk and
the snap behavior are much different only converging to $\Lambda$CDM
values at high redshifts.

Model III and IV are different from previous models because dark
energy fluid (given our CPL parameters values) has always been an
\textit{important} ingredient of universe dynamics with negative
pressure ($w_{a}$ is negative). The Model III seems to be the most
problematic showing a transition from positive to negative $q(z)$
at $z=1.52$ and from dark matter to dark energy domination at $z \sim 1.23$.
On the contrary, Model IV shows more normal properties, as a transition
from positive to negative $q(z)$ at $z=0.53$ and from dark matter
to dark energy domination at $z \sim 0.16$ also exhibiting very
similar trends for the jerk and the snap.

\section{Conclusions}
\label{sec:Conclusion}

In this paper we are intended to perform a very detailed analysis
of Cosmography, to state its degree of usability and extendibility
when used as a model to universe dynamics.

First of all we have shown that results from a cosmographic
approach are \textit{sub-judice} because of the observational
errors of cosmological data one works with. When working with
cosmography one has to choose the highest order of expansion for the
series of the $a(t)$ function so stating how many cosmographic
(i.e. cosmological) parameters it is possible to extract from the
analysis. But we have seen that differences between using a two,
three or four dimensional series are well inside the errors given
by data, so that (at the moment) any possibility of exactly stating
what kind of series is better to be used is out of question.

From statistical considerations, we have seen that working
with the series expansion approach, it is possible to derive up to
a maximum of three cosmographic parameters, the deceleration, the
jerk and the snap. Finally, the statistical method we have chosen
to work with, MCMC, gives us the possibility to sensibly narrow down
the confidence errors on cosmographic parameters with respect to
other results in the literature (coming from the same series
approach).

Then MCMCs give also the possibility to manage priors on physical
quantities. Comparing results with the literature we see that inside
the narrower errors our estimations still agree with many works;
what is more interesting, is that they also agree with a
different approach leading to numerical evaluation of an exact
analytical function of distance modulus (\cite{Rapetti07}).

What is even more interesting is what seems to come out from the
analysis of dark energy parameters (in the CPL parametrization)
related to the set of cosmographic parameters we have found out.
We also know that Cosmography could be used as an independent tool
to detect possible deviations from the $\Lambda$CDM model; but
until now nobody has checked what the obtained cosmographic sets
means in terms of dynamical information.

We have tried to give an answer, but this is still quite puzzling:
cosmographic sets can be related to DE models which are quite
slightly different from $\Lambda$CDM (at least at low redshifts).
At the same time, smaller differences in the Cosmographic parameters
values can give great differences in the dynamical cosmological model.
For example we have seen that in the case of two cosmographic parameters, when
working with only SN, we have a good agreement with literature.
But these values correspond to a constant dark energy model with a
low matter content. At the same time, this model is not able to
fit Hubble data at high redshift.

Finally, very different global dynamics properties, can produce
the same gross trends for Hubble and deceleration parameters (as much the
jerk and the snap ones) pose the question of a possible cosmological dynamical
degeneracy in the universe, which also makes more difficult to
eventually detect alternative scenarios to $\Lambda$CDM.

In this optics developing cosmography is most important giving the possibility to
discriminate among models using absolute values of cosmographic parameters.
We think that all these questions could be solved improving the
series expansion approach of Cosmography: we are studying the
possibility to define an \textit{Exact Cosmography} which could be
solve the cosmographic degeneracy. A detailed description of this approach
will be given in a forthcoming paper.

%
%

\appendix
\label{sec:appendix}

\section{Cosmological distances in Cosmography}

The physical distance traveled by a photon
emitted at time $t_{*}$ and absorbed at the current epoch $t_{0}$ is
\begin{equation}
D = c \int dt = c (t_{0} - t_{*}),
\end{equation}
so that inserting $t_{*} = t_{0} - {D}/{c}$ into Eq.~(\ref{eq: aseries})
gives us an expression for the redshift as a function of
$t_0$ and $D/c$, i.e. $z = z(D)$. Solving with respect to $D$, up
to the fifth order in $z$, gives us the desired expansion for
$D(z)$:
{\setlength\arraycolsep{0.2pt}
\begin{eqnarray}\label{eq: seriesdz}
D(z) &=& \frac{c z}{H_{0}} \left\{ \mathcal{D}_{z}^{0} +
\mathcal{D}_{z}^{1} z + \mathcal{D}_{z}^{2} z^{2} +
\mathcal{D}_{z}^{3} z^{3} \right.\nonumber\\
&&\left.+ \mathcal{D}_{z}^{4} z^{4} +\emph{O}[z^{5}]\right\}.
\end{eqnarray}}
In the latter we have defined
{\setlength\arraycolsep{0.2pt}\begin{eqnarray}
\mathcal{D}_{z}^{0} &=& 1, \nonumber
\\
\mathcal{D}_{z}^{1} &=& -\left(1 + \frac{q_0}{2}\right), \nonumber
\\
\mathcal{D}_{z}^{2} &=& 1 + q_{0} + \frac{q_{0}^{2}}{2} -
\frac{j_{0}}{6}, \nonumber
\\
\mathcal{D}_{z}^{3} &=&  - \left ( 1 + \frac{3}{2}q_{0}+
\frac{3}{2}q_{0}^{2} + \frac{5}{8} q_{0}^{3} - \frac{1}{2} j_{0} -
\frac{5}{12} q_{0} j_{0} - \frac{s_{0}}{24} \right ), \nonumber
\\
\mathcal{D}_{z}^{4} &=& 1 + 2 q_{0} + 3 q_{0}^{2} + \frac{5}{2}
q_{0}^{3} + \frac{7}{2} q_{0}^{4} - \frac{5}{3} q_{0}
j_{0} - \frac{7}{8} q_{0}^{2} j_{0}\nonumber\\& -& j_{0} + \frac{j_{0}^{2}}{12}
- \frac{1}{8} q_{0} s_{0}  - \frac{s_{0}}{6} - \frac{l_{0}}{120}.
\nonumber
\end{eqnarray}}
The Taylor series  expansion of the quantity
$D(z)$ is a building block for other quantities to be derived in the following sections as required by the cosmographic approach.

In typical applications, one is not interested in the physical
distance $D(z)$, but in other definitions. In our case, the
\textit{luminosity distance},
\begin{equation}\label{eq:dldef}
D_{L} = \frac{a(t_{0})}{a(t_{0} - D/c)} (a(t_{0}) r_{0}(D)),
\end{equation}
has a particular relevance. In Eq.~(\ref{eq:dldef}) we have used
{\setlength\arraycolsep{0.2pt}
\begin{eqnarray}
r_{0}(D) = \left\{
\begin{array}{ll}
\displaystyle{\sin{\int_{t_0 - D/c}^{t_0}{\frac{c dt}{a(t)}}}} & k = 1,\\
~ & ~ \\
\displaystyle{\int_{t_0 - D/c}^{t_0}{\frac{c dt}{a(t)}}} & k = 0, \\
~ & ~ \\
\displaystyle{\sinh{\int_{t_0 - D/c}^{t_0}{\frac{c dt}{a(t)}}}} & k = -1. \\
\end{array}
\right.
\end{eqnarray}}
Using Eq.~(\ref{eq: seriesdz}), and after some lengthy algebra, we obtain
{\setlength\arraycolsep{0.2pt}
\begin{eqnarray}
\frac{r_{0}(D)}{D/a_0} & = & {\mathcal{R}}_{D}^{0} +
{\mathcal{R}}_{D}^{1} \left ( \frac{H_{0} D}{c} \right ) +
{\mathcal{R}}_{D}^{2} \left ( \frac{H_{0} D}{c} \right )^{2}
\nonumber \\ & + & {\mathcal{R}}_{D}^{3} \left ( \frac{H_{0} D}{c}
\right )^{3} + {\mathcal{R}}_{D}^{4} \left ( \frac{H_{0} D}{c}
\right )^{4} + {\mathcal{R}}_{D}^{5} \left ( \frac{H_{0} D}{c}
\right )^{5} \nonumber,\label{eq: rzd}
\end{eqnarray}}
with
{\setlength\arraycolsep{0.2pt}
\begin{eqnarray}
\mathcal{R}_{D}^{0} &=& 1 \, ,\nonumber
\\
\mathcal{R}_{D}^{1} &=& \frac{1}{2} \, ,\nonumber
\\ \mathcal{R}_{D}^{2} &=& \frac{1}{6} \left [ 2 + q_{0} - \frac{k
c^{2}}{H_{0}^{2} a_{0}^{2}} \right ] \, ,\nonumber
\\
\mathcal{R}_{D}^{3} &=& \frac{1}{24} \left[ 6 + 6 q_{0} + j_{0} -
6 \frac{k c^{2}}{H_{0}^{2} a_{0}^{2}} \right] \, ,\nonumber
\\ \mathcal{R}_{D}^{4}  &=& \frac{1}{120} \left [ 24 + 36 q_{0} + 6
q_{0}^{2} + 8 j_{0} - s_{0} - \frac{5kc^{2}(7 + 2
q_{0})}{a_{0}^{2} H_{0}^{2}} \right ] \, ,\nonumber
\\
\mathcal{R}_{D}^{5} &=& \frac{1}{120} \left[ 24 + 48 q_{0} + 18
q_{0}^{2} + 4 q_{0} j_{0} + 12 j_{0} - 2 s_{0} + 24 l_{0} \right.
\nonumber \\  &-& \left.\frac{3kc^{2}(15 + 10 q_{0} +
j_{0})}{a_{0}^{2} H_{0}^{2}} \right ] \, .\nonumber
\end{eqnarray}}
One now must rewrite  $r_0(D)$  as a function of $z$ using
Eq.~(\ref{eq: seriesdz}). Inserting the result into
Eq.~(\ref{eq:dldef}), one obtains the fifth order approximation for
the Hubble free luminosity distance $d_L = D_L(z)/(c/H_0)$ as a
function of the redshift $z$:
\begin{equation}\label{eq: dlseries}
d_{L}(z) = \mathcal{D}_{L}^{0} z + \mathcal{D}_{L}^{1} \ z^2 +
\mathcal{D}_{L}^{2} \ z^{3} + \mathcal{D}_{L}^{3} \ z^{4} +
\mathcal{D}_{L}^{4} \ z^{5}.
\end{equation}
In the latter we are using the definitions
{\setlength\arraycolsep{0.2pt}
\begin{eqnarray}\label{eq: dlseries2}
\mathcal{D}_{L}^{0} &=& 1 \, ,\nonumber \\
\mathcal{D}_{L}^{1} &=& - \frac{1}{2} \left[ -1 + q_{0} \right] \,
, \nonumber \\
\mathcal{D}_{L}^{2} &=& - \frac{1}{6} \left[1 - q_{0} - 3q_{0}^{2}
+ j_{0} + \frac{k c^{2}}{H_{0}^{2}a_{0}^{2}}\right] \, ,\nonumber \\
\mathcal{D}_{L}^{3} &=& \frac{1}{24}\left[2 - 2 q_{0} - 15
q_{0}^{2} -
15 q_{0}^{3} + 5 j_{0} + 10 q_{0} j_{0} + s_{0}\right.\nonumber \\
&+& \left. \frac{2 k c^{2} (1 + 3 q_{0})}{H_{0}^{2} a_{0}^{2}}
\right] \, ,\nonumber \\
\mathcal{D}_{L}^{4} &=& \frac{1}{120}\left[-6 + 6 q_{0} + 81
q_{0}^{2} + 165 q_{0}^{3} + 105 q_{0}^{4} + 10 j_0^{2}\right.\nonumber\\& - & 27 j_{0}
- \left.110 q_{0} j_{0} - 105 q_{0}^{2} j_{0}  - 15 q_{0} s_{0}
-
11 s_{0} - l_{0} \right. \nonumber \\
&-& \left. \frac{5kc^{2}(1 + 8 q_{0} + 9 q_{0}^{2} - 2 j_{0})}{
a_{0}^{2} H_{0}^{2}} \right] \, .\nonumber \\
\end{eqnarray}}
Previous relations  in this section have been derived for any
value of the curvature parameter; in \cite{Vitagliano10}, it is
shown that ranging the curvature parameter in the interval $[-1,1]$ has
negligible effects on the estimation of cosmographic parameters, so that,  in the following
of the paper, we will assume a flat universe and use the simplified versions
with $k = 0$.

\section{Y redshift relations}
\label{sec:y_redshift}

We write here all the needed relations expressed in term of the
y-redshift for $k = 0$. The Hubble free luminosity distance
$d_{L}(y)$ is:
\begin{equation}
d_{L}(y) = \mathcal{D}_{L}^{0} y + \mathcal{D}_{L}^{1} \ y^2 +
\mathcal{D}_{L}^{2} \ y^{3} + \mathcal{D}_{L}^{3} \ y^{4} +
\mathcal{D}_{L}^{4} \ y^{5},
\end{equation}
where {\setlength\arraycolsep{0.2pt}
\begin{eqnarray}
\mathcal{D}_{L}^{0} &=& 1, \nonumber \\
\mathcal{D}_{L}^{1} &=& \frac{3}{2} \left[ 1 - q_{0} \right],
\nonumber \\
\mathcal{D}_{L}^{2} &=& \frac{1}{6} \left[11 - 5q_{0} + 3q_{0}^{2}
- j_{0} \right], \nonumber \\
\mathcal{D}_{L}^{3} &=& \frac{1}{24}\left[50 - 26 q_{0} + 21
q_{0}^{2} - 15 q_{0}^{3} -7 j_{0} + 10 q_{0} j_{0} + s_{0}
\right], \nonumber \\
\mathcal{D}_{L}^{4} &=& \frac{1}{120}\left[274 -154 q_{0} + 141
q_{0}^{2} - 135 q_{0}^{3} + 105 q_{0}^{4} + 10 j_0^{2}\right.\nonumber\\ & -& 47 j_{0}
+ \left.90 q_{0} j_{0} - 105 q_{0}^{2} j_{0}  - 15 q_{0} s_{0} +
9 s_{0} - l_{0} \right].\nonumber \\
\end{eqnarray}}
The distance modulus is
\begin{equation}\label{eq:museries_y}
\mu(y) = \frac{5}{\log 10} \cdot \left( \log y + \mathcal{M}^{1} y
+ \mathcal{M}^{2} y^2 + \mathcal{M}^{3} y^{3} + \mathcal{M}^{4}
y^{4} \right),
\end{equation}
with {\setlength\arraycolsep{0.2pt}
\begin{eqnarray}
\mathcal{M}^{1} &=& \frac{3}{2} \left[ 1 - q_{0} \right],
\nonumber \\
\mathcal{M}^{2} &=& \frac{1}{24} \left[17 - 2 q_{0} + 9q_{0}^{2}
- 4j_{0} \right], \nonumber \\
\mathcal{M}^{3} &=& \frac{1}{24}\left[11 - q_{0} + 2 q_{0}^{2} -
10 q_{0}^{3} - j_{0} + 8 q_{0} j_{0} + s_{0} \right], \nonumber
\\
\mathcal{M}^{4} &=& \frac{1}{2880}\left[971 -76 q_{0} + 134
q_{0}^{2} - 300 q_{0}^{3} + 1575 q_{0}^{4} + 200 j_0^{2}  \right. \nonumber \\
&-& \left. 68 j_{0} + 260 q_{0} j_{0} - 1800 q_{0}^{2} j_{0} - 300
q_{0} s_{0} + 36 s_{0}  \right. \nonumber \\
&-& \left. 24 l_{0} \right].
\end{eqnarray}}
The Hubble parameter is
\begin{eqnarray}
H(y) &=& H_{0} \cdot \left( \mathcal{H}^{0}_{0} +
\mathcal{H}^{1}_{0} \ y + \mathcal{H}^{2}_{0} \ y^{2} +
\mathcal{H}^{3}_{0} \ y^{3}\right.\nonumber\\
& +&\left. \mathcal{H}^{4}_{0} \ y^{4}\right)
\end{eqnarray}
where{\setlength\arraycolsep{0.2pt}
\begin{eqnarray}
\mathcal{H}^{0}_{0} = 1,
\end{eqnarray}
\begin{eqnarray}
\mathcal{H}^{1}_{0} = 1 + q_{0},
\end{eqnarray}
\begin{eqnarray}
\mathcal{H}^{2}_{0} = 1 + \frac{j_{0}}{2} + q_{0} -
\frac{q_{0}^2}{2},
\end{eqnarray}
\begin{eqnarray}
\mathcal{H}^{3}_{0} = \frac{1}{6} \left[ 6 + 6 q_{0} - 3 q_{0}^{2}
+ 3 q_{0}^{3}+ 3 j_{0} - 4 j_{0} q_{0}  - s_{0} \right]
\end{eqnarray}
\begin{eqnarray}
\mathcal{H}^{4}_{0} &=& \frac{1}{24} \left[ 24 + 24 q_{0} - 12
q_{0}^{2} + 12 q_{0}^{3} - 15 q_{0}^{4} + 12 j_{0} - 4 j_{0}^{2}
\right. \nonumber \\ &-& \left. 16 j_{0} q_{0} + 25 j_{0} q_{0}^2
- 4 s_{0} + 7 q_{0} s_{0} + l_{0} \right],
\end{eqnarray}}
The square of the Hubble parameter is
\begin{eqnarray}
H^{2}(y) &=& H^{2}_{0} \left( \mathcal{H}^{2,0}_{0} +
\mathcal{H}^{2,1}_{0} y + \mathcal{H}^{2,2}_{0} y^{2} +
\mathcal{H}^{2,3}_{0} y^{3} \right. \nonumber \\
&+& \left. \mathcal{H}^{2,4}_{0} y^{4}\right),
\end{eqnarray}
where{\setlength\arraycolsep{0.2pt}
\begin{eqnarray}
\mathcal{H}^{2,0}_{0} = 1,
\end{eqnarray}
\begin{eqnarray}
\mathcal{H}^{2,1}_{0} = 2 \left[ 1 + q_{0} \right],
\end{eqnarray}
\begin{eqnarray}
\mathcal{H}^{2,2}_{0} = 3 + 4 q_{0} + j_{0},
\end{eqnarray}
\begin{eqnarray}
\mathcal{H}^{2,3}_{0} = 4 + 6 q_{0} + 2 j_{0} - \frac{j_{0} \
q_{0}}{3} - \frac{s_{0}}{3},
\end{eqnarray}
\begin{eqnarray}
\mathcal{H}^{2,4}_{0} &=& \frac{1}{12} \left[60 + 96 q_{0} + 36
j_{0} - j_{0}^{2}  - 8 j_{0} q_{0} + 3 j_{0} q_{0} - 8 s_{0}  \right. \nonumber \\
&+& \left. 3 q_{0} s_{0} + l_{0} \right].
\end{eqnarray}}

\section{H derivatives}
\label{sec:derivativesH}

Here we report convenient expressions for the conversion from
higher order derivatives in time to higher order derivatives in
redshift: {\setlength\arraycolsep{0.2pt}
\begin{eqnarray}\label{eq:secondHt}
\frac{d^{2}}{dt^{2}} &=& (1+z) H \left[H+ (1+z) \frac{d
H}{dz}\right] \frac{d}{dz} \nonumber \\
&+& (1 + z)^{2} H^{2} \frac{d^{2}}{dz^{2}},
\end{eqnarray}
\begin{eqnarray}\label{eq:thirdHt}
\frac{d^{3}}{dt^{3}} = &-& (1 + z) H \left\{ H^{2} + (1 + z)^{2}
\left(\frac{dH}{dz}\right)^{2} + (1 + z) H  \right. \nonumber \\
&\times& \left. \left[ 4 \frac{dH}{dz} + (1 + z)
\frac{d^{2}H}{dz^{2}}
\right] \right\} \frac{d}{dz} - 3 (1 + z)^{2} H^{2}  \nonumber \\
&\times& \left[ H + (1+z) \frac{dH}{dz} \right]
\frac{d^{2}}{dz^{2}} - (1 + z)^{3} H^{3} \frac{d^{3}}{dz^{3}},
\end{eqnarray}
\begin{eqnarray}\label{eq:fourthHt}
\frac{d^{4}}{dt^{4}} &=& (1 + z) H \left[ H^{2} + 11 (1+z) H^{2}
\frac{dH}{dz} + 11 (1+z) H   \right. \nonumber
\\ &\times& \left. \frac{dH}{dz} + (1+z)^{3}
\left(\frac{dH}{dz}\right)^{3} + 7 (1+z)^{2} H \frac{d^{2}H}{dz^{2}}  \right. \nonumber\\
&+& \left. 4 (1+z)^{3} H \frac{dH}{dz} \frac{d^{2}H}{d^{2}z} +
(1+z)^{3} H^{2} \frac{d^{3}H}{d^{3}z}\right] \frac{d}{dz}  \nonumber \\
&+& (1 + z)^{2} H^{2} \left[ 7 H^{2} + 22 H \frac{dH}{dz} + 7 (1 +
z)^{2} \left(\frac{dH}{dz}\right)^{2} \nonumber \right. \\
&+& \left. 4 H \frac{d^{2}H}{dz^{2}} \right]
\frac{d^{2}}{dz^{2}} + 6 (1 + z)^{3} H^{3} \left[ H + (1+z) \frac{dH}{dz} \right]  \nonumber \\
&\times& \frac{d^{3}}{dz^{3}} + (1 + z)^{4} H^{4}
\frac{d^{4}}{dz^{4}} +  (1 + z)^{4} H^{4} \frac{d^{4}}{dz^{4}}.
\end{eqnarray}}

\section{Square Hubble parameter}
\label{sec:squareH}

The expansion series of the square Hubble
parameter, $H^{2}$, will be a useful tool as well. It is a simple and just a long matter of algebra to
compute its derivatives with the same procedure as above. The final
results are
\begin{equation}
\frac{d(H^{2})}{dz} = \frac{2 H^{2}}{1 + z} (1 + q)
\end{equation}
\begin{equation}
\frac{d^{2}(H^{2})}{dz^{2}} = \frac{2H^{2}}{(1 + z)^{2}} (1 + 2 q
+ j)
\end{equation}
\begin{equation}
\frac{d^{3}(H^{2})}{dz^{3}} = \frac{2H^{2}}{(1 + z)^{3}} (- q j -
s)
\end{equation}
\begin{equation}
\frac{d^{4}(H^{2})}{dz^{4}} = \frac{2H^{2}}{(1 + z)^{4}} ( 4 q j +
3 q s + 3 q^{2} j - j^{2} + 4 s + l)
\end{equation}

{\renewcommand{\arraystretch}{1.5}
\begin{table*}
\begin{minipage}{\textwidth}
\caption{Results in the two dimensional parameter space
($q_{0}$,\,$j_{0}$) and z-redshift. Column~(1):
simulation identification in the text with the used data
($S=$ SNeIa, $G=$ GRBs, $H=$ Hubble, $B=$ BAO; the index $(c)$ refers to the redshift-cut sub-sample,
with $z<1$) and the applied priors (we indicate the redshift range where positiveness of quantities described
in \S~(\ref{sec:preliminary}) is applied). Columns~(2)~-~(3)~-~(4):
best fit estimations of cosmographic parameters with $1\sigma$
confidence level. Column~(5): chi square and reduced chi square
values. Columns~(6)~-~(7)~-~(8)~-~(9): best fit estimations of dark
energy (CPL) parameters with $1\sigma$ confidence level for the constant
dark energy model and the dynamical one with fixed $\Omega_{m}$ as described
in \S~(\ref{sec:CPL}).}\label{lab:2P_Z} \resizebox*{\textwidth}{!}{
\begin{tabular}{cccccccc}
\hline
\multicolumn{1}{c}{$\mathrm{Data}$} & \multicolumn{3}{c}{$\mathrm{Best \; fit \; parameters}$} & \multicolumn{2}{c}{$\mathrm{CDE}$} & \multicolumn{2}{c}{$\mathrm{DDE} \; (\Omega_{m} = 0.245)$} \\
\hline \hline
S & $q_{0}$                    & $j_{0}$                     & $\chi^{2} \, (\chi^{2}/d.o.f.)$ & $\Omega_{m}$ & $w_{0}$ & $w_{0}$  & $w_{a}$ \\
\hline
$z>0$       & $-1.014^{+0.008}_{-0.013}$ & $1.060^{+0.063}_{-0.035}$ & $622.11 \,(1.123)$ & $0.004^{+0.005}_{-0.003}$ & $-1.013^{+0.008}_{-0.014}$ & $-1.337^{+0.007}_{-0.011}$ & $-1.295^{+0.005}_{-0.011}$ \\
$z<1.4$     & $-0.738^{+0.043}_{-0.043}$ & $0.372^{+0.099}_{-0.085}$ & $573.95 \,(1.036)$ & $0.005^{+0.009}_{-0.004}$ & $-0.831^{+0.029}_{-0.031}$ & $-1.093^{+0.039}_{-0.038}$ & $-0.858^{+0.059}_{-0.059}$ \\
$z>0^{(c)}$ & $-1.017^{+0.010}_{-0.017}$ & $1.072^{+0.082}_{-0.045}$ & $599.21 \,(1.122)$ & $0.005^{+0.006}_{-0.003}$ & $-1.016^{+0.010}_{-0.017}$ & $-1.339^{+0.009}_{-0.014}$ & $-1.295^{+0.005}_{-0.011}$ \\
$z<1^{(c)}$ & $-0.669^{+0.049}_{-0.052}$ & $0.284^{+0.151}_{-0.100}$ & $529.97 \,(0.992)$ & $0.019^{+0.027}_{-0.014}$ & $-0.796^{+0.038}_{-0.049}$ & $-1.033^{+0.043}_{-0.046}$ & $-0.723^{+0.063}_{-0.064}$ \\
\hline \hline
S-G  & $q_{0}$                    & $j_{0}$                     & $\chi^{2} \, (\chi^{2}/d.o.f.)$ & $\Omega_{m}$ & $w_{0}$ & $w_{0}$  & $w_{a} $ \\
\hline
$z>0$       & $-1.004^{+0.003}_{-0.005}$ & $1.016^{+0.016}_{-0.011}$ & $1166.07 \,(1.872)$ & $0.0004^{+0.0038}_{-0.0032}$ & $-1.003^{+0.002}_{-0.003}$ & $-1.328^{+0.003}_{-0.004}$ & $-1.293^{+0.014}_{-0.023}$ \\
$z<6.7$     & $-0.507^{+0.026}_{-0.031}$ & $0.010^{+0.034}_{-0.025}$ & $1015.14 \,(1.629)$ & $0.001^{+0.002}_{-0.001}$ & $-0.672^{+0.017}_{-0.020}$ & $-0.889^{+0.023}_{-0.027}$ & $-0.578^{+0.030}_{-0.036}$ \\
$z>0^{(c)}$ & $-1.016^{+0.009}_{-0.017}$ & $1.068^{+0.084}_{-0.042}$ & $602.03 \, (1.089)$ & $0.004^{+0.007}_{-0.003}$ & $-1.015^{+0.009}_{-0.018}$ & $-1.338^{+0.008}_{-0.015}$ & $-1.295^{+0.005}_{-0.010}$ \\
$z<1^{(c)}$ & $-0.673^{+0.045}_{-0.050}$ & $0.292^{+0.149}_{-0.096}$ & $533.20 \, (0.964)$ & $0.020^{+0.027}_{-0.015}$ & $-0.799^{+0.035}_{-0.049}$ & $-1.036^{+0.040}_{-0.044}$ & $-0.727^{+0.065}_{-0.062}$ \\
\hline \hline
S-H  & $q_{0}$                    & $j_{0}$                     & $\chi^{2} \, (\chi^{2}/d.o.f.)$ & $\Omega_{m}$ & $w_{0}$ & $w_{0}$  & $w_{a} $ \\
\hline
$z>0$       & $-1.159^{+0.059}_{-0.048}$ & $1.859^{+0.291}_{-0.337}$ & $717.88 \, (1.268)$ & $0.056^{+0.015}_{-0.019}$ & $-1.170^{+0.063}_{-0.053}$ & $-1.464^{+0.052}_{-0.042}$ & $-1.283^{+0.010}_{-0.011}$ \\
$z<1.76$    & $-0.636^{+0.052}_{-0.056}$ & $0.219^{+0.152}_{-0.095}$ & $637.70 \, (1.127)$ & $0.017^{+0.023}_{-0.013}$ & $-0.771^{+0.040}_{-0.052}$ & $-1.003^{+0.046}_{-0.049}$ & $-0.689^{+0.055}_{-0.055}$ \\
$z>0^{(c)}$ & $-1.016^{+0.010}_{-0.016}$ & $1.069^{+0.081}_{-0.041}$ & $606.06 \, (1.118)$ & $0.004^{+0.006}_{-0.003}$ & $-1.015^{+0.009}_{-0.017}$ & $-1.339^{+0.008}_{-0.014}$ & $-1.295^{+0.005}_{-0.010}$ \\
$z<1^{(c)}$ & $-0.667^{+0.048}_{-0.051}$ & $0.288^{+0.151}_{-0.103}$ & $532.80 \, (0.983)$ & $0.022^{+0.028}_{-0.016}$ & $-0.797^{+0.039}_{-0.049}$ & $-1.031^{+0.042}_{-0.046}$ & $-0.714^{+0.065}_{-0.062}$ \\
\hline \hline
S-G-H  & $q_{0}$                    & $j_{0}$                     & $\chi^{2} \, (\chi^{2}/d.o.f.)$ & $\Omega_{m}$ & $w_{0}$ & $w_{0}$  & $w_{a} $ \\
\hline
$z>0$       & $-1.004^{+0.002}_{-0.004}$ & $1.015^{+0.015}_{-0.009}$ & $1267.78 \,(1.996)$ & $0.0005^{+0.0008}_{-0.0004}$ & $-1.003^{+0.002}_{-0.003}$ & $-1.328^{+0.002}_{-0.004}$ & $-1.293^{+0.003}_{-0.006}$ \\
$z<6.7$     & $-0.476^{+0.010}_{-0.018}$ & $-0.019^{+0.017}_{-0.011}$ & $1050.51 \,(1.654)$ & $0.001^{+0.002}_{-0.001}$ & $-0.652^{+0.007}_{-0.012}$ & $-0.862^{+0.009}_{-0.016}$ & $-0.543^{+0.011}_{-0.020}$ \\
$z>0^{(c)}$ & $-1.017^{+0.010}_{-0.017}$ & $1.074^{+0.086}_{-0.046}$ & $609.44 \,(1.086)$ & $0.005^{+0.007}_{-0.004}$ & $-1.016^{+0.010}_{-0.018}$ & $-1.340^{+0.009}_{-0.014}$ & $-1.295^{+0.005}_{-0.010}$ \\
$z<1^{(c)}$ & $-0.667^{+0.050}_{-0.052}$ & $0.285^{+0.162}_{-0.104}$ & $535.94 \,(0.955)$ & $0.021^{+0.030}_{-0.015}$ & $-0.796^{+0.040}_{-0.052}$ & $-1.031^{+0.044}_{-0.046}$ & $-0.714^{+0.068}_{-0.064}$ \\
\hline \hline
S-G-H-B & $q_{0}$                    & $j_{0}$                     & $\chi^{2} \, (\chi^{2}/d.o.f.)$ & $\Omega_{m}$ & $w_{0}$ & $w_{0}$  & $w_{a} $ \\
\hline
$z>0$       & $-1.004^{+0.002}_{-0.003}$ & $1.014^{+0.016}_{-0.008}$ & $1267.61 \,(1.993)$ & $0.0005^{+0.0008}_{-0.0004}$ & $-1.003^{+0.002}_{-0.003}$ & $-1.328^{+0.002}_{-0.003}$ & $-1.293^{+0.003}_{-0.006}$ \\
$z<6.7$     & $-0.477^{+0.010}_{-0.018}$ & $-0.018^{+0.017}_{-0.011}$ & $1051.21 \,(1.653)$ & $0.001^{+0.002}_{-0.001}$ & $-0.653^{+0.007}_{-0.012}$ & $-0.863^{+0.009}_{-0.016}$ & $-0.544^{+0.011}_{-0.020}$ \\
$z>0^{(c)}$ & $-1.016^{+0.009}_{-0.016}$ & $1.072^{+0.080}_{-0.043}$ & $609.08 \,(1.084)$ & $0.005^{+0.007}_{-0.003}$ & $-1.016^{+0.009}_{-0.017}$ & $-1.339^{+0.008}_{-0.014}$ & $-1.294^{+0.004}_{-0.010}$ \\
$z<1^{(c)}$ & $-0.667^{+0.049}_{-0.050}$ & $0.282^{+0.153}_{-0.104}$ & $536.17 \,(0.954)$ &
$0.020^{+0.028}_{-0.015}$ & $-0.795^{+0.040}_{-0.050}$ & $-1.030^{+0.044}_{-0.044}$ & $-0.715^{+0.065}_{-0.063}$ \\
\hline \hline
S-H-B & $q_{0}$                    & $j_{0}$                     & $\chi^{2} \, (\chi^{2}/d.o.f.)$ & $\Omega_{m}$ & $w_{0}$ & $w_{0}$  & $w_{a} $ \\
\hline
$z>0$       & $-1.152^{+0.052}_{-0.050}$ & $1.819^{+0.308}_{-0.294}$ & $718.55 \,(1.267)$ & $0.055^{+0.015}_{-0.017}$ & $-1.165^{+0.057}_{-0.055}$ & $-1.459^{+0.046}_{-0.045}$ & $-1.283^{+0.009}_{-0.010}$ \\
$z<1.76$    & $-0.637^{+0.050}_{-0.059}$ & $0.223^{+0.163}_{-0.092}$ & $638.01 \,(1.125)$ & $0.018^{+0.023}_{-0.013}$ & $-0.773^{+0.038}_{-0.054}$ & $-1.004^{+0.044}_{-0.051}$ & $-0.691^{+0.054}_{-0.056}$ \\
$z>0^{(c)}$ & $-1.017^{+0.010}_{-0.017}$ & $1.074^{+0.083}_{-0.045}$ & $606.23 \,(1.116)$ & $0.005^{+0.007}_{-0.004}$ & $-1.016^{+0.010}_{-0.017}$ & $-1.339^{+0.009}_{-0.014}$ & $-1.295^{+0.005}_{-0.010}$ \\
$z<1^{(c)}$ & $-0.669^{+0.048}_{-0.051}$ & $0.288^{+0.152}_{-0.105}$ & $532.98 \,(0.981)$ & $0.021^{+0.028}_{-0.015}$ & $-0.797^{+0.040}_{-0.048}$ & $-1.032^{+0.042}_{-0.045}$ & $-0.718^{+0.065}_{-0.063}$ \\
\hline \hline
\end{tabular}}
\end{minipage}
\end{table*}}

{\renewcommand{\arraystretch}{1.5}
\begin{table*}
\begin{minipage}{\textwidth}
\caption{Results in the three dimensional parameter space
($q_{0}$, $j_{0}$, $s_{0}$) and z-redshift. Void rows correspond to
MCMC which do not pass the convergence test. Column~(1):
simulation identification in the text with the used data
($S=$ SNeIa, $G=$ GRBs, $H=$ Hubble, $B=$ BAO; the index $(c)$ refers to the redshift-cut sub-sample,
with $z<1$) and the applied priors
(we indicate the redshift range where positiveness of quantities described
in \S~(\ref{sec:preliminary}) is applied). Columns~(2)~-~(3)~-~(4):
best fit estimations of cosmographic parameters with $1\sigma$
confidence level. Column~(5): chi square and reduced chi square
values. Columns~(6)~-~(7)~-~(8)): best fit estimations of dark
energy (CPL) parameters with $1\sigma$ confidence level for the dynamical
dark energy model as described in \S~(\ref{sec:CPL}).}\label{lab:3P_Z} \resizebox*{\textwidth}{!}{
\begin{tabular}{cccccccc}
\hline
\multicolumn{1}{c}{$\mathrm{Data}$} & \multicolumn{4}{c}{$\mathrm{Best \; fit \; parameters}$} & \multicolumn{3}{c}{$\mathrm{DDE}$} \\
\hline \hline
S & $q_{0}$ & $j_{0}$ & $s_{0}$ & $\chi^{2} \, (\chi^{2}/d.o.f.)$ & $\Omega_{m}$ & $w_{0}$  & $w_{a} $ \\
\hline
$z>0$       & $-$ & $-$ & $-$ & $-$ & $-$ & $-$ & $-$ \\
$z<1.4$     & $-$ & $-$ & $-$ & $-$ & $-$ & $-$ & $-$ \\
$z>0^{(c)}$ & $-0.319^{+0.045}_{-0.044}$ & $-0.062^{+0.128}_{-0.275}$ & $-0.680^{+0.456}_{-0.429}$ & $541.76 \,(1.016)$ & $0.158^{+0.066}_{-0.074}$ & $-0.650^{+0.078}_{-0.073}$ & $-0.213^{+0.129}_{-0.093}$ \\
$z<1^{(c)}$ & $-$ & $-$ & $-$ & $-$ & $-$ & $-$ & $-$ \\
\hline \hline
S-G  & $q_{0}$ & $j_{0}$ & $s_{0}$ & $\chi^{2} \, (\chi^{2}/d.o.f.)$ & $\Omega_{m}$ & $w_{0}$  & $w_{a}$ \\
\hline
$z>0$       & $-$ & $-$ & $-$ & $-$ & $-$ & $-$ & $-$ \\
$z<6.7$     & $-$ & $-$ & $-$ & $-$ & $-$ & $-$ & $-$ \\
$z>0^{(c)}$ & $-0.319^{+0.045}_{-0.046}$ & $-0.052^{+0.129}_{-0.286}$ & $-0.688^{+0.464}_{-0.440}$ & $545.36 \,(0.988)$ & $0.159^{+0.067}_{-0.075}$ & $-0.653^{+0.080}_{-0.073}$ & $-0.214^{+0.128}_{-0.090}$ \\
$z<1^{(c)}$ & $-$ & $-$ & $-$ & $-$ & $-$ & $-$ & $-$ \\
\hline \hline
S-H  & $q_{0}$ & $j_{0}$ & $s_{0}$ & $\chi^{2} \, (\chi^{2}/d.o.f.)$ & $\Omega_{m}$ & $w_{0}$  & $w_{a} $ \\
\hline
$z>0$       & $-0.219^{+0.025}_{-0.025}$ & $-0.536^{+0.051}_{-0.040}$ & $-0.203^{+0.061}_{-0.114}$ & $587.99 \,(1.041)$ & $0.110^{+0.018}_{-0.017}$ & $-0.537^{+0.012}_{-0.015}$ & $-0.409^{+0.028}_{-0.015}$ \\
$z<1.76$    & $-0.944^{+0.032}_{-0.039}$ & $3.134^{+0.063}_{-0.088}$ & $4.399^{+0.463}_{-0.556}$ & $614.23 \,(1.087)$ & $0.379^{+0.017}_{-0.019}$ & $-1.553^{+0.010}_{-0.013}$ & $-0.295^{+0.027}_{-0.034}$ \\
$z>0^{(c)}$ & $-0.322^{+0.045}_{-0.045}$ & $-0.071^{+0.124}_{-0.257}$ & $-0.657^{+0.437}_{-0.434}$ & $543.97 \,(1.005)$ & $0.151^{+0.062}_{-0.071}$ & $-0.648^{+0.075}_{-0.071}$ & $-0.212^{+0.132}_{-0.089}$ \\
$z<1^{(c)}$ & $-$ & $-$ & $-$ & $-$ & $-$ & $-$ & $-$ \\
\hline \hline
S-G-H  & $q_{0}$ & $j_{0}$ & $s_{0}$ & $\chi^{2} \, (\chi^{2}/d.o.f.)$ & $\Omega_{m}$ & $w_{0}$  & $w_{a} $ \\
\hline
$z>0$       & $-$ & $-$ & $-$ & $-$ & $-$ & $-$ & $-$ \\
$z<6.7$     & $-$ & $-$ & $-$ & $-$ & $-$ & $-$ & $-$ \\
$z>0^{(c)}$ & $-$ & $-$ & $-$ & $-$ & $-$ & $-$ & $-$ \\
$z<1^{(c)}$ & $-$ & $-$ & $-$ & $-$ & $-$ & $-$ & $-$ \\
\hline \hline
S-G-H-B & $q_{0}$ & $j_{0}$ & $s_{0}$ & $\chi^{2} \, (\chi^{2}/d.o.f.)$ & $\Omega_{m}$ & $w_{0}$  & $w_{a} $ \\
\hline
$z>0$       & $0.336^{+0.021}_{-0.018}$ & $0.131^{+0.045}_{-0.039}$ & $-0.066^{+0.027}_{-0.024}$ & $1031.03 \,(1.624)$ & $0.329^{+0.018}_{-0.034}$ & $-0.163^{+0.028}_{-0.022}$ & $-0.456^{+0.024}_{-0.015}$ \\
$z<6.7$     & $-$ & $-$ & $-$ & $-$ & $-$ & $-$ & $-$ \\
$z>0^{(c)}$ & $-0.328^{+0.046}_{-0.046}$ & $-0.051^{+0.135}_{-0.292}$ & $-0.662^{+0.437}_{-0.432}$ & $549.42 \,(0.979)$ & $0.151^{+0.066}_{-0.070}$ & $-0.652^{+0.075}_{-0.075}$ & $-0.207^{+0.140}_{-0.096}$ \\
$z<1^{(c)}$ & $-$ & $-$ & $-$ & $-$ & $-$ & $-$ & $-$ \\
\hline \hline
S-H-B & $q_{0}$ & $j_{0}$ & $s_{0}$ & $\chi^{2} \, (\chi^{2}/d.o.f.)$ & $\Omega_{m}$ & $w_{0}$  & $w_{a} $ \\
\hline
$z>0$       & $-0.221^{+0.027}_{-0.025}$ & $-0.538^{+0.053}_{-0.040}$ & $-0.202^{+0.059}_{-0.108}$ & $590.19 \,(1.043)$ & $0.109^{+0.019}_{-0.016}$ & $-0.538^{+0.012}_{-0.015}$ & $-0.409^{+0.028}_{-0.015}$ \\
$z<1.76$    & $-0.958^{+0.033}_{-0.026}$ & $3.138^{+0.053}_{-0.052}$ & $4.442^{+0.373}_{-0.389}$ & $615.64 \,(1.088)$ & $0.373^{+0.018}_{-0.012}$ & $-1.551^{+0.010}_{-0.011}$ & $-0.292^{+0.026}_{-0.030}$ \\
$z>0^{(c)}$ & $-0.328^{+0.045}_{-0.045}$ & $-0.063^{+0.138}_{-0.279}$ & $-0.690^{+0.413}_{-0.452}$ & $545.50 \,(1.006)$ & $0.154^{+0.063}_{-0.073}$ & $-0.654^{+0.077}_{-0.072}$ & $-0.218^{+0.134}_{-0.089}$ \\
$z<1^{(c)}$ & $-$ & $-$ & $-$ & $-$ & $-$ & $-$ & $-$ \\
\hline \hline
\end{tabular}}
\end{minipage}
\end{table*}}

{\renewcommand{\arraystretch}{1.5}
\begin{table*}
\begin{minipage}{\textwidth}
\caption{Results in the two dimensional parameter space
($q_{0}$,\,$j_{0}$) and y-redshift. Column~(1):
simulation identification in the text with the used data
($S=$ SNeIa, $G=$ GRBs, $H=$ Hubble, $B=$ BAO; the index $(c)$ refers to the redshift-cut sub-sample,
with $z<1$) and the applied priors
(we indicate the redshift range where positiveness of quantities described
in \S~(\ref{sec:preliminary}) is applied). Columns~(2)~-~(3)~-~(4):
best fit estimations of cosmographic parameters with $1\sigma$
confidence level. Column~(5): chi square and reduced chi square
values. Columns~(6)~-~(7)~-~(8)~-~(9): best fit estimations of dark
energy (CPL) parameters with $1\sigma$ confidence level for the constant
dark energy model and the dynamical one with fixed $\Omega_{m}$ as described
in \S~(\ref{sec:CPL}).}\label{lab:2P_Y} \resizebox*{\textwidth}{!}{
\begin{tabular}{cccccccc}
\hline
\multicolumn{1}{c}{$\mathrm{Data}$} & \multicolumn{3}{c}{$\mathrm{Best \; fit \; parameters}$} & \multicolumn{2}{c}{$\mathrm{CDE}$} & \multicolumn{2}{c}{$\mathrm{DDE} \; (\Omega_{m} = 0.245)$} \\
\hline \hline
S & $q_{0}$                    & $j_{0}$                     & $\chi^{2} \, (\chi^{2}/d.o.f.)$ & $\Omega_{m}$ & $w_{0}$ & $w_{0}$  & $w_{a}$ \\
\hline
$0<y<1$         & $-0.668^{+0.047}_{-0.054}$ & $0.414^{+0.394}_{-0.184}$ & $544.39 \,(0.983)$ & $0.066^{+0.085}_{-0.048}$ & $-0.833^{+0.060}_{-0.114}$ & $-1.032^{+0.042}_{-0.048}$ & $-0.599^{+0.230}_{-0.111}$ \\
$0<y<0.584$     & $-0.664^{+0.047}_{-0.051}$ & $0.373^{+0.360}_{-0.158}$ & $544.31 \,(0.982)$ & $0.056^{+0.079}_{-0.040}$ & $-0.821^{+0.052}_{-0.103}$ & $-1.028^{+0.041}_{-0.044}$ & $-0.621^{+0.199}_{-0.097}$ \\
$0<y<1^{(c)}$   & $-0.646^{+0.052}_{-0.066}$ & $0.375^{+0.524}_{-0.184}$ & $526.08 \,(0.985)$ & $0.069^{+0.109}_{-0.051}$ & $-0.818^{+0.062}_{-0.150}$ & $-1.012^{+0.046}_{-0.058}$ & $-0.562^{+0.295}_{-0.121}$ \\
$0<y<0.5^{(c)}$ & $-0.648^{+0.053}_{-0.062}$ & $0.392^{+0.490}_{-0.196}$ & $526.10 \,(0.985)$ & $0.073^{+0.103}_{-0.054}$ & $-0.824^{+0.066}_{-0.143}$ & $-1.014^{+0.047}_{-0.054}$ & $-0.556^{+0.286}_{-0.124}$ \\
\hline \hline
S-G  & $q_{0}$                    & $j_{0}$                     & $\chi^{2} \, (\chi^{2}/d.o.f.)$ & $\Omega_{m}$ & $w_{0}$ & $w_{0}$  & $w_{a} $ \\
\hline
$0<y<1$         & $-0.671^{+0.054}_{-0.048}$ & $0.405^{+0.400}_{-0.177}$ & $551.19 \,(0.885)$ & $0.063^{+0.084}_{-0.046}$ & $-0.830^{+0.057}_{-0.114}$ & $-1.034^{+0.043}_{-0.047}$ & $-0.609^{+0.227}_{-0.107}$ \\
$0<y<0.871$     & $-0.669^{+0.046}_{-0.056}$ & $0.400^{+0.391}_{-0.170}$ & $551.18 \,(0.885)$ & $0.062^{+0.083}_{-0.045}$ & $-0.829^{+0.055}_{-0.112}$ & $-1.032^{+0.041}_{-0.049}$ & $-0.610^{+0.219}_{-0.108}$ \\
$0<y<1^{(c)}$   & $-0.643^{+0.052}_{-0.055}$ & $0.357^{+0.383}_{-0.175}$ & $529.32 \, (0.957)$ & $0.064^{+0.089}_{-0.047}$ & $-0.813^{+0.059}_{-0.114}$ & $-1.009^{+0.046}_{-0.048}$ & $-0.577^{+0.231}_{-0.108}$ \\
$0<y<0.5^{(c)}$ & $-0.645^{+0.052}_{-0.062}$ & $0.372^{+0.487}_{-0.188}$ & $529.33 \, (0.957)$ & $0.068^{+0.18}_{-0.050}$ & $-0.817^{+0.064}_{-0.144}$ & $-1.011^{+0.046}_{-0.055}$ & $-0.567^{+0.291}_{-0.115}$ \\
\hline \hline
S-H  & $q_{0}$                    & $j_{0}$                     & $\chi^{2} \, (\chi^{2}/d.o.f.)$ & $\Omega_{m}$ & $w_{0}$ & $w_{0}$  & $w_{a} $ \\
\hline
$0<y<1$         & $-0.924^{+0.250}_{-0.190}$ & $3.535^{+2.022}_{-2.478}$ & $589.03 \, (1.041)$ & $0.402^{+0.046}_{-0.171}$ & $-1.594^{+0.577}_{-0.353}$ & $-1.258^{+0.223}_{-0.169}$ & $1.263^{+0.964}_{-1.319}$ \\
$0<y<0.638$     & $-0.905^{+0.255}_{-0.202}$ & $3.360^{+2.116}_{-2.532}$ & $589.24 \, (1.041)$ & $0.397^{+0.049}_{-0.205}$ & $-1.560^{+0.610}_{-0.369}$ & $-1.241^{+0.225}_{-0.179}$ & $1.169^{+1.014}_{-1.368}$ \\
$0<y<1^{(c)}$   & $-0.640^{+0.049}_{-0.065}$ & $0.365^{+0.512}_{-0.180}$ & $528.60 \, (0.975)$ & $0.070^{+0.108}_{-0.051}$ & $-0.814^{+0.060}_{-0.148}$ & $-1.006^{+0.043}_{-0.056}$ & $-0.555^{+0.291}_{-0.121}$ \\
$0<y<0.5^{(c)}$ & $-0.642^{+0.054}_{-0.067}$ & $0.391^{+0.531}_{-0.206}$ & $528.63 \, (0.975)$ & $0.077^{+0.113}_{-0.057}$ & $-0.823^{+0.070}_{-0.159}$ & $-1.009^{+0.048}_{-0.060}$ & $-0.542^{+0.321}_{-0.127}$ \\
\hline \hline
S-G-H  & $q_{0}$                    & $j_{0}$                     & $\chi^{2} \, (\chi^{2}/d.o.f.)$ & $\Omega_{m}$ & $w_{0}$ & $w_{0}$  & $w_{a} $ \\
\hline
$0<y<1$         & $-0.878^{+0.237}_{-0.202}$ & $3.036^{+2.122}_{-2.298}$ & $597.09 \,(0.940)$ & $0.382^{+0.058}_{-0.206}$ & $-1.492^{+0.567}_{-0.388}$ & $-1.217^{+0.209}_{-0.178}$ & $0.991^{+1.032}_{-1.248}$ \\
$0<y<0.871$     & $-0.891^{+0.232}_{-0.188}$ & $3.178^{+1.947}_{-2.264}$ & $596.89 \,(0.940)$ & $0.389^{+0.052}_{-0.180}$ & $-1.521^{+0.545}_{-0.354}$ & $-1.228^{+0.204}_{-0.165}$ & $1.071^{+0.956}_{-1.215}$ \\
$0<y<1^{(c)}$   & $-0.642^{+0.051}_{-0.060}$ & $0.386^{+0.473}_{-0.197}$ & $531.88 \,(0.948)$ & $0.075^{+0.102}_{-0.054}$ & $-0.821^{+0.065}_{-0.139}$ & $-1.009^{+0.045}_{-0.053}$ & $-0.547^{+0.280}_{-0.125}$ \\
$0<y<0.5^{(c)}$ & $-0.645^{+0.053}_{-0.063}$ & $0.395^{+0.529}_{-0.201}$ & $531.88 \,(0.948)$ & $0.077^{+0.108}_{-0.055}$ & $-0.824^{+0.067}_{-0.153}$ & $-1.011^{+0.047}_{-0.055}$ & $-0.543^{+0.303}_{-0.125}$ \\
\hline \hline
S-G-H-B & $q_{0}$                    & $j_{0}$                     & $\chi^{2} \, (\chi^{2}/d.o.f.)$ & $\Omega_{m}$ & $w_{0}$ & $w_{0}$  & $w_{a} $ \\
\hline
$0<y<1$         & $-0.932^{+0.186}_{-0.169}$ & $3.580^{+1.824}_{-1.850}$ & $597.22 \,(0.939)$ & $0.402^{+0.043}_{-0.099}$ & $-1.602^{+0.407}_{-0.312}$ & $-1.265^{+0.164}_{-0.148}$ & $1.273^{+0.856}_{-0.964}$ \\
$0<y<0.871$     & $-0.923^{+0.200}_{-0.179}$ & $3.479^{+1.911}_{-1.965}$ & $597.29 \,(0.939)$ & $0.399^{+0.046}_{-0.121}$ & $-1.581^{+0.451}_{-0.333}$ & $-1.256^{+0.178}_{-0.157}$ & $1.220^{+0.916}_{-1.057}$ \\
$0<y<1^{(c)}$   & $-0.645^{+0.053}_{-0.064}$ & $0.392^{+0.527}_{-0.207}$ & $532.53 \,(0.947)$ & $0.075^{+0.105}_{-0.055}$ & $-0.823^{+0.068}_{-0.146}$ & $-1.011^{+0.046}_{-0.055}$ & $-0.548^{+0.292}_{-0.124}$ \\
$0<y<0.5^{(c)}$ & $-0.641^{+0.049}_{-0.061}$ & $0.360^{+0.479}_{-0.175}$ & $532.50 \,(0.947)$ & $0.067^{+0.106}_{-0.050}$ & $-0.813^{+0.059}_{-0.141}$ & $-1.007^{+0.043}_{-0.054}$ & $-0.563^{+0.279}_{-0.115}$ \\
\hline \hline
S-H-B & $q_{0}$                    & $j_{0}$                     & $\chi^{2} \, (\chi^{2}/d.o.f.)$ & $\Omega_{m}$ & $w_{0}$ & $w_{0}$  & $w_{a} $ \\
\hline
$0<y<1$         & $-0.940^{+0.240}_{-0.177}$ & $3.688^{+1.917}_{-2.377}$ & $589.62 \,(1.040)$ & $0.405^{+0.043}_{-0.149}$ & $-1.623^{+0.545}_{-0.323}$ & $-1.271^{+0.216}_{-0.155}$ & $1.319^{+0.911}_{-1.266}$ \\
$0<y<0.871$     & $-0.912^{+0.259}_{-0.196}$ & $3.390^{+2.097}_{-2.564}$ & $597.29 \,(0.939)$ & $0.397^{+0.050}_{-0.199}$ & $-1.564^{+0.610}_{-0.374}$ & $-1.247^{+0.228}_{-0.173}$ & $1.184^{+1.023}_{-1.366}$ \\
$0<y<1^{(c)}$   & $-0.644^{+0.052}_{-0.060}$ & $0.387^{+0.450}_{-0.198}$ & $529.27 \,(0.975)$ & $0.073^{+0.096}_{-0.054}$ & $-0.822^{+0.066}_{-0.129}$ & $-1.010^{+0.046}_{-0.052}$ & $-0.553^{+0.258}_{-0.120}$ \\
$0<y<0.5^{(c)}$ & $-0.644^{+0.053}_{-0.065}$ & $0.381^{+0.526}_{-0.196}$ & $529.26 \,(0.975)$ & $0.072^{+0.108}_{-0.054}$ & $-0.820^{+0.066}_{-0.151}$ & $-1.010^{+0.047}_{-0.057}$ & $-0.554^{+0.299}_{-0.122}$ \\
\hline \hline
\end{tabular}}
\end{minipage}
\end{table*}}

{\renewcommand{\arraystretch}{1.5}
\begin{table*}
\begin{minipage}{\textwidth}
\caption{Results in the three dimensional parameter space
($q_{0}$, $j_{0}$, $s_{0}$) and y-redshift. Void rows correspond to
MCMC which do not pass the convergence test. Column~(1):
simulation identification in the text with the used data
($S=$ SNeIa, $G=$ GRBs, $H=$ Hubble, $B=$ BAO; the index $(c)$ refers to the redshift-cut sub-sample,
with $z<1$) and the applied priors
(we indicate the redshift range where positiveness of quantities described
in \S~(\ref{sec:preliminary}) is applied). Columns~(2)~-~(3)~-~(4):
best fit estimations of cosmographic parameters with $1\sigma$
confidence level. Column~(5): chi square and reduced chi square
values. Columns~(6)~-~(7)~-~(8)): best fit estimations of dark
energy (CPL) parameters with $1\sigma$ confidence level for the dynamical
dark energy model as described in \S~(\ref{sec:CPL}).}\label{lab:3P_Y} \resizebox*{\textwidth}{!}{
\begin{tabular}{cccccccc}
\hline
\multicolumn{1}{c}{$\mathrm{Data}$} & \multicolumn{4}{c}{$\mathrm{Best \; fit \; parameters}$} & \multicolumn{3}{c}{$\mathrm{DDE}$} \\
\hline \hline
S & $q_{0}$ & $j_{0}$ & $s_{0}$ & $\chi^{2} \, (\chi^{2}/d.o.f.)$ & $\Omega_{m}$ & $w_{0}$  & $w_{a} $ \\
\hline
$0<y<1$         & $-0.498^{+0.048}_{-0.074}$ & $0.105^{+0.452}_{-0.129}$ & $-0.053^{+0.130}_{-0.464}$ & $542.51 \,(0.981)$ & $0.066^{+0.121}_{-0.046}$ & $-0.711^{+0.054}_{-0.156}$ & $0.007^{+0.073}_{-0.140}$ \\
$0<y<0.584$     & $-0.494^{+0.047}_{-0.066}$ & $0.091^{+0.373}_{-0.128}$ & $-0.080^{+0.116}_{-0.398}$ & $542.55 \,(0.981)$ & $0.059^{+0.114}_{-0.040}$ & $-0.703^{+0.049}_{-0.141}$ & $0.002^{+0.067}_{-0.126}$ \\
$0<y<1^{(c)}$   & $-0.502^{+0.054}_{-0.082}$ & $0.134^{+0.540}_{-0.160}$ & $-0.056^{+0.141}_{-0.479}$ & $525.63 \,(0.986)$ & $0.073^{+0.146}_{-0.052}$ & $-0.720^{+0.062}_{-0.192}$ & $0.007^{+0.074}_{-0.181}$ \\
$0<y<0.5^{(c)}$ & $-0.498^{+0.054}_{-0.074}$ & $0.093^{+0.439}_{-0.150}$ & $-0.068^{+0.155}_{-0.680}$ & $525.68 \,(0.986)$ & $0.069^{+0.146}_{-0.049}$ & $-0.714^{+0.058}_{-0.186}$ & $0.0009^{+0.0788}_{-0.2164}$ \\
\hline \hline
S-G  & $q_{0}$ & $j_{0}$ & $s_{0}$ & $\chi^{2} \, (\chi^{2}/d.o.f.)$ & $\Omega_{m}$ & $w_{0}$  & $w_{a}$ \\
\hline
$0<y<1$         & $-0.501^{+0.050}_{-0.079}$ & $0.130^{+0.520}_{-0.157}$ & $-0.063^{+0.151}_{-0.511}$ & $549.00 \,(0.883)$ & $0.079^{+0.121}_{-0.056}$ & $-0.724^{+0.062}_{-0.167}$ & $0.006^{+0.071}_{-0.168}$ \\
$0<y<0.871$     & $-$ & $-$ & $-$ & $-$ & $-$ & $-$ & $-$ \\
$0<y<1^{(c)}$   & $-0.501^{+0.051}_{-0.073}$ & $0.146^{+0.483}_{-0.159}$ & $-0.038^{+0.215}_{-0.335}$ & $528.94 \,(0.958)$ & $0.070^{+0.112}_{-0.049}$ & $-0.718^{+0.057}_{-0.148}$ & $0.014^{+0.071}_{-0.124}$ \\
$0<y<0.5^{(c)}$ & $-0.496^{+0.053}_{-0.079}$ & $0.098^{+0.517}_{-0.135}$ & $-0.039^{+0.191}_{-0.510}$ & $529.00 \,(0.958)$ & $0.068^{+0.143}_{-0.050}$ & $-0.711^{+0.057}_{-0.195}$ & $0.003^{+0.073}_{-0.176}$ \\
\hline \hline
S-H  & $q_{0}$ & $j_{0}$ & $s_{0}$ & $\chi^{2} \, (\chi^{2}/d.o.f.)$ & $\Omega_{m}$ & $w_{0}$  & $w_{a} $ \\
\hline
$0<y<1$         & $-$ & $-$ & $-$ & $-$ & $-$ & $-$ & $-$ \\
$0<y<0.638$     & $-$ & $-$ & $-$ & $-$ & $-$ & $-$ & $-$ \\
$0<y<1^{(c)}$   & $-0.497^{+0.051}_{-0.073}$ & $0.102^{+0.480}_{-0.134}$ & $-0.076^{+0.138}_{-0.420}$ & $527.28 \,(0.975)$ & $0.067^{+0.126}_{-0.047}$ & $-0.711^{+0.055}_{-0.157}$ & $0.002^{+0.069}_{-0.141}$ \\
$0<y<0.5^{(c)}$ & $-0.497^{+0.053}_{-0.076}$ & $0.105^{+0.497}_{-0.150}$ & $-0.066^{+0.151}_{-0.479}$ & $527.29 \,(0.975)$ & $0.070^{+0.132}_{-0.050}$ & $-0.714^{+0.058}_{-0.171}$ & $0.002^{+0.077}_{-0.162}$ \\
\hline \hline
S-G-H  & $q_{0}$ & $j_{0}$ & $s_{0}$ & $\chi^{2} \, (\chi^{2}/d.o.f.)$ & $\Omega_{m}$ & $w_{0}$  & $w_{a} $ \\
\hline
$0<y<1$         & $-$ & $-$ & $-$ & $-$ & $-$ & $-$ & $-$ \\
$0<y<0.871$     & $-$ & $-$ & $-$ & $-$ & $-$ & $-$ & $-$ \\
$0<y<1^{(c)}$   & $-0.493^{+0.048}_{-0.067}$ & $0.079^{+0.414}_{-0.118}$ & $-0.073^{+0.115}_{-0.448}$ & $530.65 \,(0.948)$ & $0.058^{+0.119}_{-0.041}$ & $-0.701^{+0.050}_{-0.140}$ & $-0.001^{+0.070}_{-0.131}$ \\
$0<y<0.5^{(c)}$ & $-0.498^{+0.052}_{-0.076}$ & $0.128^{+0.479}_{-0.157}$ & $-0.051^{+0.119}_{-0.389}$ & $530.58 \,(0.947)$ & $0.066^{+0.126}_{-0.047}$ & $-0.710^{+0.056}_{-0.163}$ & $0.012^{+0.071}_{-0.146}$ \\
\hline \hline
S-G-H-B  & $q_{0}$ & $j_{0}$ & $s_{0}$ & $\chi^{2} \, (\chi^{2}/d.o.f.)$ & $\Omega_{m}$ & $w_{0}$  & $w_{a} $ \\
\hline
$0<y<1$         & $-$ & $-$ & $-$ & $-$ & $-$ & $-$ & $-$ \\
$0<y<0.871$     & $-0.675^{+0.124}_{-0.115}$ & $1.244^{+0.842}_{-1.248}$ & $-0.098^{+0.219}_{-15.996}$& $572.41 \,(0.901)$ & $0.364^{+0.087}_{-0.117}$ & $-1.272^{+0.312}_{-0.205}$ & $-0.326^{+0.352}_{-2.416}$ \\
$0<y<1^{(c)}$   & $-0.497^{+0.050}_{-0.069}$ & $0.100^{+0.406}_{-0.131}$ & $-0.058^{+0.143}_{-0.457}$ & $531.52 \,(0.947)$ & $0.065^{+0.123}_{-0.046}$ & $-0.714^{+0.058}_{-0.151}$ & $0.004^{+0.071}_{-0.154}$ \\
$0<y<0.5^{(c)}$ & $-0.503^{+0.052}_{-0.070}$ & $0.134^{+0.455}_{-0.158}$ & $-0.041^{+0.151}_{-0.423}$ & $531.49 \,(0.947)$ & $0.067^{+0.107}_{-0.047}$ & $-0.716^{+0.056}_{-0.129}$ & $0.016^{+0.074}_{-0.144}$ \\
\hline \hline
S-H-B  & $q_{0}$ & $j_{0}$ & $s_{0}$ & $\chi^{2} \, (\chi^{2}/d.o.f.)$ & $\Omega_{m}$ & $w_{0}$  & $w_{a} $ \\
\hline
$0<y<1$         & $-$ & $-$ & $-$ & $-$ & $-$ & $-$ & $-$ \\
$0<y<0.638$     & $-$ & $-$ & $-$ & $-$ & $-$ & $-$ & $-$ \\
$0<y<1^{(c)}$   & $-0.506^{+0.055}_{-0.078}$ & $0.132^{+0.517}_{-0.169}$ & $-0.105^{+0.162}_{-0.787}$ & $528.17 \,(0.974)$ & $0.084^{+0.136}_{-0.061}$ & $-0.730^{+0.069}_{-0.183}$ & $-0.006^{+0.086}_{-0.216}$ \\
$0<y<0.5^{(c)}$ & $-0.509^{+0.054}_{-0.085}$ & $0.138^{+0.578}_{-0.170}$ & $-0.058^{+0.148}_{-0.687}$ & $528.21 \,(0.975)$ & $0.086^{+0.148}_{-0.060}$ & $-0.733^{+0.068}_{-0.205}$ & $-0.00004^{+0.08549}_{-0.23504}$ \\
\hline \hline
\end{tabular}}
\end{minipage}
\end{table*}}

\begin{figure*}
\centering
\includegraphics[width=84mm]{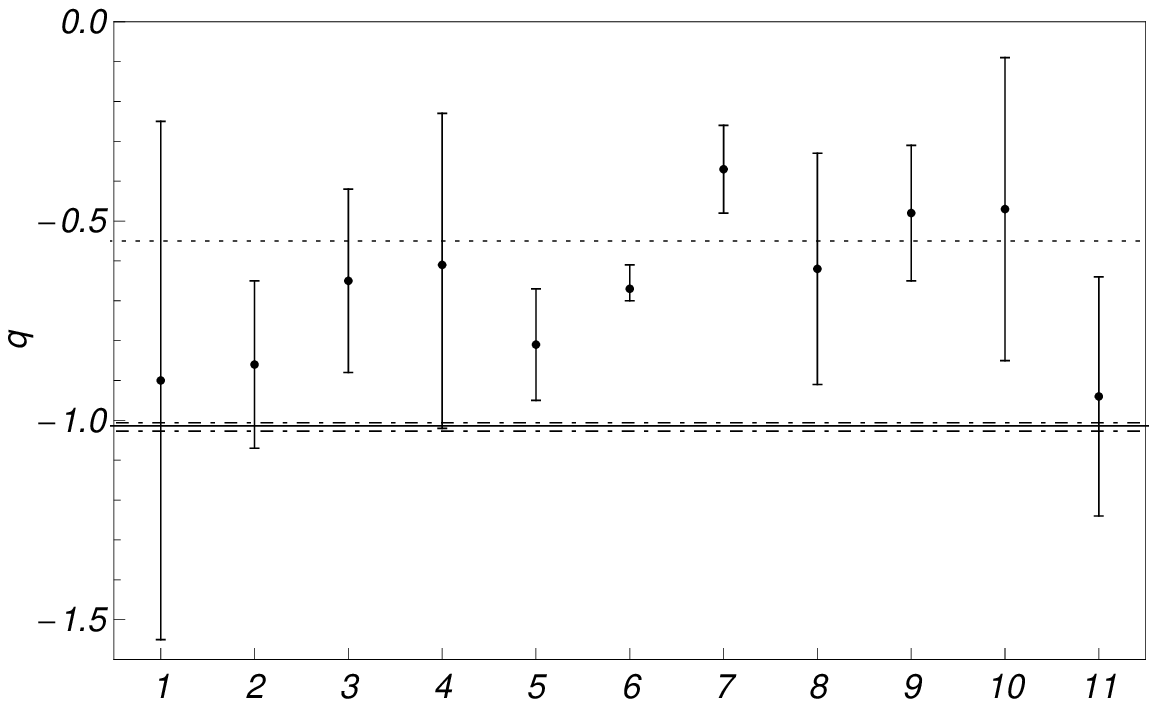}
\includegraphics[width=84mm]{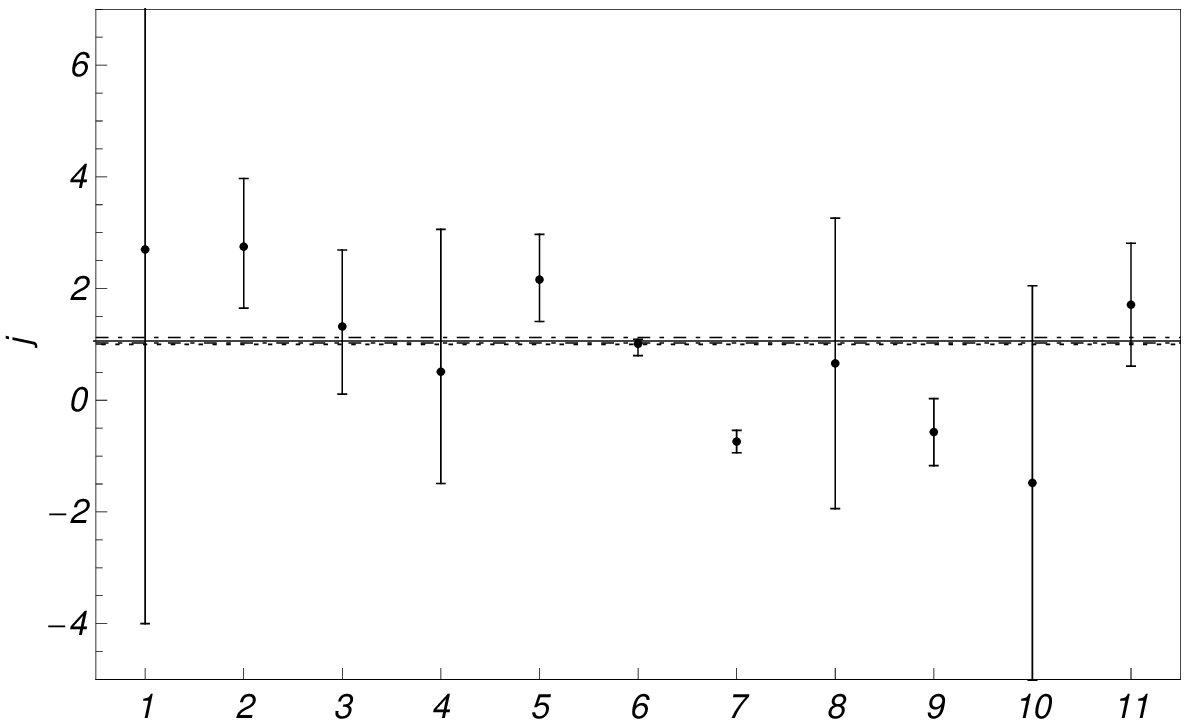}
\includegraphics[width=84mm]{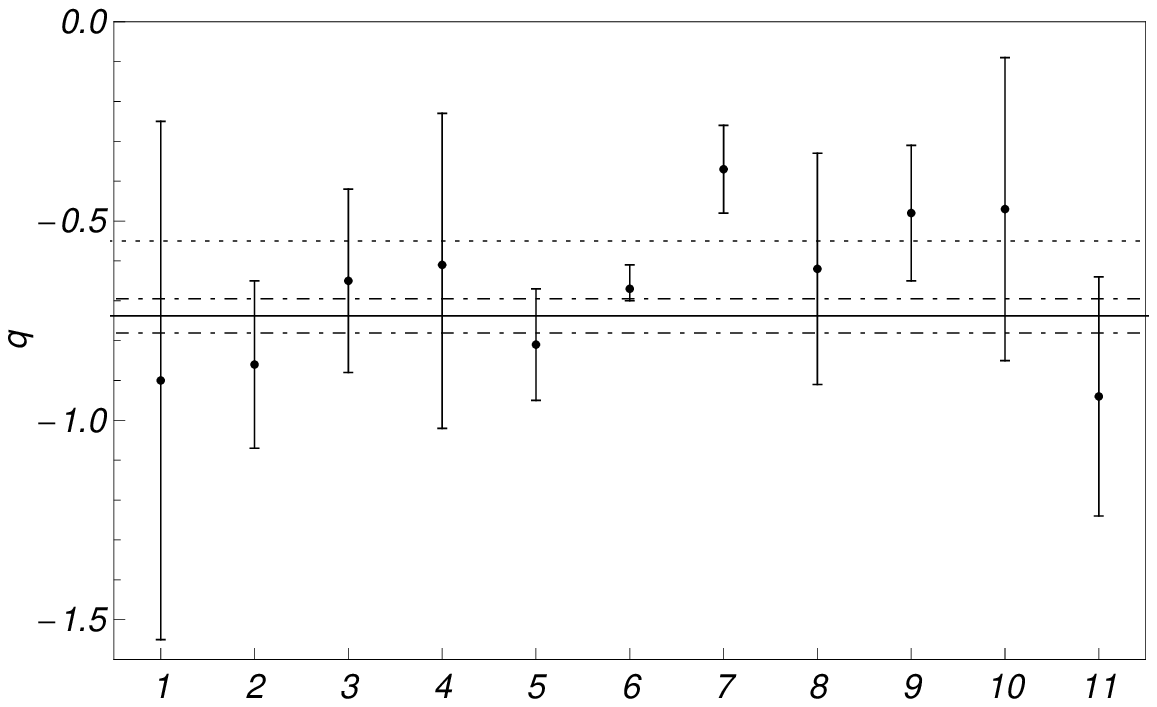}
\includegraphics[width=84mm]{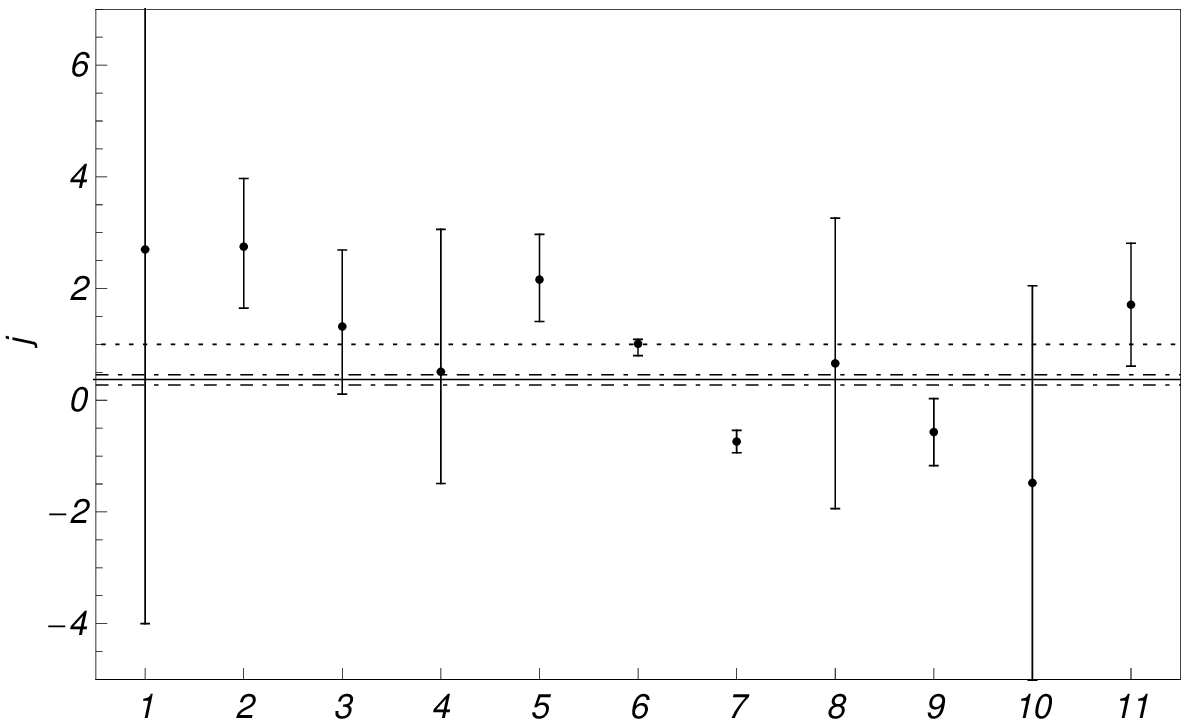}
\includegraphics[width=84mm]{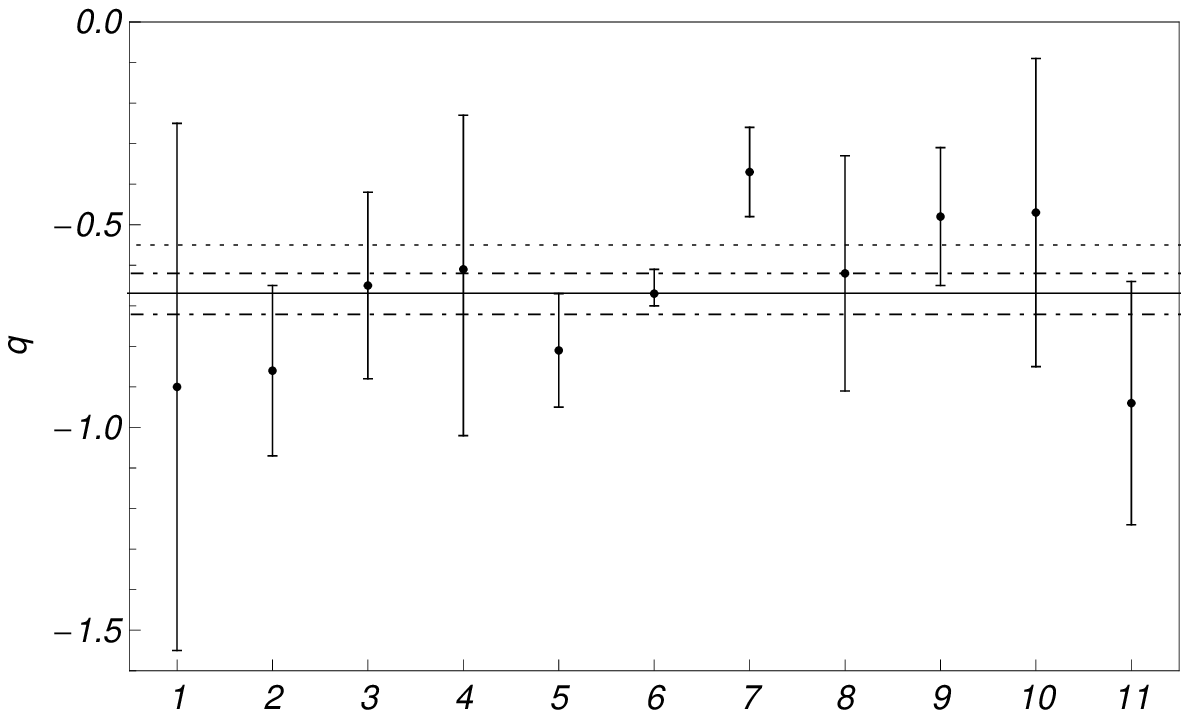}
\includegraphics[width=84mm]{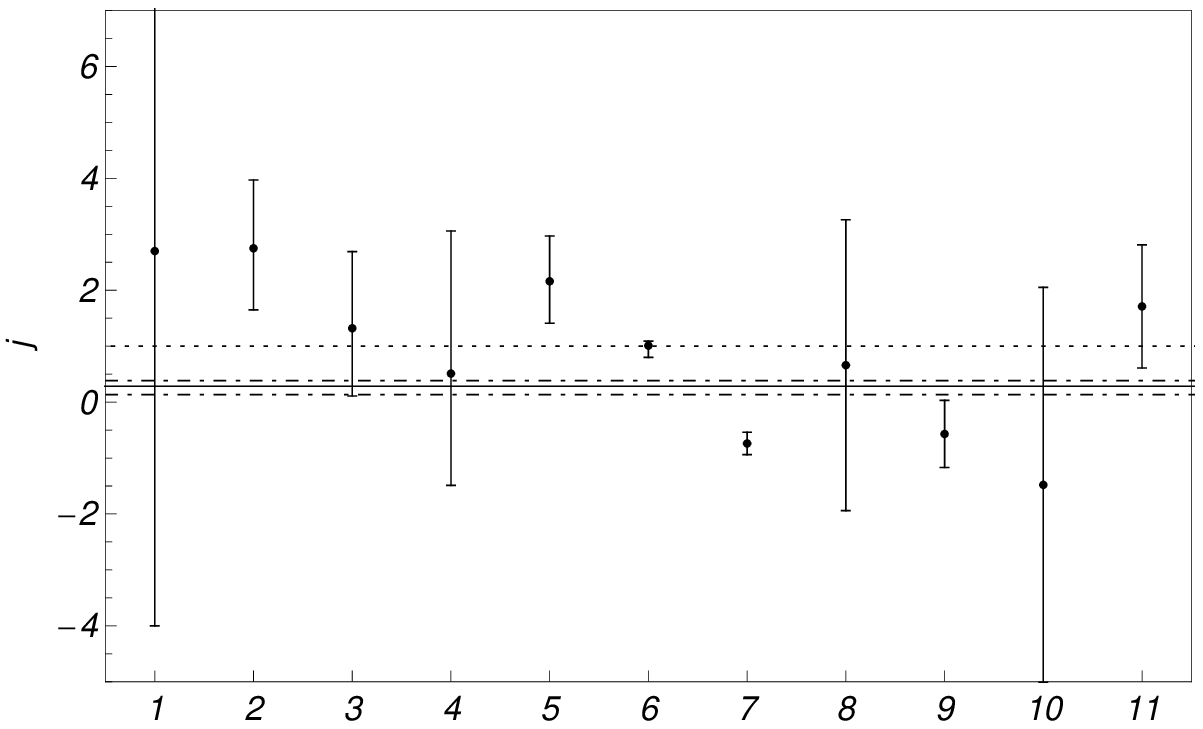}
\includegraphics[width=84mm]{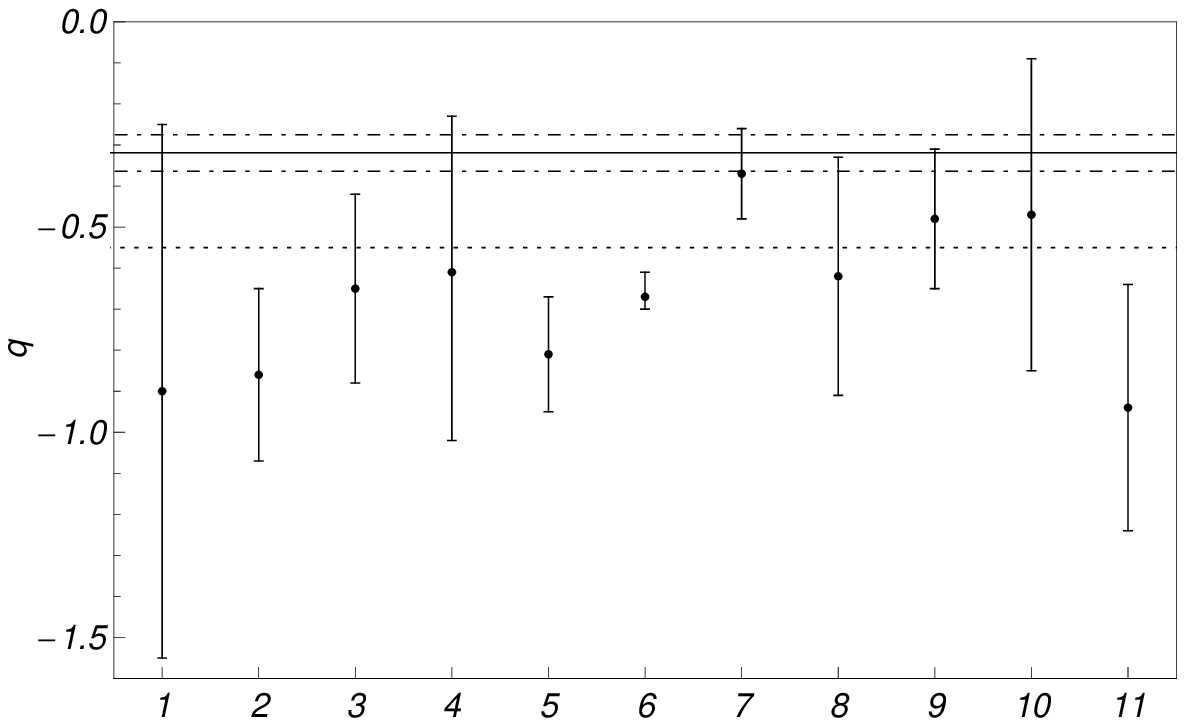}
\includegraphics[width=84mm]{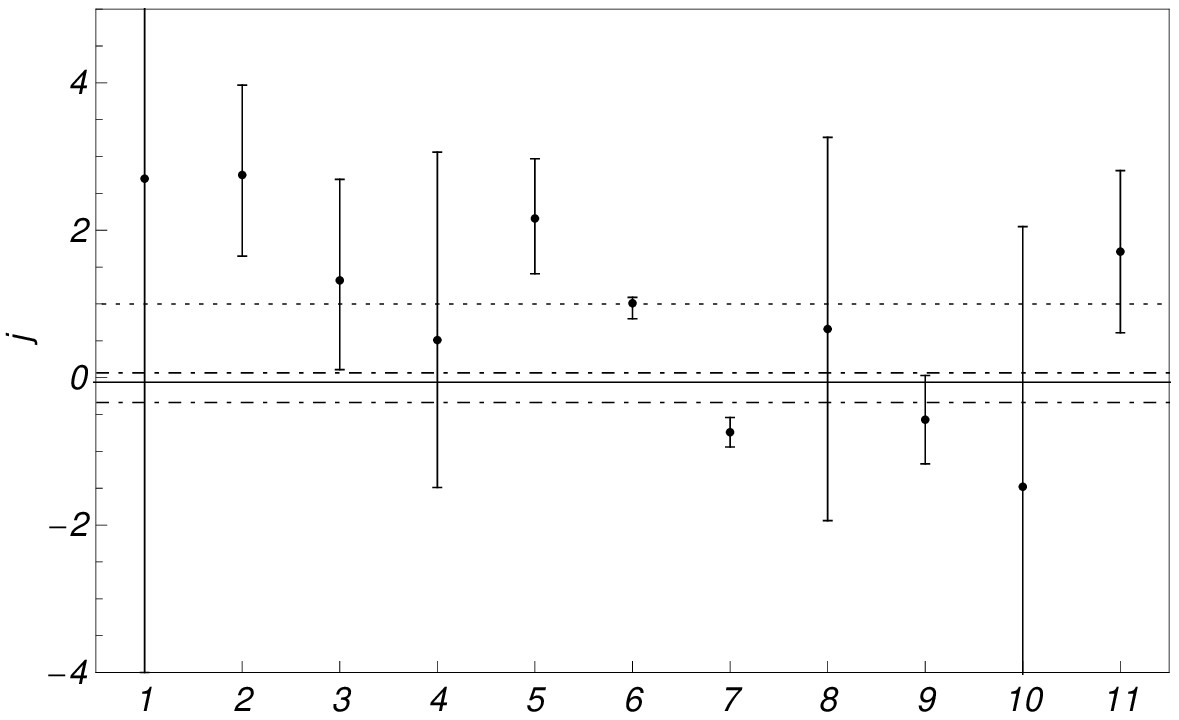}
\caption{Comparing deceleration and jerk
values from the literature with results from: (\textit{First panel.})
SNeIa only, two dimensional cosmography and prior on z-redshift range $z>0$;
(\textit{Second panel.}) SNeIa only, two
dimensional cosmography and prior on z-redshift range $z<1.4$;
(\textit{Third panel.}) cut SNeIa only, two
dimensional cosmography and prior on z-redshift range $z<1$;
(\textit{Fourth panel.}) cut SNeIa only, three
dimensional cosmography and prior on z-redshift range $z>0$.
References in numerical order as they
appear in the figure: (1)~-~John~(2004); (2)~-~Rapetti~(2006),
Gold SN subsample; (3)~-~Rapetti~(2006), Legacy SN subsample;
(4)~-~Rapetti~(2006), X-ray clusters subsample;
(5)~-~Rapetti~(2006), all subsamples; (6)~-~Poplawski~(2006);
(7)~-~Cattoen~(2007), Gold SN in z-redshift; (8)~-~Cattoen~(2007),
Gold SN in y-redshift; (9)~-~Cattoen~(2007), Legacy SN in
z-redshift; (10)~-~Cattoen~(2007), Legacy SN in y-redshift;
(11)~-~Capozziello,~Izzo~(2008). Solid horizontal line is our best
fit result; dotdashed horizontal lines show its $1\sigma$
confidence level; dotted horizontal line shows $\Lambda$CDM
expectation.} \label{fig:Q_J_1}
\end{figure*}

\begin{figure*}
\centering
\includegraphics[width=84mm]{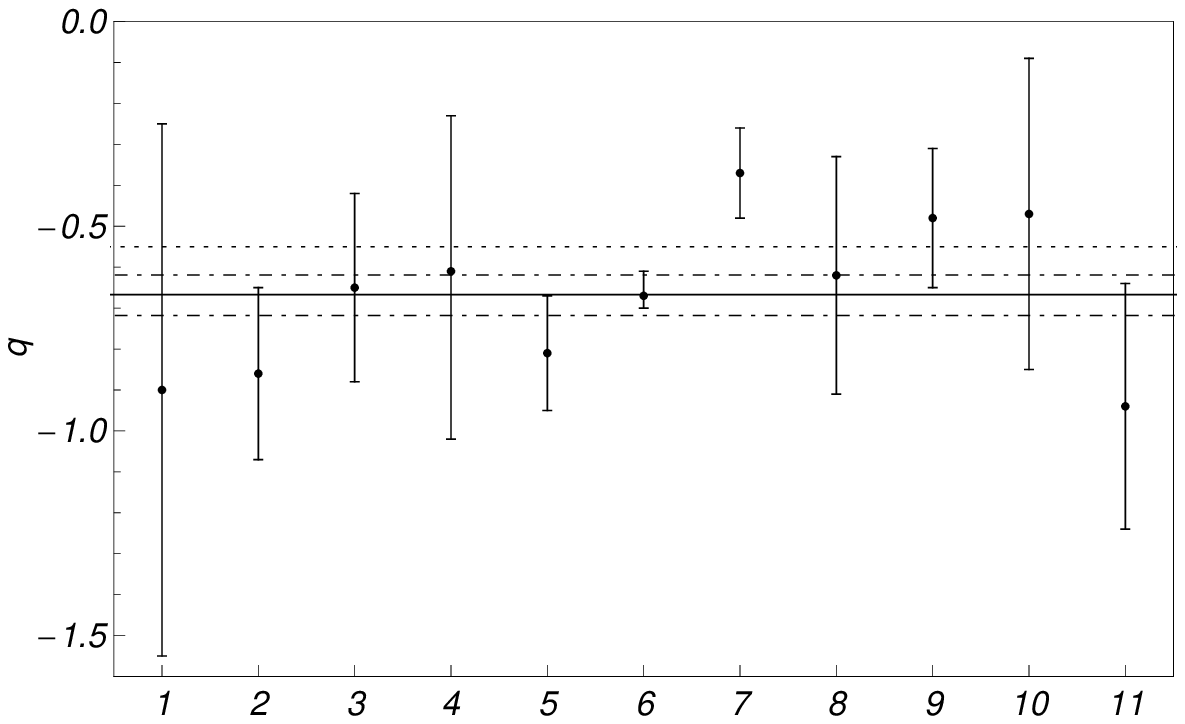}
\includegraphics[width=84mm]{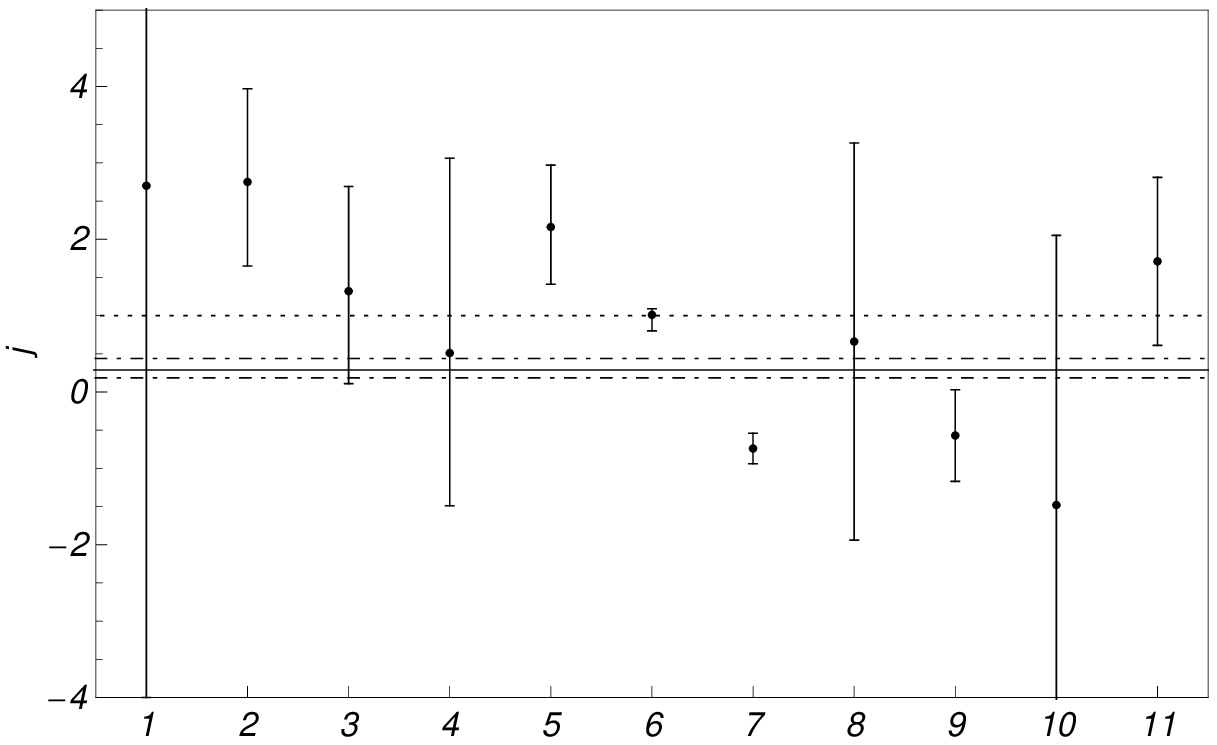}
\includegraphics[width=84mm]{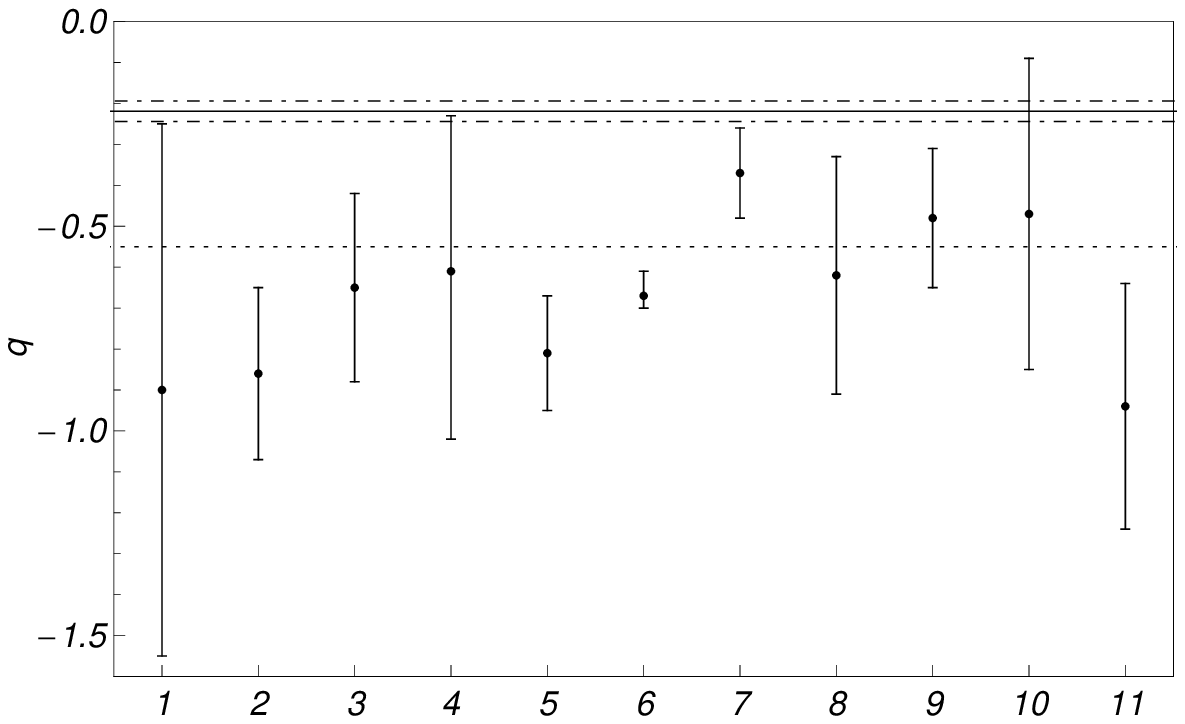}
\includegraphics[width=84mm]{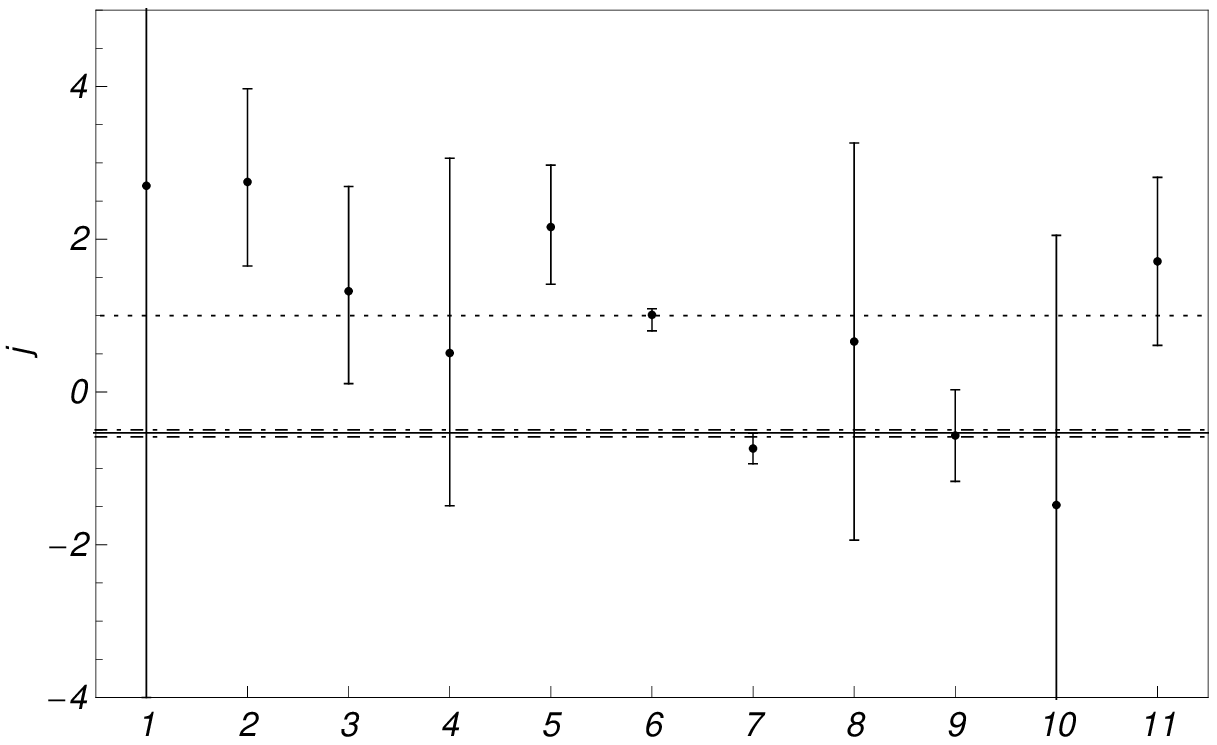}
\includegraphics[width=84mm]{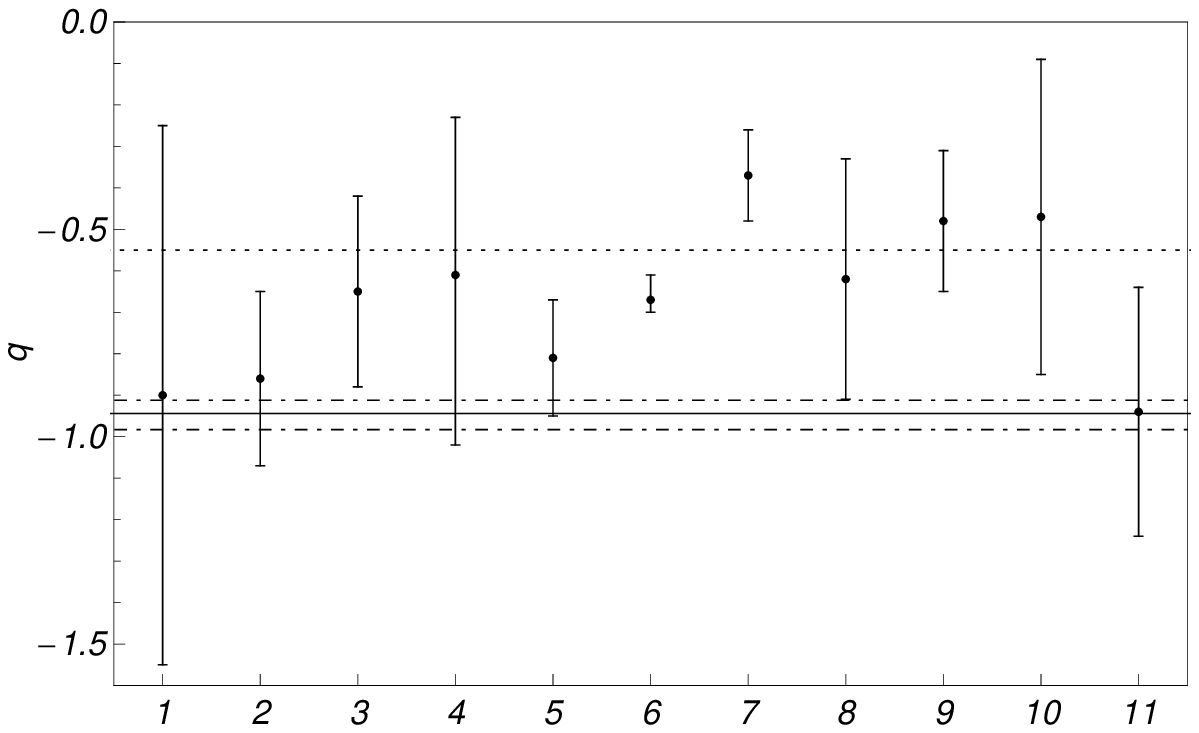}
\includegraphics[width=84mm]{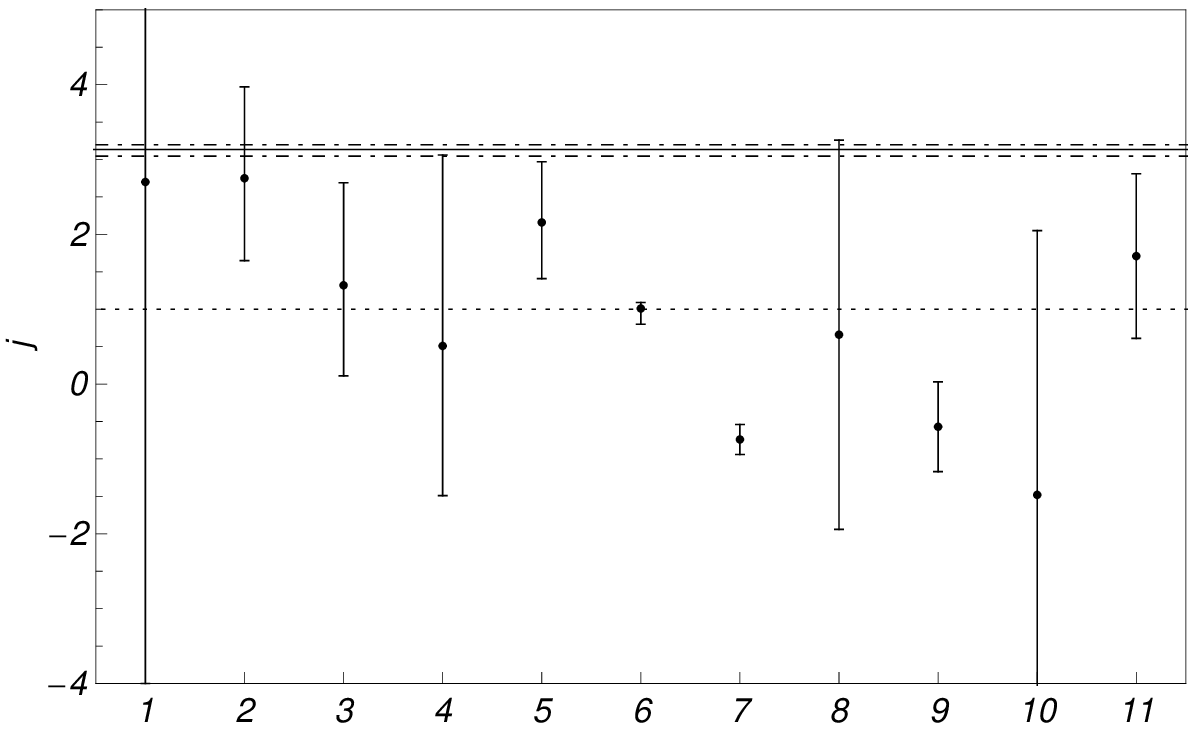}
\caption{Comparing deceleration and jerk
values from the literature with results from: (\textit{First panel.})
all cut data, two dimensional cosmography and prior on z-redshift range $z<1$;
(\textit{Second panel.}) SNeIa and Hubble data, three
dimensional cosmography and prior on z-redshift range $z>0$;
(\textit{Third panel.}) SNeIa and Hubble data, three
dimensional cosmography and prior on z-redshift range $z<1.76$.
References in numerical order as the appear in the figure: (1)~-~John~(2004);
(2)~-~Rapetti~(2006), Gold SN subsample; (3)~-~Rapetti~(2006),
Legacy SN subsample; (4)~-~Rapetti~(2006), X-ray clusters
subsample; (5)~-~Rapetti~(2006), all subsamples;
(6)~-~Poplawski~(2006); (7)~-~Cattoen~(2007), Gold SN in
z-redshift; (8)~-~Cattoen~(2007), Gold SN in y-redshift;
(9)~-~Cattoen~(2007), Legacy SN in z-redshift;
(10)~-~Cattoen~(2007), Legacy SN in y-redshift;
(11)~-~Capozziello,~Izzo~(2008). Solid horizontal line is our best
fit result; dotdashed horizontal lines show its $1\sigma$
confidence level; dotted horizontal line shows $\Lambda$CDM
expectation.} \label{fig:Q_J_2}
\end{figure*}

\begin{figure*}
\centering
\includegraphics[width=84mm]{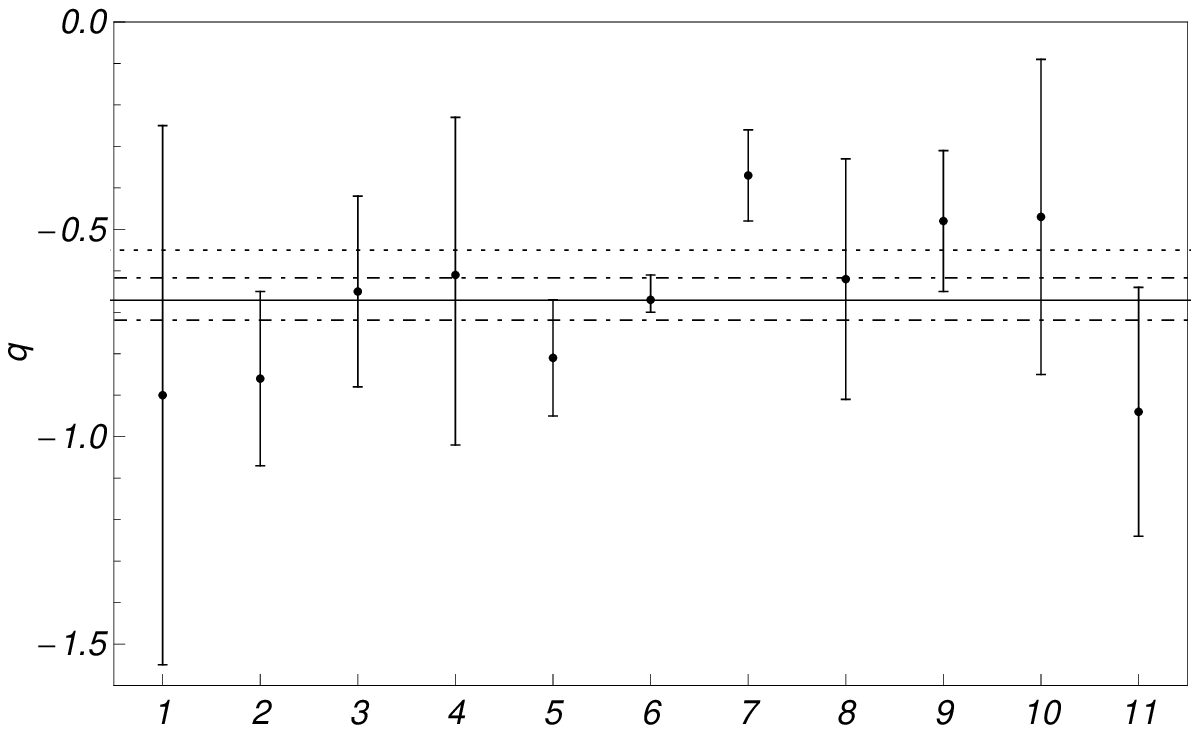}
\includegraphics[width=84mm]{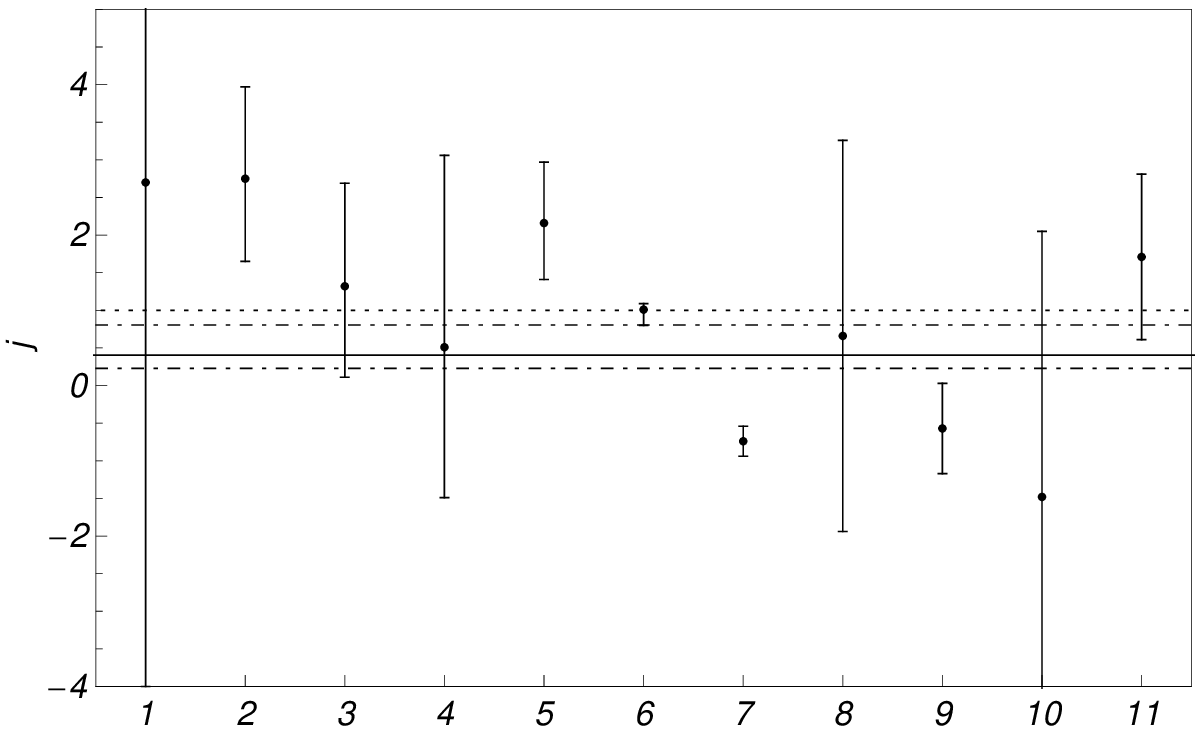}
\includegraphics[width=84mm]{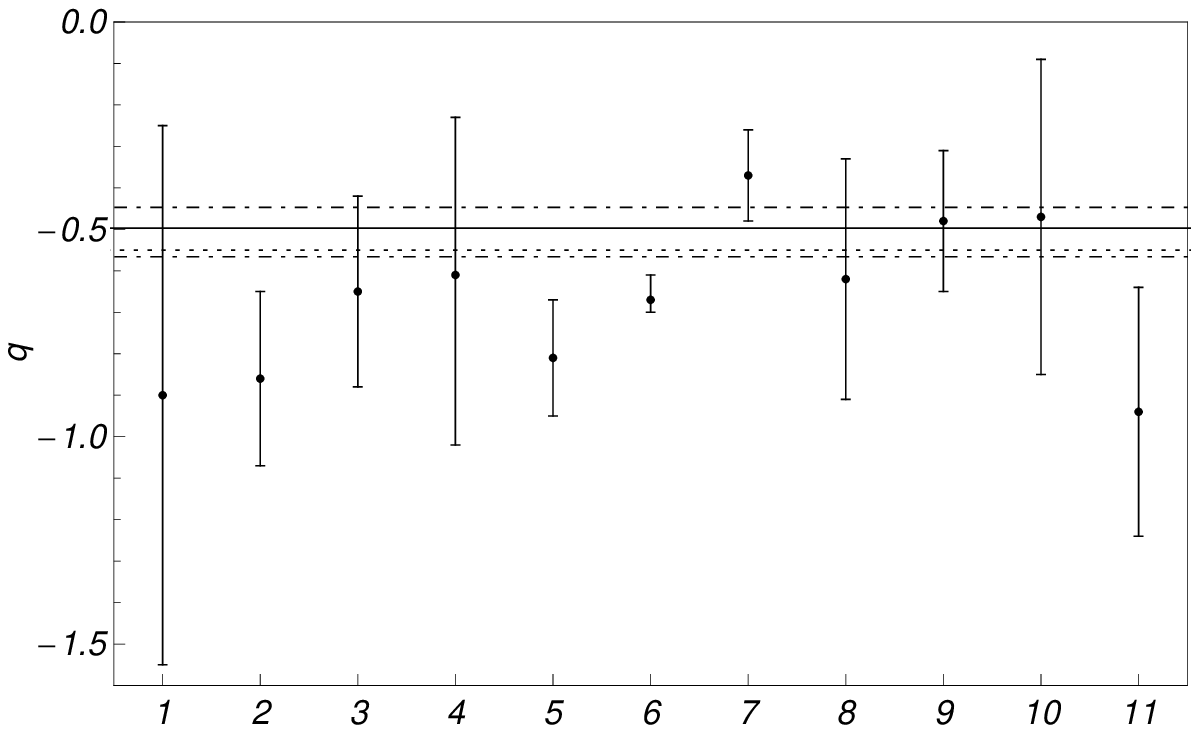}
\includegraphics[width=84mm]{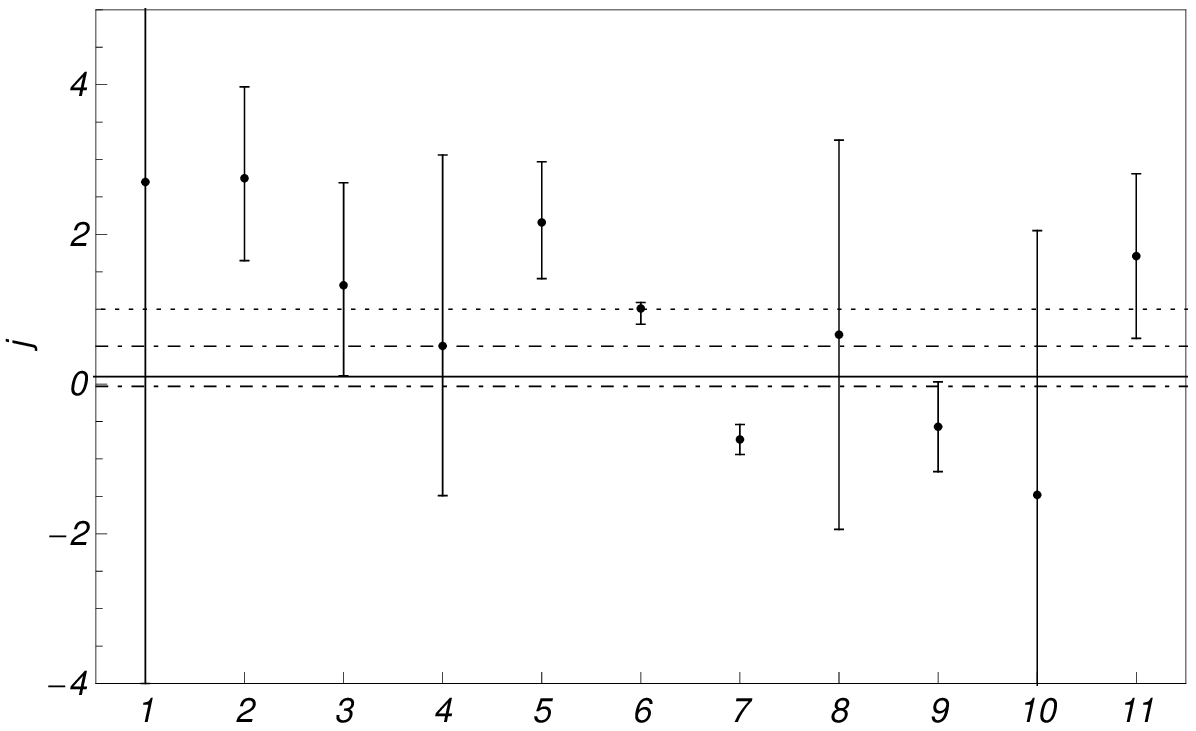}
\includegraphics[width=84mm]{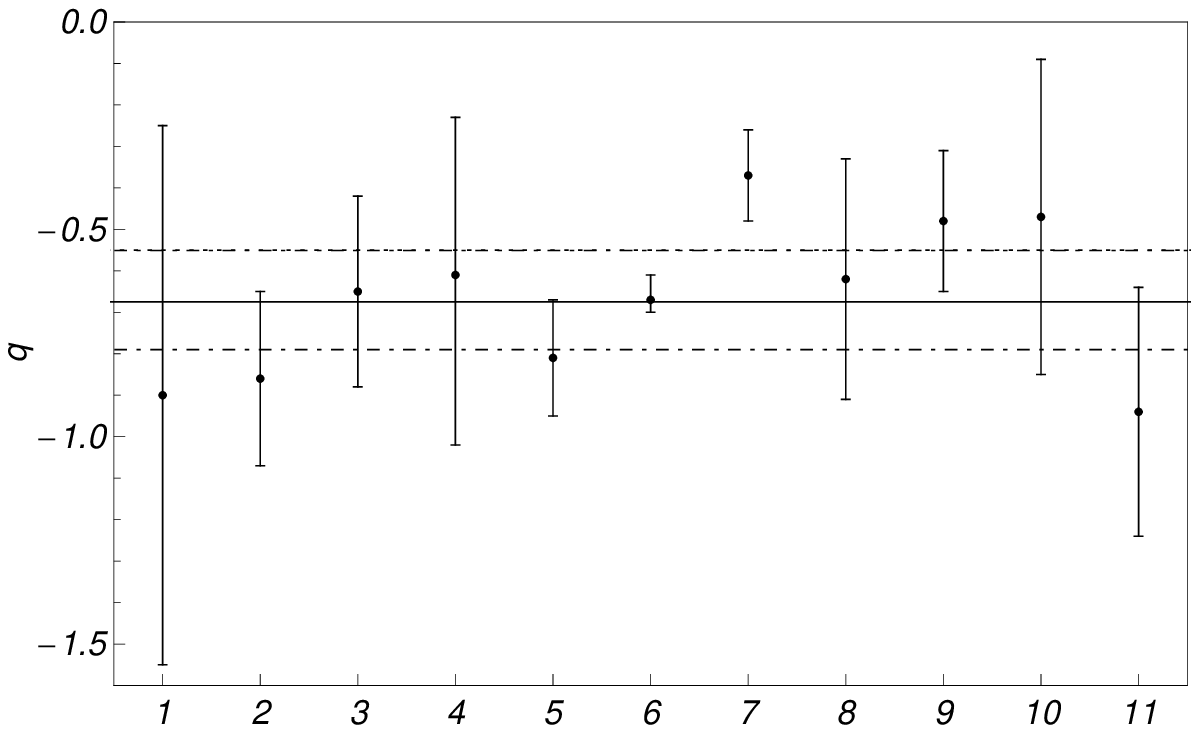}
\includegraphics[width=84mm]{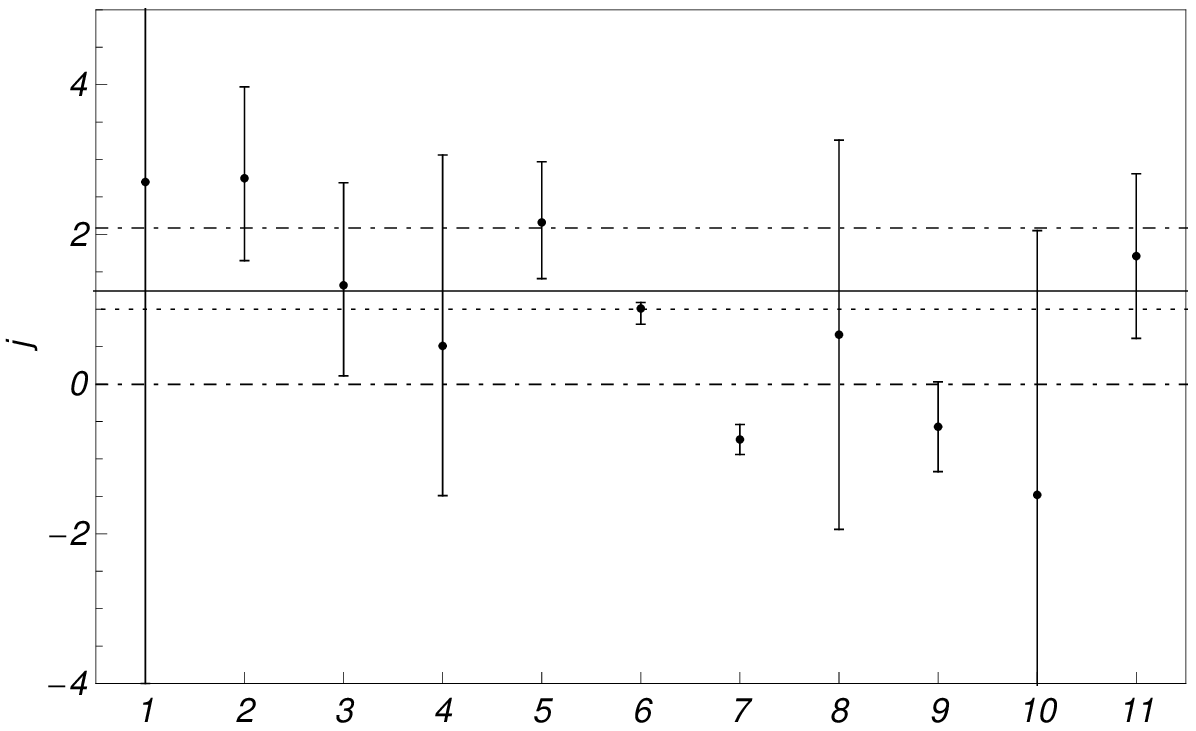}
\caption{Comparing deceleration and jerk
values from the literature with results from: (\textit{First panel.})
SNeIa data, two dimensional cosmography and prior on y-redshift range $0<y<1$;
(\textit{Second panel.}) all data, three
dimensional cosmography and prior on y-redshift range $0<y<1$;
(\textit{Third panel.}) all data, three
dimensional cosmography and prior on y-redshift range $0<y<0.871$.
References in numerical order as the
appear in the figure: (1)~-~John~(2004); (2)~-~Rapetti~(2006),
Gold SN subsample; (3)~-~Rapetti~(2006), Legacy SN subsample;
(4)~-~Rapetti~(2006), X-ray clusters subsample;
(5)~-~Rapetti~(2006), all subsamples; (6)~-~Poplawski~(2006);
(7)~-~Cattoen~(2007), Gold SN in z-redshift; (8)~-~Cattoen~(2007),
Gold SN in y-redshift; (9)~-~Cattoen~(2007), Legacy SN in
z-redshift; (10)~-~Cattoen~(2007), Legacy SN in y-redshift;
(11)~-~Capozziello,~Izzo~(2008). Solid horizontal line is our best
fit result; dotdashed horizontal lines show its $1\sigma$
confidence level; dotted horizontal line shows $\Lambda$CDM
expectation.} \label{fig:Q_J_3}
\end{figure*}


\begin{thebibliography}{99}

\bibitem[Catto$\rm{\ddot{e}}$n, 2007]{Cattoen07} Catto$\rm{\ddot{e}}$n, C., Visser, M., 2007, Class. Quant. Grav., 24, 5985
\bibitem[Catto$\rm{\ddot{e}}$n, 2007b]{Cattoen07b} Catto$\rm{\ddot{e}}$n, C., Visser, M., 2007, arXiv:gr-qc/0703122
\bibitem[Visser, 2004]{Visser04} Visser, M., 2004, Class. Quant. Grav., 21, 2603
\bibitem[John, 2005]{John05} John, M.~V., 2005, ApJ, 630, 667
\bibitem[Poplawski, 2006]{Poplawski06} Poplawski, N.~J., 2006, Phys. Lett. B, 640, 135
\bibitem[Poplawski, 2007]{Poplawski07} Poplawski, N.~J., 2007, Class. Quant. Grav., 24, 3013
\bibitem[Salzano, 2009]{Salzano09} Capozziello S., Cardone V.F.,  Salzano V., 2008,  Phys. Rev. D 78, 063504.
\bibitem[Mariam, 2010]{Mariam10}  Bouhmadi-Lopez, M., Capozziello S. ,  Cardone V.F., 2010, Phys.Rev.D 82, 103526.
\bibitem[Mariusz, 2010]{Mariusz10} Balcerzak A.,  Dabrowski M. P., 2010, Phys.\ Rev.\  D  81,  123527.
\bibitem[Capozziello, 2008]{Capozziello08} Capozziello, S., Izzo, L., 2008, A\&A, 490, 31
\bibitem[Izzo, 2009]{Izzo09} Izzo L.,  Capozziello S., Covone G.,  Capaccioli M., 2009, Astron. Astroph.  508, 63 (2009).
\bibitem[Dunajski, 2008]{Dunajski08} Dunajski, M., Gibbons, G., 2008, Class. Quant. Grav., 25, 235012


\bibitem[Chevallier, 2001]{Chevallier01} Chevallier, M., Polarski, D., Int. J. Mod. Phys. D., 2001, 10, 213
\bibitem[Linder, 2003]{Linder03} Linder, E.V., Phys. Rev. Lett., 2003, 90, 091301
\bibitem[Komatsu, 2010]{Komatsu10} Komatsu, E., et al., submitted to ApJS, arXiv:1001.4538


\bibitem[Stern, 2010]{Stern09} Stern, D., Jimenez, R., Verde, L., Kamionkowski, M., Stanford, A., 2010, JCAP, 1, 2, 8
\bibitem[Simon, 2005]{Simon05} Simon, J., Verde, L., Jimenez, R., 2005, Phys. Rev. D, 71, 123001
\bibitem[Morsell, 2011]{Mortsell11} M$\ddot{\mathrm{o}}$rtsell, E., J$\ddot{\mathrm{o}}$hnsson, J., arXiv:1102.4485
\bibitem[Jimenez, 2002]{Jimenez02} Jimenez, R., Loeb, A., 2002, ApJ, 573, 37
\bibitem[Jimenez, 2003]{Jimenez03} Jimenez, R., Verde, L., Treu, T., Stern, D., 2003, ApJ, 593, 622
\bibitem[Stern, 2010b]{Stern09B} Stern, D., Jimenez, R., Verde, L., Stanford, A., Kamionkowski, M., 2010, ApJS, 188, 280-289
\bibitem[Stern, 2000]{Stern00} Stern, D., 2000, arXiv:astro-ph/0012146
\bibitem[Le Fevre, 2005]{LeFevre05} Le F\`{e}vre, O., et al., 2005, A\&A, 439, 845
\bibitem[Jimenez, 2004]{Jimenez04} Jimenez, R., MacDonald, J., Dunlop, J.S., Padoan, P., Peacock, J.A., 2004, MNRAS, 349, 240
\bibitem[Riess, 2009]{Riess09} Riess, Adam G. and others , 2009, ApJ, 699, 539-563
\bibitem[Berg, 2004]{Berg} Berg, B.A., {\it Markov Chain Monte Carlo Simulations and Their Statistical Analysis}, World
    Scientific Publishing Co. Pte. Ltd., Singapore
\bibitem[MacKay, 2002]{MacKay} MacKay, D.~J.~C., 2003, {\it Information Theory, Inference, and Learning Algorithms}, Cambridge
    University Press
\bibitem[Neal, 1993]{Neal} Neal, R.M., {\it 25 September 1993, Technical Report CRG-TR-93-1}, Department of Computer Science,
    University of Toronto
\bibitem[Dunkley, 2005]{Dunkley05} Dunkley, J., Bucher, M., Ferreira, P.~G., Moodley, K., Skordis K., 2005, MNRAS, 356, 925


\bibitem[Amanullah, 2010]{Amanullah10} Amanullah, R., et al., 2010, ApJ, 716, 712-738
\bibitem[Kowalski, 2008]{Kowalski08} Kowalski, M., Rubin, D., Aldering, G. et al., 2008, ApJ, 686, 749
\bibitem[Amanullah, 2008]{Amanullah08} Amanullah, R., et al., 2008, A\&A, 486, 375
\bibitem[Hicken, 2009a]{Hicken09a} Hicken, M., et al., 2009a, ApJ, 700, 331-357
\bibitem[Holtzman, 2008]{Holtzman08} Holtzman, J.A., et al., 2008, AJ, 136, 2306


\bibitem[Cardone, 2009]{Cardone09} Cardone, V. F., Capozziello, S., Dainotti, M.G., 2009, MNRAS, 440, 775
\bibitem[Schaefer, 2007]{Schaefer07} Schaefer, B.E., 2007, ApJ, 660, 16


\bibitem[Percival, 2010]{Percival10} Percival, W.J., et al., 2010, MNRAS, 401, 2148


\bibitem[Rapetti, 2007]{Rapetti07} Rapetti, D., Allen, S.W., Amin, A., Blandford, R.D., 2007, MNRAS, 375, 1510
\bibitem[John, 2004]{John04} John, M.~V., 2004, ApJ, 614, 1
\bibitem[Bueno Sanchez et al., 2009]{BuenoSanchez09} Bueno Sanchez, J.C., Nesseris, S., Perivolaropoulos, L., 2009, JCAP, 11,
    29
\bibitem[Perivolaropoulos, 2009]{Perivolaropoulos08} Perivolaropoulos, L., Shafieloo, A., 2009, Phys. Rev. D, 79, 123502
\bibitem[Vitagliano, 2010]{Vitagliano10} Vitagliano, V., Xia J.-Q., Liberati, S., Viel, M., 2010, JCAP, 03, 005

\end{thebibliography}
\end{document}